\documentclass[a4paper,12pt]{article}
\renewcommand{\theequation}{\arabic{section}.\arabic{equation}}
\newcommand{\eqreset}{\setcounter{equation}{0}}
\newtheorem{theorem}{Theorem}[section] 

\newtheorem{lemma}[theorem]{Lemma}

\begin{document}
\title{Analytic Bethe ansatz and functional relations 
related to tensor-like representations of type II 
 Lie superalgebras $B(r|s)$ and $D(r|s)$}
\author{Zengo Tsuboi
\\ 
Institute of Physics,                                   
 University of Tokyo,\\ Komaba 
  3-8-1, Meguro-ku, Tokyo 153-8902, Japan}
\date{}
\maketitle
\begin{abstract}
 An analytic Bethe ansatz is carried out related to 
 tensor-like representations of 
 the type II Lie superalgebras $B(r|s)=osp(2r+1|2s)$ 
 ($r \in {\bf Z}_{\ge 0}$, $s \in {\bf Z}_{\ge 1}$) 
 and $D(r|s)=osp(2r|2s)$ 
 ($r \in {\bf Z}_{\ge 2}$, $s \in {\bf Z}_{\ge 1}$). 
 We present eigenvalue formulae of transfer matrices 
 in dressed vacuum forms 
 labeled by Young (super) diagrams.   
A class of transfer matrix functional 
 relations ($T$-system) is discussed. 
 In particular for $B(0|s)=osp(1|2s)$ ($s\in {\bf Z}_{\ge 1}$) case, 
 a complete set of functional relations is proposed 
 by using duality among dressed vacuum forms.
\end{abstract}

\noindent Journal-ref: J. Phys. A: Math. Gen. 32 (1999) 7175-7206 \\ 
DOI: 10.1088/0305-4470/32/41/311 \\\\
Short title: Analytic Bethe ansatz \\ 
\newpage
\eqreset
\section{Introduction}
Solvable lattice models related to Lie superalgebras \cite{Ka} 
have been received much attentions
\cite{KulSk,Kul,BS,KuR,DFI,Sa,ZBG,MR94}. 
To construct eigenvalue formulae of transfer matrices 
for such models is an important problem in mathematical 
physics. To achieve this program,  
 Bethe ansatz has been often used. 
 
Nowadays, there are many literatures 
(See for example \cite{Kul,EK,EKS,FK,Ma,RM,PF,MR,ZB} 
and references therein.) 
on Bethe ansatz analysis 
for solvable lattice models related to Lie superalgebras. 
However most of them deal only with models related to 
 simple representations like fundamental ones. 
 Only a few people \cite{Ma,PF,ZB} tried to deal with 
 more complicated models such as fusion models \cite{KRS} by 
Bethe ansatz; while 
there was no systematic study on this subject. 

To break through such situations, we have 
recently executed \cite{T1,T2,T3,T4} an 
analytic Bethe ansatz \cite{R1,R2,BR,KS1,KOS,S2,KS2} systematically 
for the type I Lie superalgebras $sl(r+1|s+1),C(s)$ 
 cases. Namely, we have proposed a set of 
 dressed vacuum forms (DVFs) and 
 a class of functional relations for it. 
 The purpose of this paper
  is to develop our recent works
 to type II Lie superalgebras $B(r|s)=osp(2r+1|2s)$ 
 ($r \in {\bf Z}_{\ge 0}$, $s \in {\bf Z}_{\ge 1}$) 
 and $D(r|s)=osp(2r|2s)$ 
 ($r \in {\bf Z}_{\ge 2}$, $s \in {\bf Z}_{\ge 1}$) cases. 
 
 We can express \cite{Kul,MR} the 
 Bethe ansatz equation (BAE) (\ref{BAE}) 
 by using the representation theoretical data of 
  $B(r|s)$ 
 ($r \in {\bf Z}_{\ge 1}$, $s \in {\bf Z}_{\ge 1}$) 
 or $D(r|s)$ ($r \in {\bf Z}_{\ge 2}$, $s \in {\bf Z}_{\ge 1}$), 
 as long as we adopt the distinguished simple root system \cite{Ka}.  
On the other hand, $B(0|s)=osp(1|2s)$ is a peculiar one among 
 the Lie superalgebras. 
 In contrast to other Lie superalgebras, 
 its simple root system is unique.  
 Corresponding to this fact, BAEs 
 (\ref{BAE1})-(\ref{BAE4}) associated with the root system 
 of $B(0|s)$ will be also unique.  
 Peculiarity of these BAEs is that so far 
 a naive description in terms of 
 the simple root system does not exist  
 for the $s$-th BAEs (\ref{BAE3}) and (\ref{BAE4}), 
 which correspond to the odd root $\alpha_{s}$
  with  $(\alpha_{s}|\alpha_{s})\ne 0$ (cf. \cite{Kul,MR}).  
 
 We assume, as our starting point, above-mentioned 
 BAEs (\ref{BAE})-(\ref{BAE4}) for 
 $B(r|s)$ and $D(r|s)$, and 
  then carry out an analytic Bethe ansatz systematically 
 to construct a class of DVFs . 
On constructing DVFs, 
{\em the pole-freeness under the BAE} 
and {\em the top term hypothesis} 
 \cite{KS1,KOS} play important roles.

We introduce skew-Young (super) diagrams $\lambda \subset \mu$, 
which are related to tensor-like representations
\footnote{In this paper, we do not deal with spinorial 
representations.} of $B(r|s)$ or $D(r|s)$. 
On these skew-Young (super) diagrams, we define 
 a set of admissible tableaux  $B(\lambda \subset \mu)$ 
with some semi-standard like conditions. 
There is a one-to-one correspondence between 
these conditions for $B(0|s)$ case and  the 
  conditions for $A_{2s}^{(2)}$ case \cite{KS2}.
 In addition, 
 in contrast to $B(r|s)$ case, these conditions for $D(r|s)$ case 
 have non-local nature. 
 Next, we define a function ${\cal T}_{\lambda \subset \mu}(u)$  
 (\ref{Tge1}) of a spectral parameter $u$ 
 as summations over $B(\lambda \subset \mu)$. 
 It will provide us the spectra of a set of transfer 
matrices for various fusion $B(r|s)$ 
or $D(r|s)$ 
vertex models.  It contains the top term \cite{KS1,KOS},  
 which carries the highest weight of the irreducible 
 representation of $B(r|s)$ or $D(r|s)$ 
 labeled by a skew-Young (super) diagram 
 $\lambda \subset \mu$. 
  In particular, the simplest example of 
 ${\cal T}_{\lambda \subset \mu}(u)$, that is, 
 ${\cal T}^{1}(u)={\cal T}_{1}(u)={\cal T}_{(1^{1})}(u)$
  reduces to the eigenvalue formula of the transfer matrix 
  \cite{MR} of 
  some vertex model related to the fundamental  
 representation of  $B(r|s)$ or $D(r|s)$ 
 after some redefinitions.   
  The BAEs (\ref{BAE})-(\ref{BAE4}) 
  are assumed common to all the DVFs 
 for transfer matrices with various fusion types in the auxiliary space 
  as long as they act on a common quantum space. 
 Therefore, we can prove the pole-freeness of 
 ${\cal T}^{a}(u)={\cal T}_{(1^{a})}(u)$
  for any $a \in {\bf Z}_{\ge 0}$ under the common 
 BAEs (\ref{BAE})-(\ref{BAE4}). 
  We further mention a determinant formula, 
  by which ${\cal T}_{\lambda \subset \mu}(u)$ can be expressed  
  only by the fundamental functions $\{ {\cal T}^{a} \} $ 
  and then pole freeness follows immediately. 
  A set of transfer matrix functional relations among DVFs 
   also follows 
  from this  formula.
 It will be a kind of the $T$-system \cite{KNS1} 
(see also \cite{BR,KP,KNS2,KS2,KOS,KNH,
TK,KLWZ,Ma,PF,ZB,T1,T2,T3,T4,T5}). 
 In particular for $B(0|s)$ case, 
  there is remarkable duality among DVFs  
 (see, Theorem \ref{dual-th} and (\ref{dual2})). 
On constructing above-mentioned functional relations, 
this duality among DVFs plays an important role.  

The outline of this paper is as follows.
In section 2, we briefly mention the Lie superalgebras 
$B(r|s)$ and $D(r|s)$. 
In section 3, we execute an analytic Bethe ansatz 
based on the BAEs (\ref{BAE})-(\ref{BAE4}) associated 
with distinguished simple root systems. 
 In Section 4, we discuss transfer matrix functional 
 relations. 
 Section 5 is devoted to summary and discussion. 
 In appendix A.1-A.3, we prove the pole-freeness of DVFs. 
Appendix B provides generating series of 
${\cal T}^{a}(u)$ and ${\cal T}_{m}(u)={\cal T}_{(m)}(u)$. 
In this paper, we adopt similar notation in \cite{KS1,KOS,T1,T2,T3,T4}. 
 Finally we note that we can recover many formulae
  in \cite{KS1,KOS}  
 for $B_{r}$ or $D_{r}$, if we set $s=0$ and redefine the vacuum parts. 
\eqreset
\section{Lie superalgebras}
In this section, we briefly mention the 
Lie superalgebras $B(r|s)$ and $D(r|s)$  
(see for example \cite{Ka,Ka2,BB,FJ,MSS,Je}). 

There are several choices of simple root systems and 
the simplest one is the distinguished simple root system.
 They read as follows: \\ 
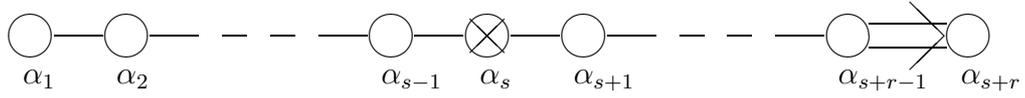
\begin{figure}
    \setlength{\unitlength}{0.9pt}
    \begin{center}
    \begin{picture}(410,50) 
      \put(10,20){\circle{20}}
      \put(20,20){\line(1,0){20}}
      \put(50,20){\circle{20}}
      \put(60,20){\line(1,0){20}}
      \put(90,20){\line(1,0){10}}
      \put(110,20){\line(1,0){10}}
      \put(130,20){\line(1,0){20}}
      \put(160,20){\circle{20}}
      \put(170,20){\line(1,0){20}}
      \put(7,0){$\alpha_{1}$}
      \put(46,0){$\alpha_{2}$}
      \put(156,0){$\alpha_{s-1}$}
      \put(192.929,12.9289){\line(1,1){14.14214}}
      \put(192.929,27.07107){\line(1,-1){14.14214}}
      \put(200,20){\circle{20}}
      \put(210,20){\line(1,0){20}}
      \put(240,20){\circle{20}}
      \put(250,20){\line(1,0){20}}
      \put(280,20){\line(1,0){10}}
      \put(300,20){\line(1,0){10}}
      \put(320,20){\line(1,0){20}}
      \put(350,20){\circle{20}}
      \put(358.8,25){\line(1,0){32.4}}
      \put(358.8,15){\line(1,0){32.4}}
      \put(400,20){\circle{20}}
      \put(197,0){$\alpha_{s}$}
      \put(236,0){$\alpha_{s+1}$}
      \put(346,0){$\alpha_{s+r-1}$}
      \put(397,0){$\alpha_{s+r}$}
      \put(390,20){\line(-1,1){14.14214}}
      \put(390,20){\line(-1,-1){14.14214}}
  \end{picture}
  \end{center}
  \caption{Dynkin diagram for the Lie superalgebra 
  $B(r|s)=osp(2r+1|2s)$ ($r \in {\bf Z}_{\ge 1}, 
  s \in {\bf Z}_{\ge 1}$)
   corresponding to the distinguished simple 
  root system: white circles denote even roots; 
  a gray (a cross) circle denotes an odd root $\alpha$ with 
   $(\alpha|\alpha)=0$.}
  \label{dynkin-brs}
\end{figure}
\begin{figure}
    \setlength{\unitlength}{0.8pt}
    \begin{center}
    \begin{picture}(250,50) 
      \put(10,20){\circle{20}}
      \put(20,20){\line(1,0){20}}
      \put(50,20){\circle{20}}
      \put(60,20){\line(1,0){20}}
      \put(90,20){\line(1,0){10}}
      \put(110,20){\line(1,0){10}}
      \put(130,20){\line(1,0){20}}
      \put(160,20){\circle{20}}
      \put(170,20){\line(1,0){20}}
      \put(7,0){$\alpha_{1}$}
      \put(46,0){$\alpha_{2}$}
      \put(156,0){$\alpha_{s-2}$}
      \put(200,20){\circle{20}}
      \put(240,20){\circle*{20}}
      \put(208.8,25){\line(1,0){32.4}}
      \put(208.8,15){\line(1,0){32.4}}
      \put(197,0){$\alpha_{s-1}$}
      \put(236,0){$\alpha_{s}$}
      \put(230,20){\line(-1,1){14.14214}}
      \put(230,20){\line(-1,-1){14.14214}}
  \end{picture}
  \end{center}
  \caption{Dynkin diagram for the Lie superalgebra 
  $B(0|s)=osp(1|2s)$ ($s \in {\bf Z}_{\ge 1}$):
   a black circle denotes an odd root $\alpha$ 
   with $(\alpha|\alpha)\ne 0$.}
  \label{dynkin-b0s}
\end{figure}
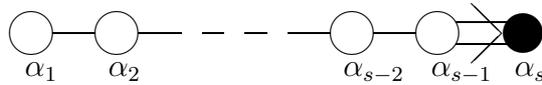
\begin{figure}
    \setlength{\unitlength}{0.9pt}
    \begin{center}
    \begin{picture}(420,50) 
      \put(10,20){\circle{20}}
      \put(20,20){\line(1,0){20}}
      \put(50,20){\circle{20}}
      \put(60,20){\line(1,0){20}}
      \put(90,20){\line(1,0){10}}
      \put(110,20){\line(1,0){10}}
      \put(130,20){\line(1,0){20}}
      \put(160,20){\circle{20}}
      \put(170,20){\line(1,0){20}}
      \put(7,0){$\alpha_{1}$}
      \put(46,0){$\alpha_{2}$}
      \put(156,0){$\alpha_{s-1}$}
      \put(192.929,12.9289){\line(1,1){14.14214}}
      \put(192.929,27.07107){\line(1,-1){14.14214}}
      \put(200,20){\circle{20}}
      \put(210,20){\line(1,0){20}}
      \put(240,20){\circle{20}}
      \put(250,20){\line(1,0){20}}
      \put(280,20){\line(1,0){10}}
      \put(300,20){\line(1,0){10}}
      \put(320,20){\line(1,0){20}}
      \put(350,20){\circle{20}}
      \put(398,44){\circle{20}}
      \put(359,25){\line(2,1){30}}
      \put(398,-4){\circle{20}}
      \put(359,16){\line(2,-1){30}}
      \put(390,24){$\alpha_{s+r-1}$}
      \put(393,-24){$\alpha_{s+r}$}
      \put(197,0){$\alpha_{s}$}
      \put(236,0){$\alpha_{s+1}$}
      \put(343,0){$\alpha_{s+r-2}$}
  \end{picture}
  \end{center}
  \caption{Dynkin diagram for the Lie superalgebra 
  $D(r|s)=osp(2r|2s)$ ($r \in {\bf Z}_{\ge 2}, 
  s \in {\bf Z}_{\ge 1}$)
   corresponding to the distinguished simple 
  root system.}
  \label{dynkin-drs}
\end{figure}
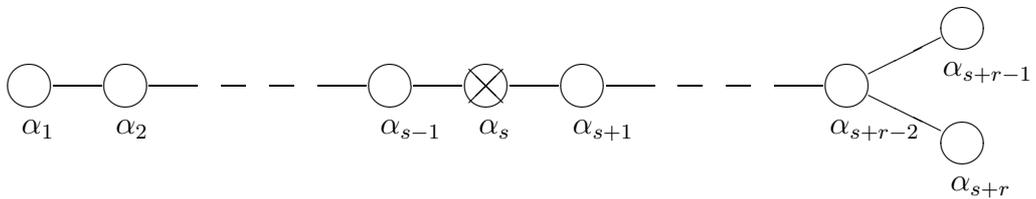
$B(r|s)$ ($r,s \in {\bf Z}_{\ge 1}$) case 
(see Figure \ref{dynkin-brs}):
\begin{eqnarray}
   && \alpha_{i} = \delta_{i}-\delta_{i+1} 
   \quad i=1,2,\dots,s-1, \nonumber  \\ 
   && \alpha_{s} =\delta_{s}-\epsilon_{1}, \nonumber  \\  
   && \alpha_{s+j} = \epsilon_{j}-\epsilon_{j+1} ,
    \quad j=1,2,\dots,r-1, \\ 
   && \alpha_{s+r} =\epsilon_{r}; \nonumber
 \end{eqnarray}
$B(0|s)$ ($s \in {\bf Z}_{\ge 1}$) case (see Figure \ref{dynkin-b0s}):
\begin{eqnarray}
   && \alpha_{i} = \delta_{i}-\delta_{i+1} 
   \quad {\rm for} \quad i=1,2,\dots,s-1, \nonumber  \\ 
   && \alpha_{s} =\delta_{s}; \label{root-B0s}
\end{eqnarray}
$D(r|s)$ ($r \in  {\bf Z}_{\ge 2}$, $s \in {\bf Z}_{\ge 1}$) case 
(see Figure \ref{dynkin-drs}):
\begin{eqnarray}
   && \alpha_{i} = \delta_{i}-\delta_{i+1} 
   \quad i=1,2,\dots,s-1, \nonumber  \\ 
   && \alpha_{s} =\delta_{s}-\epsilon_{1}, \nonumber  \\  
   && \alpha_{s+j} = \epsilon_{j}-\epsilon_{j+1} ,
    \quad j=1,2,\dots,r-2, \\ 
   && \alpha_{s+r-1} =\epsilon_{r-1}-\epsilon_{r}, \nonumber \\ 
   && \alpha_{s+r} =\epsilon_{r-1}+\epsilon_{r}, \nonumber
 \end{eqnarray}
where  
 $\epsilon_{1},\dots, \epsilon_{r};\delta_{1},\dots,\delta_{s}$ 
are the bases of the dual space of the Cartan subalgebra 
with the bilinear 
form $(\ |\ )$ such that 
\footnote{We normalized the longest simple root as 
$|(\alpha|\alpha)|=2$.}
\begin{equation}
 (\epsilon_{i}|\epsilon_{j})=\delta_{i\, j}, \quad 
 (\epsilon_{i}|\delta_{j})=(\delta_{i}|\epsilon_{j})=0 , \quad 
 (\delta_{i}|\delta_{j})=-\delta_{i\, j} . 
\end{equation}
 $\{\alpha_i \}_{i \ne s}$ are even roots and $\alpha_{s}$ 
is an odd root. Note that $(\alpha_{s} | \alpha_{s})=0$ 
for $B(r|s)$ ($r,s \in {\bf Z}_{\ge 1}$) and $D(r|s)$ 
($r \in  {\bf Z}_{\ge 2}$, $s \in {\bf Z}_{\ge 1}$) cases, 
while $(\alpha_{s} | \alpha_{s}) \ne 0$ for 
$B(0|s)$ ($s \in {\bf Z}_{\ge 1}$) case.
 
Let $\lambda \subset \mu$ be a skew-Young (super) diagram labeled by 
the sequences of non-negative integers 
$\lambda =(\lambda_{1},\lambda_{2},\dots)$ and 
$\mu =(\mu_{1},\mu_{2},\dots)$ such that
$\mu_{i} \ge \lambda_{i}: i=1,2,\dots;$  
$\lambda_{1} \ge \lambda_{2} \ge \dots \ge 0$;  
$\mu_{1} \ge \mu_{2} \ge \dots \ge 0$ and 
$\lambda^{\prime}=(\lambda_{1}^{\prime},\lambda_{2}^{\prime},\dots)$ 
be the conjugate of $\lambda $. 
In particular, for $B(r|s)$ ($r,s \in {\bf Z}_{\ge 1}$), 
$\lambda=\phi $, $\mu_{r+1}\le s$ case, 
the Kac-Dynkin label 
$[b_{1},b_{2},\dots , b_{s+r}]$ is related to 
 the Young (super) diagram with shape 
 $\mu=(\mu_{1},\mu_{2},\dots)$ as follows:  \\ 
\begin{eqnarray}
 && b_{i} = \mu_{i}^{\prime}-\mu_{i+1}^{\prime} 
  \qquad {\rm for} \qquad i\in \{1,2,\dots,s-1\}, \nonumber \\ 
 && b_{s} = \mu_{s}^{\prime}+\eta_{1}, \qquad  \label{Kac-Dynkin} \\ 
 && b_{s+j} = \eta_{j}-\eta_{j+1} 
  \qquad {\rm for} \qquad j \in \{1,2,\dots,r-1\}, \nonumber \\
 && b_{s+r} = 2\eta_{r} \nonumber,
\end{eqnarray} 
  where $\eta_{i}=Max\{\mu_{i}-s,0 \} $. 
For $B(0|s)$ ($s \in {\bf Z}_{\ge 1}$), 
$\lambda=\phi $ case, 
the Kac-Dynkin label 
$[b_{1},b_{2},\dots , b_{s}]$ is related to 
 the Young (super) diagram with shape 
 $\mu=(\mu_{1},\mu_{2},\dots)$ as follows:
\begin{eqnarray}
 && b_{i} = \mu_{i}^{\prime}-\mu_{i+1}^{\prime} 
  \qquad {\rm for} \qquad i\in \{1,2,\dots,s-1\}, \nonumber \\ 
 && b_{s} = 2\mu_{s}^{\prime}.  \label{Kac-Dynkin-b0s}
\end{eqnarray}  
For $D(r|s)$ case, we use only the Young (super) diagram with 
shape $\mu=(1^{a})$ or $\mu=(m^{1})$. 
The Young (super) diagram with shape $\mu=(1^{a})$ is 
related to the Kac-Dynkin label 
$[b_{1},b_{2},\dots ,b_{s+r}]$ as follows:
\begin{eqnarray}
 && b_{j} = a\delta_{j 1}. \label{KD-tate}
\end{eqnarray} 
And the Young (super) diagram with shape 
 $\mu=(m^{1})$ is related to 
 the Kac-Dynkin label 
$[b_{1},b_{2},\dots ,  b_{s+r}]$ as follows:
\begin{eqnarray}
b_{j}=
 \left\{
  \begin{array}{lll}
   \! \delta_{j m} 
     & \! {\rm if} \! & m \in \{1,2,\dots, s \},  \\ 
   \! (m-s+1)\delta_{j s}+(m-s)\delta_{j s+1} 
     & \! {\rm if} \! & r \in {\bf Z}_{\ge 3},  
         m \in {\bf Z}_{\ge s+1}, \\ 
   \! (m-s+1)\delta_{j s}+(m-s)(\delta_{j s+1}+\delta_{j s+2}) 
     & \! {\rm if} \! & r=2,  m \in {\bf Z}_{\ge s+1}. 
  \end{array}
 \right. 
 \label{KD-yoko}
\end{eqnarray}
An irreducible representation of $B(0|s)$
with the Kac-Dynkin label $[b_{1},b_{2},\dots ,  b_{s}]$
 is finite dimensional \cite{Ka2} if and only if 
\begin{eqnarray}
&& b_{j} \in {\bf Z}_{\ge 0} \qquad {\rm for} \qquad 
j \in \{1,2,\dots, s-1\}, \nonumber \\ 
&& b_{s} \in  2{\bf Z}_{\ge 0}. \label{finite}
\end{eqnarray}
The dimensionality of the irreducible representation 
 $V[b_{1},b_{2},\dots ,  b_{s}]$ of 
 $B(0|s)$ with the highest weight 
 labeled by the Kac-Dynkin label  $[b_{1},b_{2},\dots ,  b_{s}]$ 
 is given \cite{Ka2,Je} as follows
\footnote{We assume that 
$b_{j}+b_{j+1}+\cdots +b_{s-1}=0$ if $j=s$.} 
 \begin{eqnarray}
\hspace{-30pt} &&  {\rm dim}V[b_{1},b_{2},\dots,b_{s}]=
   \prod_{1\le i < j \le s}
  \frac{b_{i}+b_{i+1}+\cdots +b_{j-1}+j-i}{j-i}\nonumber \\
\hspace{-30pt}&& \hspace{10pt} \times  
  \frac{b_{i}+b_{i+1}+\cdots +b_{j-1}+
  2(b_{j}+b_{j+1}+\cdots +b_{s-1})+b_{s}+2s-i-j+1}
         {2s-i-j+1} \nonumber \\
\hspace{-30pt}&& \hspace{10pt}\times  \prod_{1 \le k \le s}
  \frac{2(b_{k}+b_{k+1}+\cdots +b_{s-1})+b_{s}+2s-2k+1}
         {2s-2k+1}.
  \label{dim}
 \end{eqnarray}
\eqreset
\section{Analytic Bethe ansatz} 
We assume, as our starting point, the 
following type of the Bethe ansatz equations
\footnote{In this paper, we deal with the case,
 as an example,  
 that the quantum spaces of the transfer matrices are fundamental 
representations. } 
\cite{RW,OW,Kul,RM,KOS}. \\ 
$B(r|s)$ ($r,s \in {\bf Z}_{\ge 1}$) or $D(r|s)$ ($r \in  {\bf Z}_{\ge 2}, s \in {\bf Z}_{\ge 1}$) case:
\begin{eqnarray}
- \left\{
 \prod_{j=1}^{N}
 \frac{\Phi (u_k^{(a)}-w_{j}-1)}
      {\Phi(u_k^{(a)}-w_{j}+1)}
   \right\}^{\delta_{a 1}}
   &=&(-1)^{{\rm deg}(\alpha_a)} 
    \prod_{b=1}^{s+r}\frac{Q_{b}(u_k^{(a)}+(\alpha_a|\alpha_b))}
           {Q_{b}(u_k^{(a)}-(\alpha_a|\alpha_b))}. \label{BAE} 
\end{eqnarray}
$B(0|s)$ ($s \in {\bf Z}_{\ge 2}$) case: 
\begin{eqnarray}
\hspace{-40pt}
&& - \prod_{j=1}^{N}
\frac{ \Phi (u_k^{(1)}-w_{j}-1)}
     { \Phi (u_k^{(1)}-w_{j}+1)} 
 = 
\frac{Q_{1}(u_k^{(1)}-2)Q_{2}(u_k^{(1)}+1)}
{Q_{1}(u_k^{(1)}+2)Q_{2}(u_k^{(1)}-1)}, \label{BAE1}
\\
\hspace{-40pt}
&&-1= \frac{Q_{a-1}(u_k^{(a)}+1)Q_{a}(u_k^{(a)}-2)Q_{a+1}(u_k^{(a)}+1)}
      {Q_{a-1}(u_k^{(a)}-1)Q_{a}(u_k^{(a)}+2)Q_{a+1}(u_k^{(a)}-1)}
      \quad{\rm for}
       \quad 2 \le a \le s-1, \label{BAE2}\\ 
\hspace{-40pt}
&&1= \frac{Q_{s-1}(u_k^{(s)}+1)Q_{s}(u_k^{(s)}+1)Q_{s}(u_k^{(s)}-2)}
        {Q_{s-1}(u_k^{(s)}-1)Q_{s}(u_k^{(s)}-1)Q_{s}(u_k^{(s)}+2)}
       \label{BAE3} .
\end{eqnarray}
$B(0|1)$ case: 
\begin{eqnarray}
\hspace{-40pt}
&& \prod_{j=1}^{N}
\frac{ \Phi (u_k^{(1)}-w_{j}-1)}
     { \Phi (u_k^{(1)}-w_{j}+1)} 
 = 
\frac{Q_{1}(u_k^{(1)}+1)Q_{1}(u_k^{(1)}-2)}
     {Q_{1}(u_k^{(1)}-1)Q_{1}(u_k^{(1)}+2)}
       \label{BAE4}.
\end{eqnarray}
Here $ Q_{a}(u)= \prod_{j=1}^{N_{a}}\Phi(u-u_j^{(a)})$; 
$ N \in {\bf Z }_{\ge 0}$ is the number of 
the lattice sites; $N_{a} \in {\bf Z }_{\ge 0}$; 
$u_{j}^{(a)}, w_{j}\in {\bf C}$; $a,k \in {\bf Z}$ 
($a \in \{1,2,\dots,s+r \}$ ($r=0$ for $B(0|s)$ case); 
$\ k \in \{1,2,\dots, N_{a} \}$); 
\begin{eqnarray}
    {\rm deg}(\alpha_a)&=&\left\{
              \begin{array}{@{\,}ll}
                0  & \mbox{for even root} \\ 
                1 & \mbox{for odd root} 
              \end{array}
              \right. \\ 
             &=& \delta_{a,s}. \nonumber 
\end{eqnarray}
 $\Phi$ is a function, which has zero at $u=0$. 
For example, $\Phi(u)$ has the following form
\begin{equation}
\Phi(u)=u.
\end{equation}
Remarkable enough, Bethe ansatz equations can be written 
in terms of root systems of  Lie algebras \cite{RW,OW} or 
Lie superalgebras \cite{Kul,MR}. 
Martins and Ramos \cite{MR} pointed out that 
$B(0|s)$ is an exception
 to this observation (see also \cite{Kul}).  
To put it more precisely, an exception lies in 
the right hand side of  (\ref{BAE3}) and (\ref{BAE4}), 
which correspond to 
the odd root $\alpha_{s}$ with $(\alpha_{s}|\alpha_{s}) \ne 0$.  
In fact, one can derive (\ref{BAE1}) and (\ref{BAE2}) from 
(\ref{BAE}) and (\ref{root-B0s}); 
while cannot derive (\ref{BAE3}) and (\ref{BAE4}). \\ 
{\em Remark:} 
There are compact expressions of BAEs 
for twisted quantum affine algebras \cite{RW}. 
 Moreover 
the BAEs (\ref{BAE1})-(\ref{BAE4}) 
resemble to the BAEs for $A_{2s}^{(2)}$. 
This resemblance will originate from resemblance between 
$B(0|s)^{(1)}$ and $A_{2s}^{(2)}$. 
Thus there is a possibility that the BAEs 
(\ref{BAE1})-(\ref{BAE4}) are also 
compactly written in terms of root system of the 
Lie superalgebra $B(0|s)$ ($ s \in {\bf Z}_{\ge 1} $). 
We also point out that the expression (\ref{BAE}) 
is not always valid for non-distinguished simple root systems.  
In fact we have confirmed for several cases that the Bethe 
ansatz equations corresponding to the 
 odd roots $\alpha $ with $(\alpha|\alpha)\ne 0$ 
have similar structure to 
(\ref{BAE3}) or (\ref{BAE4}) 
by using the correspondence \cite{T2} between 
the particle-hole transformation 
 and the (super) Weyl reflection. 

We define the set 
\begin{eqnarray}
    &&  J=J_{+} \cup J_{-},
\end{eqnarray}
where 
\begin{eqnarray}
J_{-}= \{ 1,2,\dots, s,\overline{s},\dots,\overline{2},
                  \overline{1} \} \label{set} 
\end{eqnarray}
is common for $B(r|s)$ and $D(r|s)$; 
while $J_{+}$ is not:
\begin{eqnarray}
    && J_{+}=\{s+1,s+2,\dots,s+r,\overline{s+r},\dots,
       \overline{s+2},\overline{s+1} \} \cup \{ 0 \} 
       \quad {\rm for } \quad B(r|s), \nonumber \\ 
    && J_{+}=\{s+1,s+2,\dots,s+r,\overline{s+r},\dots,
       \overline{s+2},\overline{s+1} \} 
       \quad {\rm for } \quad D(r|s). \nonumber 
\end{eqnarray}
On this set $J$, we define the total order 
\begin{eqnarray} 
 1\prec 2 \prec \cdots \prec s+r \prec 0 
 \prec \overline{s+r} \prec \cdots \prec \overline{2} 
 \prec \overline{1} \label{order}
\end{eqnarray}
 for $B(r|s)$ case, and the partial order 
\begin{eqnarray} 
 1\prec 2 \prec \cdots \prec s+r-1 \prec 
 \begin{array}{c} 
  s+r \\ 
      \\ 
  \overline{s+r} 
 \end{array}
 \prec \overline{s+r-1} \prec \cdots \prec \overline{2} 
 \prec \overline{1} \label{order2}
\end{eqnarray}
for $D(r|s)$ case. In contrast to $B(r|s)$ case,
 there is no order between 
$s+r$ and $\overline{s+r}$ for $D(r|s)$ case. 
We also define the grading parameter as follows: 
\begin{equation}
      p(a)=\left\{
              \begin{array}{@{\,}ll}
                0  & \mbox{for $a \in J_{+}$},  \\ 
                1 & \mbox{for $a \in J_{-}$ }.
              \end{array}
            \right. \label{grading}
\end{equation}
For $a \in J $, we define 
\footnote{
In this paper, we often abbreviate the spectral parameter $u$.}
 the following functions. \\ 
 $B(r|s)$ ($r \in {\bf Z}_{\ge 0},s \in {\bf Z}_{\ge 1}$) case:
\begin{eqnarray} 
\hspace{-20pt} && \framebox{$a$}_{u}=\psi_{a}(u)
   \frac{Q_{a-1}(u-a-1)Q_{a}(u-a+2)}
        {Q_{a-1}(u-a+1)Q_{a}(u-a)}
      \quad  {\rm for} \quad 1 \le a \le s,\nonumber \\
\hspace{-20pt} && \framebox{$a$}_{u}=\psi_{a}(u)
   \frac{Q_{a-1}(u-2s+a+1)Q_{a}(u-2s+a-2)}
        {Q_{a-1}(u-2s+a-1)Q_{a}(u-2s+a)} \nonumber \\ 
\hspace{-20pt} && \hspace{170pt}   
{\rm for} \quad s+1 \le a \le s+r,\nonumber \\
\hspace{-20pt} && \framebox{$0$}_{u}=\psi_{0}(u)
   \frac{Q_{s+r}(u-s+r+1)Q_{s+r}(u-s+r-2)}
        {Q_{s+r}(u-s+r-1)Q_{s+r}(u-s+r)},
      \label{z+}   \\
\hspace{-20pt} && \framebox{$\overline{a}$}_{u}=\psi_{\overline{a}}(u)
   \frac{Q_{a-1}(u+2r-a-2)Q_{a}(u+2r-a+1)}
        {Q_{a-1}(u+2r-a)Q_{a}(u+2r-a-1)}
        \nonumber \\ 
\hspace{-20pt} && \hspace{170pt} {\rm for} \quad 
s+1 \le a \le s+r, \nonumber \\ 
\hspace{-20pt} && \framebox{$\overline{a}$}_{u}=\psi_{\overline{a}}(u)
   \frac{Q_{a-1}(u-2s+2r+a)Q_{a}(u-2s+2r+a-3)}
        {Q_{a-1}(u-2s+2r+a-2)Q_{a}(u-2s+2r+a-1)} 
  \nonumber \\ 
\hspace{-20pt} && \hspace{170pt}  
{\rm for} \quad 1 \le a \le s. \nonumber 
\end{eqnarray}
$D(r|s)$ ($r \in {\bf Z}_{\ge 2},s \in {\bf Z}_{\ge 1}$) case: 
\begin{eqnarray} 
\hspace{-20pt} && \framebox{$a$}_{u}=\psi_{a}(u)
   \frac{Q_{a-1}(u-a-1)Q_{a}(u-a+2)}
        {Q_{a-1}(u-a+1)Q_{a}(u-a)}
      \quad {\rm for} \quad 1 \le a \le s,\nonumber \\
\hspace{-20pt} && \framebox{$a$}_{u}=\psi_{a}(u)
   \frac{Q_{a-1}(u-2s+a+1)Q_{a}(u-2s+a-2)}
        {Q_{a-1}(u-2s+a-1)Q_{a}(u-2s+a)} 
        \nonumber \\ 
\hspace{-20pt} && \hspace{160pt} 
 {\rm for} \quad s+1 \le a \le s+r-2,\nonumber \\
\hspace{-20pt} && \framebox{$r+s-1$}_{u}=\psi_{r+s-1}(u)
   \frac{Q_{s+r-2}(u-s+r)Q_{s+r-1}(u-s+r-3)}
        {Q_{s+r-2}(u-s+r-2)Q_{s+r-1}(u-s+r-1)}
        \nonumber \\ 
\hspace{-20pt} && \hspace{160pt} \times 
   \frac{Q_{s+r}(u-s+r-3)}{Q_{s+r}(u-s+r-1)}
 ,\nonumber \\
\hspace{-20pt} && \framebox{$r+s$}_{u}=\psi_{r+s}(u)
   \frac{Q_{s+r-1}(u-s+r+1)Q_{s+r}(u-s+r-3)}
        {Q_{s+r-1}(u-s+r-1)Q_{s+r}(u-s+r-1)},
      \label{z++}   \\
\hspace{-20pt} && 
 \framebox{$\overline{r+s}$}_{u}=\psi_{\overline{r+s}}(u)
   \frac{Q_{s+r-1}(u-s+r-3)Q_{s+r}(u-s+r+1)}
        {Q_{s+r-1}(u-s+r-1)Q_{s+r}(u-s+r-1)},
      \nonumber   \\
\hspace{-20pt} && 
\framebox{$\overline{r+s-1}$}_{u}=\psi_{\overline{r+s-1}}(u)
   \frac{Q_{s+r-2}(u-s+r-2)Q_{s+r-1}(u-s+r+1)}
        {Q_{s+r-2}(u-s+r)Q_{s+r-1}(u-s+r-1)}
        \nonumber \\ 
\hspace{-20pt} && \hspace{160pt} 
 \times \frac{Q_{s+r}(u-s+r+1)}{Q_{s+r}(u-s+r-1)},
        \nonumber \\
\hspace{-20pt} && \framebox{$\overline{a}$}_{u}=\psi_{\overline{a}}(u)
   \frac{Q_{a-1}(u+2r-a-3)Q_{a}(u+2r-a)}
        {Q_{a-1}(u+2r-a-1)Q_{a}(u+2r-a-2)}
        \nonumber \\ 
\hspace{-20pt} && \hspace{160pt}
    {\rm for} \quad s+1 \le a \le s+r-2, \nonumber \\ 
\hspace{-20pt} && \framebox{$\overline{a}$}_{u}=\psi_{\overline{a}}(u)
   \frac{Q_{a-1}(u-2s+2r+a-1)Q_{a}(u-2s+2r+a-4)}
        {Q_{a-1}(u-2s+2r+a-3)Q_{a}(u-2s+2r+a-2)} 
   \nonumber \\ 
\hspace{-20pt} && \hspace{160pt}
   {\rm for} \quad 1 \le a \le s.\nonumber 
\end{eqnarray}
Here we assume $Q_{0}(u)=1$. 
The vacuum parts of the functions 
$\framebox{a}_{u}$ (\ref{z+}) and (\ref{z++})
 are given as follows. \\
For $B(r|s)$ ($r \in {\bf Z}_{\ge 0},s \in {\bf Z}_{\ge 1}$) case:
\begin{eqnarray}
 \psi_{1}(u)&=&\phi(u-2)\phi(u-2s+2r-1), \nonumber \\ 
 \psi_{a}(u)&=&\phi(u)\phi(u-2s+2r-1) 
  \quad {\rm for} \quad 
   2 \preceq a \preceq \overline{2}, \nonumber \\
 \psi_{\overline{1}}(u)&=&\phi(u)\phi(u-2s+2r+1)
   \label{psi}.
\end{eqnarray}
$D(r|s)$ ($r \in {\bf Z}_{\ge 2},s \in {\bf Z}_{\ge 1}$) case:
\begin{eqnarray}
 \psi_{1}(u)&=&\phi(u-2)\phi(u-2s+2r-2), \nonumber \\ 
 \psi_{a}(u)&=&\phi(u)\phi(u-2s+2r-2) 
 \quad {\rm for} \quad
   2 \preceq a \preceq \overline{2}, \nonumber \\
 \psi_{\overline{1}}(u)&=&\phi(u)\phi(u-2s+2r) .
   \label{psi-d}
\end{eqnarray}
Here 
\begin{eqnarray}
 \phi(u)=\prod_{j=1}^{N}\Phi(u-w_{j}).
\end{eqnarray}
Under the BAEs (\ref{BAE})-(\ref{BAE4}),
 we have: 
 \footnote{
 Here $Res_{u=a}f(u)$ denotes the residue of a function $f(u)$ at $u=a $.
 }\\  
 $B(r|s)$ ($r,s \in {\bf Z}_{\ge 1}$) case:
\begin{eqnarray}
\hspace{-45pt}&& Res_{u=d+u_{k}^{(d)}}
 (\framebox{$d$}_{u}+\framebox{$d+1$}_{u})=0 
  \quad {\rm for} \quad 1\le d \le s-1 , \label{res1} \\
\hspace{-45pt}&& Res_{u=s+u_{k}^{(s)}}
(\framebox{$s$}_{u}-\framebox{$s+1$}_{u})=0  ,
      \\
\hspace{-45pt}&& Res_{u=2s-d+u_{k}^{(d)}}
 (\framebox{$d$}_{u}+\framebox{$d+1$}_{u})=0 
   \quad {\rm for} \quad s+1\le d \le s+r-1 , \label{res3} \\
\hspace{-45pt}&& Res_{u=s-r+u_{k}^{(s+r)}}
(\framebox{$s+r$}_{u}+\framebox{$0$}_{u})=0  ,
      \\
\hspace{-45pt}&& Res_{u=s-r+1+u_{k}^{(s+r)}}
(\framebox{$0$}_{u}+\framebox{$\overline{s+r}$}_{u})=0  ,
      \\
\hspace{-45pt}&& Res_{u=d-2r+1+u_{k}^{(d)}}
   (\framebox{$\overline{d+1}$}_{u}+\framebox{$\overline{d}$}_{u})=0 
   \quad {\rm for} \quad s+1 \le d \le s+r-1  ,\label{res6} \\
\hspace{-45pt}&& Res_{u=s-2r+1+u_{k}^{(s)}}
   (\framebox{$\overline{s+1}$}_{u}-\framebox{$\overline{s}$}_{u})=0,
    \\
\hspace{-45pt}&& Res_{u=-d+2s-2r+1+u_{k}^{(d)}}
   (\framebox{$\overline{d+1}$}_{u}+\framebox{$\overline{d}$}_{u})=0 
 \quad {\rm for} \quad 1 \le d \le s-1  .  \label{res8} 
\end{eqnarray}
$B(0|s)$ ($s \in {\bf Z}_{\ge 1}$) case:
\begin{eqnarray}
\hspace{-45pt}&& Res_{u=d+u_{k}^{(d)}}
 (\framebox{$d$}_{u}+\framebox{$d+1$}_{u})=0 
  \quad {\rm for} \quad 1\le d \le s-1 , \label{res1-b0s} \\
\hspace{-45pt}&& Res_{u=s+u_{k}^{(s)}}
(\framebox{$s$}_{u}-\framebox{$0$}_{u})=0  ,
 \\
 \hspace{-45pt}&& Res_{u=s+1+u_{k}^{(s)}}
   (\framebox{$0$}_{u}-\framebox{$\overline{s}$}_{u})=0,
    \\
\hspace{-45pt}&& Res_{u=-d+2s+1+u_{k}^{(d)}}
   (\framebox{$\overline{d+1}$}_{u}+\framebox{$\overline{d}$}_{u})=0 
\quad {\rm for} \quad 1 \le d \le s-1  .  \label{res4-b0s} 
\end{eqnarray}
$D(r|s)$ ($r \in {\bf Z}_{\ge 2},s \in {\bf Z}_{\ge 1}$) case: 
\begin{eqnarray}
\hspace{-45pt}&& Res_{u=d+u_{k}^{(d)}}
 (\framebox{$d$}_{u}+\framebox{$d+1$}_{u})=0 
  \quad {\rm for} \quad 1\le d \le s-1 , \label{res1-d} \\
\hspace{-45pt}&& Res_{u=s+u_{k}^{(s)}}
(\framebox{$s$}_{u}-\framebox{$s+1$}_{u})=0  ,
      \\
\hspace{-45pt}&& Res_{u=2s-d+u_{k}^{(d)}}
 (\framebox{$d$}_{u}+\framebox{$d+1$}_{u})=0 
  \quad {\rm for} \quad s+1\le d \le s+r-1 , \label{res3-d} \\
\hspace{-45pt}&& Res_{u=s-r+1+u_{k}^{(s+r)}}
(\framebox{$s+r-1$}_{u}+\framebox{$\overline{s+r}$}_{u})=0  ,
      \\
\hspace{-45pt}&& Res_{u=s-r+1+u_{k}^{(s+r)}}
(\framebox{$s+r$}_{u}+\framebox{$\overline{s+r-1}$}_{u})=0  ,
      \\
\hspace{-45pt}&& Res_{u=d-2r+2+u_{k}^{(d)}}
   (\framebox{$\overline{d+1}$}_{u}+\framebox{$\overline{d}$}_{u})=0 
 \quad {\rm for} \quad s+1 \le d \le s+r-1  ,\label{res6-8} \\
\hspace{-45pt}&& Res_{u=s-2r+2+u_{k}^{(s)}}
   (\framebox{$\overline{s+1}$}_{u}-\framebox{$\overline{s}$}_{u})=0,
    \\
\hspace{-45pt}&& Res_{u=-d+2s-2r+2+u_{k}^{(d)}}
   (\framebox{$\overline{d+1}$}_{u}+\framebox{$\overline{d}$}_{u})=0 
    \quad {\rm for} \quad 1 \le d \le s-1  .  \label{res8-d} 
\end{eqnarray}
We assign coordinates $(i,j)\in {\bf Z}^{2}$ 
on the skew-Young  superdiagram $\lambda \subset \mu$ 
such that the row index $i$ increases as we go downwards and the column 
index $j$ increases as we go from the left to the right and that 
$(1,1)$ is on the top left corner of $\mu$.
We define an admissible tableau $b$ 
on the skew-Young superdiagram 
$\lambda \subset \mu$ as a set of elements $b(i,j)\in J$ 
 labeled by the coordinates 
$(i,j)$ mentioned above, with the following rule. \\ 
The admissible condition for $B(r|s)$ 
($r \in {\bf Z}_{\ge 0}$, $s \in {\bf Z}_{\ge 1}$): 
\begin{enumerate}
\item
\begin{eqnarray*}
  b(i,j) \preceq b(i,j+1),   
\end{eqnarray*}
\item 
\begin{eqnarray*}
  b(i,j) \preceq b(i+1,j), 
\end{eqnarray*}
\item 
\begin{eqnarray*} 
  b(i,j) \prec b(i+1,j) \quad {\rm if}\quad  b(i,j) 
 \in  J_{+} \setminus \{0 \},  
\end{eqnarray*} 
\item 
\begin{eqnarray*}
  b(i,j) \prec b(i,j+1)  \quad {\rm if} \quad  b(i,j) 
 \in  J_{-} \cup \{0 \}.  
\end{eqnarray*}
\end{enumerate}
The admissible condition for $D(r|s)$ 
($r \in {\bf Z}_{\ge 2}$, $s \in {\bf Z}_{\ge 1}$); 
$\lambda=\phi$; $\mu=(1^{a})$:
\begin{enumerate}
\item 
\begin{eqnarray*}
 b(k,1) \preceq b(k+1,1) \quad 
 {\rm if} \quad b(k+1,1) \in J_{-} 
\end{eqnarray*}

\item 
\begin{eqnarray*}   
b(k,1) \prec b(k+1,1) \quad 
 {\rm if} \quad b(k+1,1) \in J_{+} 
\end{eqnarray*}
unless 
\begin{eqnarray*}
 (b(k,1),b(k+1,1))=(\overline{s+r},s+r) \quad {\rm or} 
 \quad (s+r,\overline{s+r}). 
\end{eqnarray*}
\end{enumerate}
The admissible condition for $D(r|s)$ 
($r \in {\bf Z}_{\ge 2}$, $s \in {\bf Z}_{\ge 1}$); 
$\lambda=\phi$; $\mu=(m^{1})$:
\begin{enumerate}
\item 
\begin{eqnarray*}   
b(1,k) \preceq b(1,k+1) \quad 
 {\rm if} \quad b(1,k+1) \in J_{+},  
\end{eqnarray*}

\item  
\begin{eqnarray*}
 b(1,k) \prec b(1,k+1) \quad 
 {\rm if} \quad b(1,k+1) \in J_{-},  
\end{eqnarray*}

\item 
$s+r$ and $\overline{s+r}$ do not appear simultaneously.
\end{enumerate}
Let $B(\lambda \subset \mu)$ be 
the set of admissible tableaux
\footnote{
In contrast to $B(r|s)$ case, the 
admissible condition for $D(r|s)$ 
 case has non-local nature. This property makes it 
 difficult to extend the 
admissible condition for $D(r|s)$ 
 to more general skew-Young (super) diagrams.
} 
 on $\lambda \subset \mu$. 
We shall present a function ${\cal T}_{\lambda \subset \mu}(u)$ 
with a spectral parameter $u\in {\rm C}$ and 
 skew-Young superdiagrams $\lambda \subset \mu$, 
 which is a candidate of a set of DVFs 
 for various fusion types in the 
 auxiliary spaces
 \footnote{We assume that they are finite dimensional modules of 
 quantum affine superalgebras (or super Yangians) \cite{Y1,Y2}. 
 Thus ${\cal T}_{\lambda \subset \mu}(u)$ is expected to be a 
  kind of a (super) character of such algebras. 
  At present, we can not justify this speculations mathematically 
  in general, 
 since we luck systematic representation theory of 
 such algebras. 
 We hope that mathematically satisfactory 
 account on our formulae appear after the development of 
 representation theory in the future. }
  of transfer matrices of 
 $B(r|s)$ or $D(r|s)$ vertex models. 
For the skew-Young (super) diagrams $\lambda \subset \mu$, 
define ${\cal T}_{\lambda \subset \mu}(u)$ as follows
\begin{equation}
 {\cal T}_{\lambda \subset \mu}(u)=
\sum_{b \in B(\lambda \subset \mu)}
\prod_{(i,j) \in (\lambda \subset \mu)}
(-1)^{p(b(i,j))}
\framebox{$b(i,j)$}_{u-\mu_{1}+\mu_{1}^{\prime}-2i+2j}	
\label{Tge1},
\end{equation}
where the product is taken over the coordinates $(i,j)$ on
 $\lambda \subset \mu$.

We can express ${\cal T}_{\lambda \subset \mu}(u)$ 
as determinants over matrices, whose matrix elements are 
${\cal T}^{a}$ or ${\cal T}_{m}$ 
\footnote{${\cal T}_{m}^{a}(u):={\cal T}_{(m^{a})}(u)$; 
${\cal T}_{m}(u):={\cal T}_{m}^{1}(u)$; 
${\cal T}^{a}(u):={\cal T}_{1}^{a}(u)$; 
${\cal T}_{m}^{0}(u)={\cal T}_{0}^{a}(u)=1$ for 
$m,a\in {\bf Z}_{\ge 0}$; 
${\cal T}_{m}^{a}(u)=0$ if 
$m\in {\bf Z}_{< 0}$ or $a\in {\bf Z}_{< 0}$. 
See also Appendix B.}
(cf. \cite{KOS,BR}). \\ 
For $B(r|s)$ ($r \in {\bf Z}_{\ge 0}$, $s \in {\bf Z}_{\ge 1}$) case, we have 
\begin{eqnarray}
\hspace{-30pt}
{\cal T}_{\lambda \subset \mu}(u)&=&{\rm det}_{1 \le i,j \le \mu_{1}}
    ({\cal T}^{\mu_{i}^{\prime}-\lambda_{j}^{\prime}-i+j}
    (u-\mu_{1}+\mu_{1}^{\prime}-\mu_{i}^{\prime}-\lambda_{j}^{\prime}+i+j-1))
	\label{Jacobi-Trudi1} \\ 
\hspace{-30pt}	
 &=& {\rm det}_{1 \le i,j \le \mu_{1}^{\prime}}
    ({\cal T}_{\mu_{j}-\lambda_{i}+i-j}
    (u-\mu_{1}+\mu_{1}^{\prime}+\mu_{j}+\lambda_{i}-i-j+1))	.
	\label{Jacobi-Trudi2} 
\end{eqnarray}
For $D(r|s)$ ($r \in  {\bf Z}_{\ge 2}$, $s \in {\bf Z}_{\ge 1}$) case, we have 
\begin{eqnarray}
\hspace{-60pt}&& 
{\cal T}_{m}(u)={\rm det}_{1 \le i,j \le m}
    ({\cal T}^{1-i+j}
    (u-m+i+j-1))
	\label{Jacobi-Trudi}. 
\end{eqnarray}
Note that the function ${\cal T}^{1}(u)={\cal T}_{1}(u)$ 
 coincides with  
the eigenvalue formula of a $B(r|s)$ or $D(r|s)$ vertex model 
by the algebraic Bethe ansatz
 \cite{MR} after some redefinitions. 

We remark that if $\Phi (-u) = \pm \Phi (u)$, 
$\framebox{$a$}_{u}$ is transformed to 
$ \framebox{$\overline{a}$}_{u}$
\footnote{Here we interpret $\overline{0}$ as $0$.}
under the following transformation. \\ 
$B(r|s)$ ($r \in {\bf Z}_{\ge 0}, s \in {\bf Z}_{\ge 1} $) case: 
\begin{eqnarray}
&& u \to -(u+2r-2s-1), \nonumber \\ 
&& u_{j}^{(a)} \to -u_{j}^{(a)}, \label{cross1} \\ 
&& w_{j} \to -w_{j}. \nonumber 
\end{eqnarray}
$D(r|s)$ ($r \in  {\bf Z}_{\ge 2}, s \in {\bf Z}_{\ge 1} $) case:
\begin{eqnarray}
&& u \to -(u+2r-2s-2), \nonumber \\ 
&& u_{j}^{(a)} \to -u_{j}^{(a)}, \label{cross2} \\ 
&& w_{j} \to -w_{j}. \nonumber 
\end{eqnarray}
${\cal T}_{m}(u)$ and ${\cal T}^{a}(u)$ 
are  invariant under the transformations (\ref{cross1}) 
or (\ref{cross2}). This invariance may be
 viewed as a kind of crossing symmetry.

Now we shall present examples of (\ref{Tge1})  
for  $B(2|1), J_{-}=\{1,\overline{1} \},
J_{+}=\{2,3,0,\overline{3},\overline{2} \}$ case: 
\begin{eqnarray}
{\cal T}^{1}(u) &=&
  -\begin{array}{|c|}\hline 
     	1  \\ \hline
  \end{array}
 +\begin{array}{|c|}\hline 
  	2  \\ \hline
  \end{array}
 +\begin{array}{|c|}\hline 
  	3  \\ \hline
  \end{array}
 +\begin{array}{|c|}\hline 
  	0  \\ \hline
  \end{array}
 +\begin{array}{|c|}\hline 
  	\stackrel{\ }{\overline{3}}  \\ \hline
  \end{array}
 +\begin{array}{|c|}\hline 
  	\stackrel{\ }{\overline{2}}  \\ \hline
  \end{array}
 -\begin{array}{|c|}\hline 
  	\stackrel{\ }{\overline{1}}  \\ \hline
  \end{array}
\nonumber \\ 
&=&
- \phi(-2 + u)\phi(1 + u)\frac{Q_{1}(1 + u)}{Q_{1}(-1 + u)} 
 \nonumber \\ 
  &+& 
  \phi(u)\phi(1 + u)\frac{Q_{1}(1 + u)Q_{2}(-2 + u)}{Q_{1}(-1 + u)Q_{2}(u)}
   \nonumber \\ 
   &+& 
  \phi(u)\phi(1 + u)\frac{Q_{1}(u)Q_{2}(3 + u)}{Q_{1}(2 + u)Q_{2}(1 + u)} 
  \nonumber \\
  &+& 
  \phi(u)\phi(1 + u)\frac{Q_{2}(2 + u)Q_{3}(-1 + u)}{Q_{2}(u)Q_{3}(1 + u)} 
  \nonumber \\
  &+&
  \phi(u)\phi(1 + u)\frac{Q_{2}(-1 + u)Q_{3}(2 + u)}{Q_{2}(1 + u)Q_{3}(u)} 
  \nonumber \\
  &+& 
  \phi(u)\phi(1 + u)\frac{Q_{3}(-1 + u)Q_{3}(2 + u)}{Q_{3}(u)Q_{3}(1 + u)}
  \nonumber \\ 
  &-& 
  \phi(u)\phi(3 + u)\frac{Q_{1}(u)}{Q_{1}(2 + u)} 
, 
\label{t1-ex}
\end{eqnarray}
\begin{eqnarray}
{\cal T}^{2}(u) &=&
   \begin{array}{|c|}\hline 
     	1  \\ \hline
     	1  \\ \hline 
   \end{array}
  -\begin{array}{|c|}\hline 
     	1  \\ \hline
     	2  \\ \hline 
   \end{array}
  -\begin{array}{|c|}\hline 
     	1  \\ \hline
     	3  \\ \hline 
   \end{array}
  -\begin{array}{|c|}\hline 
     	1  \\ \hline
     	0  \\ \hline 
   \end{array}
  -\begin{array}{|c|}\hline 
     	1  \\ \hline
     	\stackrel{\ }{\overline{3}}  \\ \hline 
   \end{array}
  -\begin{array}{|c|}\hline 
     	1  \\ \hline
     	\stackrel{\ }{\overline{2}}  \\ \hline 
   \end{array}
  +\begin{array}{|c|}\hline 
     	1  \\ \hline
     	\stackrel{\ }{\overline{1}}  \\ \hline 
   \end{array}
  +\begin{array}{|c|}\hline 
     	2  \\ \hline
     	3  \\ \hline 
   \end{array}
  +\begin{array}{|c|}\hline 
     	2  \\ \hline
     	0  \\ \hline 
   \end{array}
  +\begin{array}{|c|}\hline 
     	2  \\ \hline
     	\stackrel{\ }{\overline{3}} \\ \hline 
   \end{array}
   \nonumber \\
  &+& \begin{array}{|c|}\hline 
     	2  \\ \hline
     	\stackrel{\ }{\overline{2}}  \\ \hline 
   \end{array}
  -\begin{array}{|c|}\hline 
     	2  \\ \hline
     	\stackrel{\ }{\overline{1}}  \\ \hline 
   \end{array}
   +\begin{array}{|c|}\hline 
     	3  \\ \hline
     	0  \\ \hline 
   \end{array}
  +\begin{array}{|c|}\hline 
     	3  \\ \hline
     	\stackrel{\ }{\overline{3}}  \\ \hline 
   \end{array}
  +\begin{array}{|c|}\hline 
     	3  \\ \hline
     	\stackrel{\ }{\overline{2}}  \\ \hline 
   \end{array}
  -\begin{array}{|c|}\hline 
     	3  \\ \hline
     	\stackrel{\ }{\overline{1}}  \\ \hline 
   \end{array}
  +\begin{array}{|c|}\hline 
     	0  \\ \hline
     	0  \\ \hline 
   \end{array}
  +\begin{array}{|c|}\hline 
     	0  \\ \hline
     	\stackrel{\ }{\overline{3}}  \\ \hline 
   \end{array}
   +\begin{array}{|c|}\hline 
     	0  \\ \hline
     	\stackrel{\ }{\overline{2}}  \\ \hline 
   \end{array}
   -\begin{array}{|c|}\hline 
     	0  \\ \hline
     	\stackrel{\ }{\overline{1}}  \\ \hline 
   \end{array}
   \nonumber \\
   &+&\begin{array}{|c|}\hline 
     	\stackrel{\ }{\overline{3}}  \\ \hline
     	\stackrel{\ }{\overline{2}}  \\ \hline 
   \end{array}
  -\begin{array}{|c|}\hline 
     	\stackrel{\ }{\overline{3}}  \\ \hline
     	\stackrel{\ }{\overline{1}}  \\ \hline 
   \end{array}
  -\begin{array}{|c|}\hline 
     	\stackrel{\ }{\overline{2}}  \\ \hline
     	\stackrel{\ }{\overline{1}}  \\ \hline 
   \end{array}  
  +\begin{array}{|c|}\hline 
     	\stackrel{\ }{\overline{1}}  \\ \hline
     	\stackrel{\ }{\overline{1}}  \\ \hline 
   \end{array}
   \nonumber \\
&=& \phi(-1+ u)\phi(2+ u) 
\Bigg(
\phi(-3 + u)\phi(u)\frac{Q_{1}(2 + u)}{Q_{1}(-2 + u)} \nonumber \\ &+& 
 \phi(-1 + u)\phi(2 + u)\frac{Q_{1}(-1 + u)Q_{1}(2 + u)}
    {Q_{1}(u)Q_{1}(1 + u)} \nonumber \\ &+& 
 \phi(1 + u)\phi(4 + u)\frac{Q_{1}(-1 + u)}{Q_{1}(3 + u)} \nonumber \\ &-& 
  \phi(-1 + u)\phi(u)\frac{Q_{1}(2 + u)Q_{2}(-3 + u)}
    {Q_{1}(-2 + u)Q_{2}(-1 + u)} \nonumber \\ &-& 
  \phi(1 + u)\phi(2 + u)\frac{Q_{1}(-1 + u)Q_{1}(2 + u)
      Q_{2}(-1 + u)}{Q_{1}(u)Q_{1}(1 + u)Q_{2}(1 + u)} \nonumber \\ &-& 
  \phi(-1 + u)\phi(u)\frac{Q_{1}(-1 + u)Q_{1}(2 + u)Q_{2}(2 + u)}
    {Q_{1}(u)Q_{1}(1 + u)Q_{2}(u)} \nonumber \\ &+& 
  \phi(u)\phi(1 + u)\frac{Q_{1}(-1 + u)Q_{1}(2 + u)Q_{2}(-1 + u)
      Q_{2}(2 + u)}{Q_{1}(u)Q_{1}(1 + u)Q_{2}(u)Q_{2}(1 + u)} \nonumber \\ &-& 
  \phi(1 + u)\phi(2 + u)\frac{Q_{1}(-1 + u)Q_{2}(4 + u)}
    {Q_{1}(3 + u)Q_{2}(2 + u)} \nonumber \\ &+& 
  \phi(u)\phi(1 + u)\frac{Q_{1}(2 + u)Q_{3}(-2 + u)}
    {Q_{1}(u)Q_{3}(u)} \nonumber \\ 
    &-& \phi(-1 + u)\phi(u)\frac{Q_{1}(2 + u)
      Q_{2}(1 + u)Q_{3}(-2 + u)}{Q_{1}(u)Q_{2}(-1 + u)Q_{3}(u)}
       \nonumber \\ &-& 
  \phi(-1 + u)\phi(u)\frac{Q_{1}(2 + u)Q_{2}(-2 + u)Q_{3}(1 + u)}
    {Q_{1}(u)Q_{2}(u)Q_{3}(-1 + u)} \nonumber \\ &+& 
  \phi(u)\phi(1 + u)\frac{Q_{1}(2 + u)Q_{2}(-2 + u)Q_{2}(-1 + u)
      Q_{3}(1 + u)}{Q_{1}(u)Q_{2}(u)Q_{2}(1 + u)Q_{3}(-1 + u)} \nonumber \\ 
      &-& 
  \phi(-1 + u)\phi(u)\frac{Q_{1}(2 + u)Q_{3}(-2 + u)Q_{3}(1 + u)}
    {Q_{1}(u)Q_{3}(-1 + u)Q_{3}(u)} \nonumber \\ &+& 
  \phi(u)\phi(1 + u)\frac{Q_{1}(2 + u)Q_{2}(-1 + u)Q_{3}(-2 + u)
      Q_{3}(1 + u)}{Q_{1}(u)Q_{2}(1 + u)Q_{3}(-1 + u)Q_{3}(u)} \nonumber \\ &-& 
  \phi(1 + u)\phi(2 + u)\frac{Q_{1}(-1 + u)Q_{2}(3 + u)Q_{3}(u)}
    {Q_{1}(1 + u)Q_{2}(1 + u)Q_{3}(2 + u)} \nonumber \\ &+& 
  \phi(u)\phi(1 + u)\frac{Q_{1}(-1 + u)Q_{2}(2 + u)Q_{2}(3 + u)
      Q_{3}(u)}{Q_{1}(1 + u)Q_{2}(u)Q_{2}(1 + u)Q_{3}(2 + u)} \nonumber \\ &+& 
  \phi(u)\phi(1 + u)\frac{Q_{2}(3 + u)Q_{3}(-2 + u)Q_{3}(1 + u)}
    {Q_{2}(1 + u)Q_{3}(-1 + u)Q_{3}(2 + u)} \nonumber \\ &+& 
  \phi(u)\phi(1 + u)\frac{Q_{2}(-2 + u)Q_{2}(3 + u)Q_{3}(u)
      Q_{3}(1 + u)}{Q_{2}(u)Q_{2}(1 + u)Q_{3}(-1 + u)Q_{3}(2 + u)}
  \nonumber \\ 
   &+& \phi(u)\phi(1 + u)\frac{Q_{1}(-1 + u)Q_{3}(3 + u)}
    {Q_{1}(1 + u)Q_{3}(1 + u)} \nonumber \\ &-& 
  \phi(1 + u)\phi(2 + u)\frac{Q_{1}(-1 + u)Q_{2}(u)Q_{3}(3 + u)}
    {Q_{1}(1 + u)Q_{2}(2 + u)Q_{3}(1 + u)} \nonumber \\ &+& 
  \phi(u)\phi(1 + u)\frac{Q_{3}(-2 + u)Q_{3}(3 + u)}
    {Q_{3}(-1 + u)Q_{3}(2 + u)} \nonumber \\ &+& 
  \phi(u)\phi(1 + u)\frac{Q_{2}(-2 + u)Q_{3}(u)Q_{3}(3 + u)}
    {Q_{2}(u)Q_{3}(-1 + u)Q_{3}(2 + u)} \nonumber \\ &-& 
  \phi(1 + u)\phi(2 + u)\frac{Q_{1}(-1 + u)Q_{3}(u)Q_{3}(3 + u)}
    {Q_{1}(1 + u)Q_{3}(1 + u)Q_{3}(2 + u)} \nonumber \\ &+& 
  \phi(u)\phi(1 + u)\frac{Q_{1}(-1 + u)Q_{2}(2 + u)Q_{3}(u)
      Q_{3}(3 + u)}{Q_{1}(1 + u)Q_{2}(u)Q_{3}(1 + u)Q_{3}(2 + u)}
      \Bigg),
     \label{t2-ex}
\end{eqnarray}
\begin{eqnarray}
{\cal T}_{2}(u) &=&
-\begin{array}{|c|c|} \hline 
    1 & 2 \\ \hline 
\end{array}
-
\begin{array}{|c|c|}\hline
    1 & 3  \\ \hline
\end{array}
-
\begin{array}{|c|c|}\hline
    1 & 0  \\ \hline
\end{array}
-
\begin{array}{|c|c|}\hline
    1 & \stackrel{\ }{\overline{3}}  \\ \hline
\end{array}
-
\begin{array}{|c|c|}\hline
    1 & \stackrel{\ }{\overline{2}}  \\ \hline
\end{array}
+
\begin{array}{|c|c|}\hline
    1 & \stackrel{\ }{\overline{1}}  \\ \hline
\end{array} 
\nonumber \\ 
&+&
\begin{array}{|c|c|}\hline
    2 & 2  \\ \hline
\end{array}
+
\begin{array}{|c|c|}\hline
    2 & 3  \\ \hline
\end{array}
+
\begin{array}{|c|c|}\hline
    2 & 0  \\ \hline
\end{array}
+
\begin{array}{|c|c|}\hline
    2 & \stackrel{\ }{\overline{3}}  \\ \hline
\end{array} 
+
\begin{array}{|c|c|}\hline
    2 & \stackrel{\ }{\overline{2}}  \\ \hline
\end{array}
-
\begin{array}{|c|c|}\hline
    2 & \stackrel{\ }{\overline{1}}  \\ \hline
\end{array}
\nonumber \\ 
&+&
\begin{array}{|c|c|}\hline
    3 & 3  \\ \hline
\end{array}
+
\begin{array}{|c|c|}\hline
    3 & 0  \\ \hline
\end{array}
+
\begin{array}{|c|c|}\hline
    3 & \stackrel{\ }{\overline{3}}  \\ \hline
\end{array}
+
\begin{array}{|c|c|}\hline
    3 & \stackrel{\ }{\overline{2}}  \\ \hline
\end{array}
-
\begin{array}{|c|c|}\hline
    3 & \stackrel{\ }{\overline{1}}  \\ \hline
\end{array}
+
\begin{array}{|c|c|}\hline
    0 & \stackrel{\ }{\overline{3}}  \\ \hline
\end{array}
\nonumber \\ 
&+&
\begin{array}{|c|c|}\hline
    0 & \stackrel{\ }{\overline{2}}  \\ \hline
\end{array}
-
\begin{array}{|c|c|}\hline
    0 & \stackrel{\ }{\overline{1}}  \\ \hline
\end{array}
+
\begin{array}{|c|c|}\hline
    \stackrel{\ }{\overline{3}} & 
    \stackrel{\ }{\overline{3}}  \\ \hline
\end{array}
+
\begin{array}{|c|c|}\hline
    \stackrel{\ }{\overline{3}} & 
    \stackrel{\ }{\overline{2}}  \\ \hline
\end{array}
-
\begin{array}{|c|c|}\hline
    \stackrel{\ }{\overline{3}} & 
    \stackrel{\ }{\overline{1}}  \\ \hline
\end{array}
+
\begin{array}{|c|c|}\hline
    \stackrel{\ }{\overline{2}} & 
    \stackrel{\ }{\overline{2}}  \\ \hline
\end{array}
-
\begin{array}{|c|c|}\hline
    \stackrel{\ }{\overline{2}} & 
    \stackrel{\ }{\overline{1}}  \\ \hline
\end{array}
\nonumber \\ 
&=& \phi(u)\phi(1+ u)
\Bigg(
\phi(-3 + u)\phi(4 + u)\frac{Q_{1}(u)Q_{1}(1 + u)}{Q_{1}(-2 + u)Q_{1}(3 + u)}
 \nonumber \\ 
 &-& 
  \phi(-1 + u)\phi(4 + u)\frac{Q_{1}(u)Q_{1}(1 + u)Q_{2}(-3 + u)}
    {Q_{1}(-2 + u)Q_{1}(3 + u)Q_{2}(-1 + u)} \nonumber \\ &+& 
  \phi(-1 + u)\phi(2 + u)\frac{Q_{1}(2 + u)Q_{2}(-3 + u)}
    {Q_{1}(-2 + u)Q_{2}(1 + u)} \nonumber \\ &-& 
  \phi(-3 + u)\phi(2 + u)\frac{Q_{1}(2 + u)Q_{2}(-1 + u)}
    {Q_{1}(-2 + u)Q_{2}(1 + u)} \nonumber \\ &-& 
  \phi(-1 + u)\phi(4 + u)\frac{Q_{1}(-1 + u)Q_{2}(2 + u)}
    {Q_{1}(3 + u)Q_{2}(u)} \nonumber \\ &+& 
    \phi(-1 + u)\phi(2 + u)\frac{Q_{1}(-1 + u)
      Q_{2}(4 + u)}{Q_{1}(3 + u)Q_{2}(u)} \nonumber \\ &-& 
  \phi(-3 + u)\phi(2 + u)\frac{Q_{1}(u)Q_{1}(1 + u)Q_{2}(4 + u)}
    {Q_{1}(-2 + u)Q_{1}(3 + u)Q_{2}(2 + u)} \nonumber \\ &+& 
  \phi(-1 + u)\phi(2 + u)\frac{Q_{1}(u)Q_{1}(1 + u)Q_{2}(-3 + u)Q_{2}(4 + u)}
    {Q_{1}(-2 + u)Q_{1}(3 + u)Q_{2}(-1 + u)Q_{2}(2 + u)} \nonumber \\ &-& 
  \phi(-1 + u)\phi(4 + u)\frac{Q_{1}(1 + u)Q_{2}(1 + u)Q_{3}(-2 + u)}
    {Q_{1}(3 + u)Q_{2}(-1 + u)Q_{3}(u)} \nonumber \\ &+& 
  \phi(-1 + u)\phi(2 + u)\frac{Q_{1}(1 + u)Q_{2}(1 + u)Q_{2}(4 + u)
      Q_{3}(-2 + u)}{Q_{1}(3 + u)Q_{2}(-1 + u)Q_{2}(2 + u)Q_{3}(u)} \nonumber \\ &-& 
  \phi(-1 + u)\phi(4 + u)\frac{Q_{1}(1 + u)Q_{2}(-2 + u)Q_{3}(1 + u)}
    {Q_{1}(3 + u)Q_{2}(u)Q_{3}(-1 + u)} \nonumber \\ &+& 
  \phi(-1 + u)\phi(2 + u)\frac{Q_{1}(1 + u)Q_{2}(-2 + u)Q_{2}(4 + u)
      Q_{3}(1 + u)}{Q_{1}(3 + u)Q_{2}(u)Q_{2}(2 + u)Q_{3}(-1 + u)} \nonumber \\ &-& 
  \phi(-1 + u)\phi(4 + u)\frac{Q_{1}(1 + u)Q_{3}(-2 + u)Q_{3}(1 + u)}
    {Q_{1}(3 + u)Q_{3}(-1 + u)Q_{3}(u)} \nonumber \\ &+& 
  \phi(-1 + u)\phi(2 + u)\frac{Q_{1}(1 + u)Q_{2}(4 + u)Q_{3}(-2 + u)
      Q_{3}(1 + u)}{Q_{1}(3 + u)Q_{2}(2 + u)Q_{3}(-1 + u)Q_{3}(u)} \nonumber \\ &+& 
  \phi(-1 + u)\phi(2 + u)\frac{Q_{2}(3 + u)Q_{3}(-2 + u)}
    {Q_{2}(-1 + u)Q_{3}(2 + u)} \nonumber \\ &-& 
  \phi(-3 + u)\phi(2 + u)\frac{Q_{1}(u)Q_{2}(3 + u)Q_{3}(u)}
    {Q_{1}(-2 + u)Q_{2}(1 + u)Q_{3}(2 + u)} \nonumber \\ &+& 
  \phi(-1 + u)\phi(2 + u)\frac{Q_{1}(u)Q_{2}(-3 + u)Q_{2}(3 + u)Q_{3}(u)}
    {Q_{1}(-2 + u)Q_{2}(-1 + u)Q_{2}(1 + u)Q_{3}(2 + u)} \nonumber \\ &+& 
  \phi(-1 + u)\phi(2 + u)\frac{Q_{2}(-2 + u)Q_{3}(3 + u)}
    {Q_{2}(2 + u)Q_{3}(-1 + u)} \nonumber \\ &+& 
  \phi(-1 + u)\phi(2 + u)\frac{Q_{2}(u)Q_{3}(-2 + u)Q_{3}(3 + u)}
    {Q_{2}(2 + u)Q_{3}(-1 + u)Q_{3}(u)} \nonumber \\ &-& 
  \phi(-3 + u)\phi(2 + u)\frac{Q_{1}(u)Q_{2}(u)Q_{3}(3 + u)}
    {Q_{1}(-2 + u)Q_{2}(2 + u)Q_{3}(1 + u)} \nonumber \\ &+& 
  \phi(-1 + u)\phi(2 + u)\frac{Q_{1}(u)Q_{2}(-3 + u)Q_{2}(u)Q_{3}(3 + u)}
    {Q_{1}(-2 + u)Q_{2}(-1 + u)Q_{2}(2 + u)Q_{3}(1 + u)} \nonumber \\ &+& 
  \phi(-1 + u)\phi(2 + u)\frac{Q_{2}(u)Q_{2}(1 + u)Q_{3}(-2 + u)Q_{3}(3 + u)}
    {Q_{2}(-1 + u)Q_{2}(2 + u)Q_{3}(u)Q_{3}(1 + u)} \nonumber \\ &+& 
  \phi(-1 + u)\phi(2 + u)\frac{Q_{2}(1 + u)Q_{3}(-2 + u)Q_{3}(3 + u)}
    {Q_{2}(-1 + u)Q_{3}(1 + u)Q_{3}(2 + u)} \nonumber \\ &-& 
  \phi(-3 + u)\phi(2 + u)\frac{Q_{1}(u)Q_{3}(u)Q_{3}(3 + u)}
    {Q_{1}(-2 + u)Q_{3}(1 + u)Q_{3}(2 + u)} 
    \label{t21-ex}  \\ &+& 
  \phi(-1 + u)\phi(2 + u)\frac{Q_{1}(u)Q_{2}(-3 + u)Q_{3}(u)Q_{3}(3 + u)}
    {Q_{1}(-2 + u)Q_{2}(-1 + u)Q_{3}(1 + u)Q_{3}(2 + u)} 
\Bigg)
.  \nonumber
\end{eqnarray}
%
Thanks to Theorem \ref{th-tate} (see later) and 
the relation (\ref{Jacobi-Trudi1}),
 these DVFs are free of 
poles under the following BAE: 
\begin{eqnarray}
 && \frac{\phi(u_{k}^{(1)}-1)}{\phi(u_{k}^{(1)}+1)}
 =\frac{Q_{2}(u_{k}^{(1)}-1)}
       {Q_{2}(u_{k}^{(1)}+1)} \quad {\rm for } \quad 
  1 \le k \le N_{1}, \nonumber \\ 
 && -1=\frac{Q_{1}(u_{k}^{(2)}-1)
       Q_{2}(u_{k}^{(2)}+2)Q_{3}(u_{k}^{(2)}-1)}
         {Q_{1}(u_{k}^{(2)}+1)
         Q_{2}(u_{k}^{(2)}-2)Q_{3}(u_{k}^{(2)}+1)} 
         \quad {\rm for } \quad 
  1 \le k \le N_{2}, \nonumber \\ 
 && -1=\frac{Q_{2}(u_{k}^{(3)}-1)Q_{3}(u_{k}^{(3)}+1)}
         {Q_{2}(u_{k}^{(3)}+1)Q_{3}(u_{k}^{(3)}-1)},
          \quad {\rm for } \quad  
    1 \le k \le N_{3}.
\label{BAE2-b21}
\end{eqnarray}
Note that DVFs have so called 
{\em Bethe-strap} structures 
\cite{KS1,S2}, which bear resemblance to 
weight space diagrams.
 We have observed for many examples 
 that ${\cal T}_{\lambda \subset \mu}(u)$ coincides with 
 the Bethe-strap of the minimal connected component (cf \cite{K2}) 
 which
  include the top term \cite{KS1,KOS} 
 as the examples in 
 Figure \ref{best1}, Figure \ref{best2} and Figure \ref{best3}
\footnote{Recently we have found curious terms 
 (pseudo-top terms) in many Bethe-straps (cf. \cite{T4}). 
 However we have confirmed for several examples the fact that 
 the pseudo-top terms do not influence on  
 connectivity of the Bethe straps (cf \cite{KS1,KOS,K2}) 
 in the whole.}.  
 The top term of ${\cal T}_{\lambda \subset \mu}(u)$ carries 
 a $B(r|s)$ or $D(r|s)$ weight. 
 For example, for $B(r|s)$, $\lambda =\phi$, $\mu_{r+1}\le s$ 
 case, the term corresponding to the tableau 
\begin{equation}
 b(i,j)=
 \left\{
  \begin{array}{llll}
  j & {\rm for } & 1 \le i \le \mu_{j}^{\prime} & 1 \le j \le s \\
  i+s & {\rm for } & 1 \le i \le \mu_{j}^{\prime} & s+1 \le j \le \mu_{1}
  \end{array}
 \right.
\end{equation}
carries the 
$B(r|s)$ weight with the Kac-Dynkin label (\ref{Kac-Dynkin}) 
or (\ref{Kac-Dynkin-b0s}). 
The top term
\footnote{Here we omit the vacuum part.} 
\cite{KS1} of the DVF (\ref{Tge1}) for 
 $D(r|s)$, $\lambda =\phi, \mu=(1^{a})$ will be 
\begin{eqnarray}
 \left.
  (-1)^{a}
  \begin{array}{|c|}\hline 
     	1  \\ \hline
        1  \\ \hline
     	\vdots \\ \hline 
     	1  \\ \hline
  \end{array}
  \, 
 \right\}
 \! 
 {\tiny a} \; 
  =(-1)^{a}
\frac{Q_{1}(u+a)}{Q_{1}(u-a)},
\label{top-tate}
\end{eqnarray}
which carries the $D(r|s)$ weight with the Kac-Dynkin label 
 in (\ref{KD-tate}). 
The top term 
\footnote{Here we omit the vacuum part.} 
 \cite{KS1} of the DVF (\ref{Tge1}) for 
 $D(r|s)$, $\lambda =\phi, \mu=(m^{1})$ will be 
\begin{eqnarray}
&&
(-1)^{m}
\underbrace{
 \begin{array}{|c|c|c|c|}\hline 
     	1 & 2 & \cdots & m \\ \hline
  \end{array}
 }_{m}
=(-1)^{m} \frac{Q_{m}(u+1)}{Q_{m}(u-1)} 
\qquad {\rm if} \quad 1 \le m \le s,  
\nonumber \\ 
&& 
(-1)^{s}
\underbrace{
 \begin{array}{|c|c|c|c|c|c|c|}\hline 
     	1 & 2 & \cdots & s & s+1 & \cdots & s+1 \\ \hline
  \end{array}
 }_{m} \label{top-yoko} 
  \\ 
&& \quad =(-1)^{s} \frac{Q_{s}(u+m-s+1)Q_{s+1}(u-m+s)}
               {Q_{s}(u-m+s-1)Q_{s+1}(u+m-s)} \nonumber \\ 
&& \hspace{130pt} {\rm if} \quad r \ge 3 \quad 
{\rm and} \quad m \ge s+1, \nonumber \\ 
&& \quad =(-1)^{s} \frac{Q_{s}(u+m-s+1)Q_{s+1}(u-m+s)Q_{s+2}(u-m+s)}
               {Q_{s}(u-m+s-1)Q_{s+1}(u+m-s)Q_{s+2}(u+m-s)} 
 \nonumber \\ 
&& \hspace{130pt} {\rm if} \quad r =2 \quad 
{\rm and} \quad m \ge s+1, \nonumber
\end{eqnarray}
which carries the $D(r|s)$ weight    
with the Kac-Dynkin label in (\ref{KD-yoko}). 
 
{\em Remark}: 
There is a supposition (cf \cite{KOS,K2}) that the  
auxiliary space of a transfer matrix  
is a irreducible one as a representation space of the Yangian 
 (or quantum affine algebra) 
 if the Bethe strap of the DVF is connected
\footnote{Here the word "connected" means that any terms in DVF 
 are connected directly (or indirectly) each other 
by the arrows like graphs in Figures \ref{best1}-\ref{best3}.}
  in the whole. 
 Then a natural question arise: 
 "Is the Bethe strap of 
  ${\cal T}_{\lambda \subset \mu}(u)$ (\ref{Tge1}) always 
 connected in the whole ?"  The answer is no. 
 In fact for $D(r|s)$ case, 
 the Bethe strap of ${\cal T}^{a}(u)$  
 is not connected 
 if $0\le r-s-1 \le a \le 2(r-s-1)$.
  So it is desirable to 
 extract the minimal connected component of the Bethe strap 
 which contains the top term (\ref{top-tate})  
 from ${\cal T}^{a}(u)$.  
A candidate is as follows: 
\begin{eqnarray}
 {\cal T}^{a}(u)-h^{a}(u){\cal T}^{-a+2(r-s-1)}(u), 
\end{eqnarray}
where $h^{a}(u)=\prod_{j=1}^{a+1-r+s}
\psi_{1}(u+a-2j+1)\psi_{\overline{1}}(u-a+2j-1)$. 
 
For example, for $D(3|1)$ case, ${\cal T}^{2}(u)$ consists 
of 31 terms and they divide into 30 terms whose Bethe strap 
is connected in the whole
\footnote{In this case, 
$
-\begin{array}{|c|}\hline 
    2 \\ \hline 
    \overline{1} \\ \hline 
\end{array}
$ 
is a pseudo-top term (cf \cite{T4}).}
 and 1 isolated term 
$
h^{2}(u)=
\begin{array}{|c|}\hline 
    1 \\ \hline 
    \overline{1} \\ \hline 
\end{array}
=\{\phi(u-1)\phi(u+3)\}^{2}.
$ Thus  Bethe strap of ${\cal T}^{2}(u)-h^{2}(u)$ 
is connected in the whole. 
 On the other hand for $D(2|2)$ case, ${\cal T}^{2}(u)$ 
does not have such an isolated term and that the Bethe strap 
is connected in the whole 
(in this case, ${\cal T}^{2}(u)$ has 33 terms).  
So far this kind of an isolated term is peculiar to $D(r|s)$ case. 
In fact, we have never yet observed such an isolated term 
in $B(r|s)$ case.
Similarly, 
 the Bethe strap of ${\cal T}_{m}(u)$  for $D(r|s)$ 
 seems not to be connected 
 if $0\le s-r+1 \le m \le 2(s-r+1)$. 
 A candidate for the minimal connected component of the Bethe strap 
 which contains the top term (\ref{top-yoko}) is 
\begin{eqnarray}
 {\cal T}_{m}(u)-h_{m}(u){\cal T}_{-m+2(s-r+1)}(u), 
\end{eqnarray}
where $h_{m}(u)=\prod_{j=1}^{m+r-s-1}
\psi_{j}(u-m+2j-1)\psi_{\overline{j}}(u+m-2j+1)$.
 \rule{5pt}{10pt} \\ 
 A remarkable resemblance between 
 Bethe-straps for vector representations and crystal graphs 
 \cite{KN1,KN2} was pointed out in \cite{KS1}. 
 Whether such resemblance holds true for the Lie superalgebras 
 in general or not will be an interesting question.   
There is a remarkable coincidence between 
  currents of deformed Virasoro algebra  and  DVFs \cite{FR}. 
  Whether such coincidence holds true for the Lie superalgebras 
 in general or not will be another interesting question. 

\begin{figure}
    \setlength{\unitlength}{1.2pt}
    \begin{center}
    \begin{picture}(250,40) 
      \put(-8,3){$-$}
      \put(0,0){\line(1,0){10}}
      \put(10,0){\line(0,1){10}}
      \put(10,10){\line(-1,0){10}}
      \put(0,10){\line(0,-1){10}}
      \put(3,3){\scriptsize{$1$}}
      \put(12,5){\vector(1,0){20}}
      \put(15,9){\scriptsize{$(1,1)$}}
      \put(33,3){$$}
      \put(40,0){\line(1,0){10}}
      \put(50,0){\line(0,1){10}}
      \put(50,10){\line(-1,0){10}}
      \put(40,10){\line(0,-1){10}}
      \put(43,3){\scriptsize{$2$}}
      \put(52,5){\vector(1,0){20}}
      \put(55,9){\scriptsize{$(2,0)$}}
      \put(73,3){$$}
      \put(80,0){\line(1,0){10}}
      \put(90,0){\line(0,1){10}}
      \put(90,10){\line(-1,0){10}}
      \put(80,10){\line(0,-1){10}}
      \put(83,3){\scriptsize{$3$}}
      \put(92,5){\vector(1,0){20}}
      \put(95,9){\scriptsize{$(3,-1)$}}
      \put(113,3){$$}
      \put(120,0){\line(1,0){10}}
      \put(130,0){\line(0,1){10}}
      \put(130,10){\line(-1,0){10}}
      \put(120,10){\line(0,-1){10}}
      \put(123,3){\scriptsize{$0$}}
      \put(132,5){\vector(1,0){20}}
      \put(135,9){\scriptsize{$(3,0)$}}
      \put(153,3){$$}
      \put(160,0){\line(1,0){10}}
      \put(170,0){\line(0,1){10}}
      \put(170,10){\line(-1,0){10}}
      \put(160,10){\line(0,-1){10}}
      \put(163,3){\scriptsize{$\overline{3}$}}
      \put(172,5){\vector(1,0){20}}
      \put(175,9){\scriptsize{$(2,-1)$}}
      \put(197,0){\line(1,0){10}}
      \put(207,0){\line(0,1){10}}
      \put(207,10){\line(-1,0){10}}
      \put(197,10){\line(0,-1){10}}
      \put(200,3){\scriptsize{$\overline{2}$}}
      \put(212,5){\vector(1,0){20}}
      \put(215,9){\scriptsize{$(1,-2)$}}
      \put(232,3){$-$}
      \put(240,0){\line(1,0){10}}
      \put(250,0){\line(0,1){10}}
      \put(250,10){\line(-1,0){10}}
      \put(240,10){\line(0,-1){10}}
      \put(243,3){\scriptsize{$\overline{1}$}}
  \end{picture}
  \end{center}
  \caption{The  Bethe-strap structure of  
   ${\cal T}^{1}(u)$  (\ref{t1-ex}) for 
   $B(2|1)=osp(5|2)$:  
 The pair $(a,b)$ denotes the common pole $u_{k}^{(a)}+b$ of the pair   
 of the tableaux connected by the arrow.   
 This common pole vanishes under the BAE (\ref{BAE2-b21}).
 The leftmost tableau corresponds to the 
 \symbol{96}highest weight \symbol{39}, 
 which is called the {\em top term}. 
 Such a  correspondence between certain term in the DVF and a highest 
 weight (to be more precise, a kind of Drinfel'd polynomial 
 (cf \cite{D,KOS}))
  may be called {\em top term hypothesis} \cite{KS1,KOS}.}
  \label{best1}
\end{figure}
\begin{figure}
    \setlength{\unitlength}{1pt}
    \begin{center}
    \begin{picture}(250,470) 
      \put(120,0){\line(1,0){10}}
      \put(130,0){\line(0,1){20}}
      \put(130,20){\line(-1,0){10}}
      \put(120,20){\line(0,-1){20}}
      \put(120,10){\line(1,0){10}}
      \put(123,12){\scriptsize{$\overline{1}$}}
      \put(123,2){\scriptsize{$\overline{1}$}}
      \put(114,9){$$}
      \put(120,40){\line(1,0){10}}
      \put(130,40){\line(0,1){20}}
      \put(130,60){\line(-1,0){10}}
      \put(120,60){\line(0,-1){20}}
      \put(120,50){\line(1,0){10}}
      \put(123,52){\scriptsize{$\overline{2}$}}
      \put(123,42){\scriptsize{$\overline{1}$}}
      \put(110,47){$-$}
      \put(120,80){\line(1,0){10}}
      \put(130,80){\line(0,1){20}}
      \put(130,100){\line(-1,0){10}}
      \put(120,100){\line(0,-1){20}}
      \put(120,90){\line(1,0){10}}
      \put(123,92){\scriptsize{$\overline{3}$}}
      \put(123,82){\scriptsize{$\overline{1}$}}
      \put(110,87){$-$}
      \put(80,120){\line(1,0){10}}
      \put(90,120){\line(0,1){20}}
      \put(90,140){\line(-1,0){10}}
      \put(80,140){\line(0,-1){20}}
      \put(80,130){\line(1,0){10}}
      \put(83,132){\scriptsize{$\overline{3}$}}
      \put(83,122){\scriptsize{$\overline{2}$}}
      \put(160,120){\line(1,0){10}}
      \put(170,120){\line(0,1){20}}
      \put(170,140){\line(-1,0){10}}
      \put(160,140){\line(0,-1){20}}
      \put(160,130){\line(1,0){10}}
      \put(163,132){\scriptsize{$0$}}
      \put(163,122){\scriptsize{$\overline{1}$}}
      \put(149,126){$-$}
      \put(80,160){\line(1,0){10}}
      \put(90,160){\line(0,1){20}}
      \put(90,180){\line(-1,0){10}}
      \put(80,180){\line(0,-1){20}}
      \put(80,170){\line(1,0){10}}
      \put(83,172){\scriptsize{$0$}}
      \put(83,162){\scriptsize{$\overline{2}$}}
      \put(160,160){\line(1,0){10}}
      \put(170,160){\line(0,1){20}}
      \put(170,180){\line(-1,0){10}}
      \put(160,180){\line(0,-1){20}}
      \put(160,170){\line(1,0){10}}
      \put(163,172){\scriptsize{$3$}}
      \put(163,162){\scriptsize{$\overline{1}$}}
      \put(149,166){$-$}
      \put(40,200){\line(1,0){10}}
      \put(50,200){\line(0,1){20}}
      \put(50,220){\line(-1,0){10}}
      \put(40,220){\line(0,-1){20}}
      \put(40,210){\line(1,0){10}}
      \put(43,212){\scriptsize{$0$}}
      \put(43,202){\scriptsize{$\overline{3}$}}
      \put(120,200){\line(1,0){10}}
      \put(130,200){\line(0,1){20}}
      \put(130,220){\line(-1,0){10}}
      \put(120,220){\line(0,-1){20}}
      \put(120,210){\line(1,0){10}}
      \put(123,212){\scriptsize{$3$}}
      \put(123,202){\scriptsize{$\overline{2}$}}
      \put(200,200){\line(1,0){10}}
      \put(210,200){\line(0,1){20}}
      \put(210,220){\line(-1,0){10}}
      \put(200,220){\line(0,-1){20}}
      \put(200,210){\line(1,0){10}}
      \put(203,212){\scriptsize{$2$}}
      \put(203,202){\scriptsize{$\overline{1}$}}
      \put(190,206){$-$}
      \put(0,240){\line(1,0){10}}
      \put(10,240){\line(0,1){20}}
      \put(10,260){\line(-1,0){10}}
      \put(0,260){\line(0,-1){20}}
      \put(0,250){\line(1,0){10}}
      \put(3,252){\scriptsize{$0$}}
      \put(3,242){\scriptsize{$0$}}
      \put(80,240){\line(1,0){10}}
      \put(90,240){\line(0,1){20}}
      \put(90,260){\line(-1,0){10}}
      \put(80,260){\line(0,-1){20}}
      \put(80,250){\line(1,0){10}}
      \put(83,252){\scriptsize{$3$}}
      \put(83,242){\scriptsize{$\overline{3}$}}
      \put(160,240){\line(1,0){10}}
      \put(170,240){\line(0,1){20}}
      \put(170,260){\line(-1,0){10}}
      \put(160,260){\line(0,-1){20}}
      \put(160,250){\line(1,0){10}}
      \put(163,252){\scriptsize{$2$}}
      \put(163,242){\scriptsize{$\overline{2}$}}
      \put(240,240){\line(1,0){10}}
      \put(250,240){\line(0,1){20}}
      \put(250,260){\line(-1,0){10}}
      \put(240,260){\line(0,-1){20}}
      \put(240,250){\line(1,0){10}}
      \put(243,252){\scriptsize{$1$}}
      \put(243,242){\scriptsize{$\overline{1}$}}
      \put(40,280){\line(1,0){10}}
      \put(50,280){\line(0,1){20}}
      \put(50,300){\line(-1,0){10}}
      \put(40,300){\line(0,-1){20}}
      \put(40,290){\line(1,0){10}}
      \put(43,292){\scriptsize{$3$}}
      \put(43,282){\scriptsize{$0$}}
      \put(120,280){\line(1,0){10}}
      \put(130,280){\line(0,1){20}}
      \put(130,300){\line(-1,0){10}}
      \put(120,300){\line(0,-1){20}}
      \put(120,290){\line(1,0){10}}
      \put(123,292){\scriptsize{$2$}}
      \put(123,282){\scriptsize{$\overline{3}$}}
      \put(200,280){\line(1,0){10}}
      \put(210,280){\line(0,1){20}}
      \put(210,300){\line(-1,0){10}}
      \put(200,300){\line(0,-1){20}}
      \put(200,290){\line(1,0){10}}
      \put(203,292){\scriptsize{$1$}}
      \put(203,282){\scriptsize{$\overline{2}$}}
      \put(190,286){$-$}
      \put(80,320){\line(1,0){10}}
      \put(90,320){\line(0,1){20}}
      \put(90,340){\line(-1,0){10}}
      \put(80,340){\line(0,-1){20}}
      \put(80,330){\line(1,0){10}}
      \put(83,332){\scriptsize{$2$}}
      \put(83,322){\scriptsize{$0$}}
      \put(160,320){\line(1,0){10}}
      \put(170,320){\line(0,1){20}}
      \put(170,340){\line(-1,0){10}}
      \put(160,340){\line(0,-1){20}}
      \put(160,330){\line(1,0){10}}
      \put(163,332){\scriptsize{$1$}}
      \put(163,322){\scriptsize{$\overline{3}$}}
      \put(150,326){$-$}
      \put(80,360){\line(1,0){10}}
      \put(90,360){\line(0,1){20}}
      \put(90,380){\line(-1,0){10}}
      \put(80,380){\line(0,-1){20}}
      \put(80,370){\line(1,0){10}}
      \put(83,372){\scriptsize{$2$}}
      \put(83,362){\scriptsize{$3$}}
      \put(160,360){\line(1,0){10}}
      \put(170,360){\line(0,1){20}}
      \put(170,380){\line(-1,0){10}}
      \put(160,380){\line(0,-1){20}}
      \put(160,370){\line(1,0){10}}
      \put(163,372){\scriptsize{$1$}}
      \put(163,362){\scriptsize{$0$}}
      \put(150,366){$-$}
      \put(120,400){\line(1,0){10}}
      \put(130,400){\line(0,1){20}}
      \put(130,420){\line(-1,0){10}}
      \put(120,420){\line(0,-1){20}}
      \put(120,410){\line(1,0){10}}
      \put(123,412){\scriptsize{$1$}}
      \put(123,402){\scriptsize{$3$}}
      \put(110,406){$-$}
      \put(120,440){\line(1,0){10}}
      \put(130,440){\line(0,1){20}}
      \put(130,460){\line(-1,0){10}}
      \put(120,460){\line(0,-1){20}}
      \put(120,450){\line(1,0){10}}
      \put(123,452){\scriptsize{$1$}}
      \put(123,442){\scriptsize{$2$}}
      \put(110,446){$-$}
      \put(120,480){\line(1,0){10}}
      \put(130,480){\line(0,1){20}}
      \put(130,500){\line(-1,0){10}}
      \put(120,500){\line(0,-1){20}}
      \put(120,490){\line(1,0){10}}
      \put(123,492){\scriptsize{$1$}}
      \put(123,482){\scriptsize{$1$}}
      \put(114,489){$$}
      %
      \put(92,118){\vector(3,-2){26}}
      \put(102,113){\tiny{$(1,-1)$}}
      \put(52,198){\vector(3,-2){26}}
      \put(62,193){\tiny{$(2,0)$}}
      \put(132,198){\vector(3,-2){26}}
      \put(142,193){\tiny{$(1,-1)$}}
      \put(12,238){\vector(3,-2){26}}
      \put(22,233){\tiny{$(3,1)$}}
      \put(92,238){\vector(3,-2){26}}
      \put(102,233){\tiny{$(2,0)$}}
      \put(172,238){\vector(3,-2){26}}
      \put(182,233){\tiny{$(1,-1)$}}
      \put(52,278){\vector(3,-2){26}}
      \put(62,273){\tiny{$(3,1)$}}
      \put(132,278){\vector(3,-2){26}}
      \put(142,273){\tiny{$(2,0)$}}
      \put(212,278){\vector(3,-2){26}}
      \put(222,273){\tiny{$(1,-1)$}}
      \put(172,318){\vector(3,-2){26}}
      \put(182,313){\tiny{$(2,0)$}}
      \put(92,318){\vector(3,-2){26}}
      \put(102,313){\tiny{$(3,1)$}}
      \put(132,398){\vector(3,-2){26}}
      \put(142,393){\tiny{$(3,0)$}}
      \put(158,118){\vector(-3,-2){26}}
      \put(128,114){\tiny{$(3,-1)$}}
      \put(198,198){\vector(-3,-2){26}}
      \put(168,195){\tiny{$(2,-1)$}}
      \put(118,198){\vector(-3,-2){26}}
      \put(88,194){\tiny{$(3,-2)$}}
      \put(238,238){\vector(-3,-2){26}}
      \put(211,234){\tiny{$(1,0)$}}
      \put(158,238){\vector(-3,-2){26}}
      \put(128,235){\tiny{$(2,-1)$}}
      \put(78,238){\vector(-3,-2){26}}
      \put(49,235){\tiny{$(3,-2)$}}
      \put(38,278){\vector(-3,-2){26}}
      \put(8,275){\tiny{$(3,-2)$}}
      \put(118,278){\vector(-3,-2){26}}
      \put(88,275){\tiny{$(2,-1)$}}
      \put(198,278){\vector(-3,-2){26}}
      \put(170,273){\tiny{$(1,0)$}}
      \put(78,318){\vector(-3,-2){26}}
      \put(48,316){\tiny{$(2,-1)$}}
      \put(160,318){\vector(-3,-2){26}}
      \put(130,313){\tiny{$(1,0)$}}
      \put(118,398){\vector(-3,-2){26}}
      \put(89,394){\tiny{$(1,0)$}}
      %
      \put(125,77){\vector(0,-1){14}}
      \put(127,69){\tiny{$(2,-2)$}}
      \put(125,37){\vector(0,-1){14}}
      \put(127,29){\tiny{$(1,-3)$}}
      \put(85,157){\vector(0,-1){14}}
      \put(87,149){\tiny{$(3,-1)$}}
      \put(165,157){\vector(0,-1){14}}
      \put(167,149){\tiny{$(3,-2)$}}
      \put(85,357){\vector(0,-1){14}}
      \put(87,349){\tiny{$(3,0)$}}
      \put(165,357){\vector(0,-1){14}}
      \put(167,349){\tiny{$(3,1)$}}
      \put(125,437){\vector(0,-1){14}}
      \put(127,429){\tiny{$(2,1)$}}
      \put(125,477){\vector(0,-1){14}}
      \put(127,469){\tiny{$(1,2)$}}
      \put(94,160){\vector(3,-1){62}}
      \put(124,152){\tiny{$(1,-1)$}}
      \put(156,360){\vector(-3,-1){62}}
      \put(117,357){\tiny{$(1,0)$}}
  \end{picture}
  \end{center}
  \caption{The Bethe-strap structure of  
  ${\cal T}^{2}(u)$ (\ref{t2-ex}) for 
   $B(2|1)$:  
 The topmost tableau corresponds to 
 the {\em top term}.}
  \label{best2}
\end{figure}
\begin{figure}
    \setlength{\unitlength}{1.5pt}
    \begin{center}
    \begin{picture}(100,310) 
      \put(40,0){\line(1,0){20}}
      \put(60,0){\line(0,1){10}}
      \put(60,10){\line(-1,0){20}}
      \put(40,10){\line(0,-1){10}}
      \put(50,0){\line(0,1){10}}
      \put(43,2){$\overline{2}$}
      \put(53,2){$\overline{1}$}
      \put(32,3){$-$}
      \put(0,30){\line(1,0){20}}
      \put(20,30){\line(0,1){10}}
      \put(20,40){\line(-1,0){20}}
      \put(0,40){\line(0,-1){10}}
      \put(10,30){\line(0,1){10}}
      \put(3,32){$\overline{2}$}
      \put(13,32){$\overline{2}$}
      \put(80,30){\line(1,0){20}}
      \put(100,30){\line(0,1){10}}
      \put(100,40){\line(-1,0){20}}
      \put(80,40){\line(0,-1){10}}
      \put(90,30){\line(0,1){10}}
      \put(83,32){$\overline{3}$}
      \put(93,32){$\overline{1}$}
      \put(72,33){$-$}
      \put(0,60){\line(1,0){20}}
      \put(20,60){\line(0,1){10}}
      \put(20,70){\line(-1,0){20}}
      \put(0,70){\line(0,-1){10}}
      \put(10,60){\line(0,1){10}}
      \put(3,62){$\overline{3}$}
      \put(13,62){$\overline{2}$}
      \put(80,60){\line(1,0){20}}
      \put(100,60){\line(0,1){10}}
      \put(100,70){\line(-1,0){20}}
      \put(80,70){\line(0,-1){10}}
      \put(90,60){\line(0,1){10}}
      \put(83,62){$0$}
      \put(93,62){$\overline{1}$}
      \put(72,63){$-$}
      \put(-40,90){\line(1,0){20}}
      \put(-20,90){\line(0,1){10}}
      \put(-20,100){\line(-1,0){20}}
      \put(-40,100){\line(0,-1){10}}
      \put(-30,90){\line(0,1){10}}
      \put(-37,92){$\overline{3}$}
      \put(-27,92){$\overline{3}$}
      \put(40,90){\line(1,0){20}}
      \put(60,90){\line(0,1){10}}
      \put(60,100){\line(-1,0){20}}
      \put(40,100){\line(0,-1){10}}
      \put(50,90){\line(0,1){10}}
      \put(43,92){$0$}
      \put(53,92){$\overline{2}$}
      \put(120,90){\line(1,0){20}}
      \put(140,90){\line(0,1){10}}
      \put(140,100){\line(-1,0){20}}
      \put(120,100){\line(0,-1){10}}
      \put(130,90){\line(0,1){10}}
      \put(123,92){$3$}
      \put(133,92){$\overline{1}$}
      \put(112,93){$-$}
      \put(-40,120){\line(1,0){20}}
      \put(-20,120){\line(0,1){10}}
      \put(-20,130){\line(-1,0){20}}
      \put(-40,130){\line(0,-1){10}}
      \put(-30,120){\line(0,1){10}}
      \put(-37,122){$0$}
      \put(-27,122){$\overline{3}$}
      \put(40,120){\line(1,0){20}}
      \put(60,120){\line(0,1){10}}
      \put(60,130){\line(-1,0){20}}
      \put(40,130){\line(0,-1){10}}
      \put(50,120){\line(0,1){10}}
      \put(43,122){$3$}
      \put(53,122){$\overline{2}$}
      \put(120,120){\line(1,0){20}}
      \put(140,120){\line(0,1){10}}
      \put(140,130){\line(-1,0){20}}
      \put(120,130){\line(0,-1){10}}
      \put(130,120){\line(0,1){10}}
      \put(123,122){$2$}
      \put(133,122){$\overline{1}$}
      \put(112,123){$-$}
      \put(-40,150){\line(1,0){20}}
      \put(-20,150){\line(0,1){10}}
      \put(-20,160){\line(-1,0){20}}
      \put(-40,160){\line(0,-1){10}}
      \put(-30,150){\line(0,1){10}}
      \put(-37,152){$3$}
      \put(-27,152){$\overline{3}$}
      \put(40,150){\line(1,0){20}}
      \put(60,150){\line(0,1){10}}
      \put(60,160){\line(-1,0){20}}
      \put(40,160){\line(0,-1){10}}
      \put(50,150){\line(0,1){10}}
      \put(43,152){$2$}
      \put(53,152){$\overline{2}$}
      \put(120,150){\line(1,0){20}}
      \put(140,150){\line(0,1){10}}
      \put(140,160){\line(-1,0){20}}
      \put(120,160){\line(0,-1){10}}
      \put(130,150){\line(0,1){10}}
      \put(123,152){$1$}
      \put(133,152){$\overline{1}$}
      \put(114,153){$$}
      \put(-40,180){\line(1,0){20}}
      \put(-20,180){\line(0,1){10}}
      \put(-20,190){\line(-1,0){20}}
      \put(-40,190){\line(0,-1){10}}
      \put(-30,180){\line(0,1){10}}
      \put(-37,182){$3$}
      \put(-27,182){$0$}
      \put(40,180){\line(1,0){20}}
      \put(60,180){\line(0,1){10}}
      \put(60,190){\line(-1,0){20}}
      \put(40,190){\line(0,-1){10}}
      \put(50,180){\line(0,1){10}}
      \put(43,182){$2$}
      \put(53,182){$\overline{3}$}
      \put(120,180){\line(1,0){20}}
      \put(140,180){\line(0,1){10}}
      \put(140,190){\line(-1,0){20}}
      \put(120,190){\line(0,-1){10}}
      \put(130,180){\line(0,1){10}}
      \put(123,182){$1$}
      \put(133,182){$\overline{2}$}
      \put(112,183){$-$}
      \put(-40,210){\line(1,0){20}}
      \put(-20,210){\line(0,1){10}}
      \put(-20,220){\line(-1,0){20}}
      \put(-40,220){\line(0,-1){10}}
      \put(-30,210){\line(0,1){10}}
      \put(-37,212){$3$}
      \put(-27,212){$3$}
      \put(40,210){\line(1,0){20}}
      \put(60,210){\line(0,1){10}}
      \put(60,220){\line(-1,0){20}}
      \put(40,220){\line(0,-1){10}}
      \put(50,210){\line(0,1){10}}
      \put(43,212){$2$}
      \put(53,212){$0$}
      \put(120,210){\line(1,0){20}}
      \put(140,210){\line(0,1){10}}
      \put(140,220){\line(-1,0){20}}
      \put(120,220){\line(0,-1){10}}
      \put(130,210){\line(0,1){10}}
      \put(123,212){$1$}
      \put(133,212){$\overline{3}$}
      \put(112,213){$-$}
      \put(0,240){\line(1,0){20}}
      \put(20,240){\line(0,1){10}}
      \put(20,250){\line(-1,0){20}}
      \put(0,250){\line(0,-1){10}}
      \put(10,240){\line(0,1){10}}
      \put(3,242){$2$}
      \put(13,242){$3$}
      \put(80,240){\line(1,0){20}}
      \put(100,240){\line(0,1){10}}
      \put(100,250){\line(-1,0){20}}
      \put(80,250){\line(0,-1){10}}
      \put(90,240){\line(0,1){10}}
      \put(83,242){$1$}
      \put(93,242){$0$}
      \put(72,243){$-$}
      \put(0,270){\line(1,0){20}}
      \put(20,270){\line(0,1){10}}
      \put(20,280){\line(-1,0){20}}
      \put(0,280){\line(0,-1){10}}
      \put(10,270){\line(0,1){10}}
      \put(3,272){$2$}
      \put(13,272){$2$}
      \put(80,270){\line(1,0){20}}
      \put(100,270){\line(0,1){10}}
      \put(100,280){\line(-1,0){20}}
      \put(80,280){\line(0,-1){10}}
      \put(90,270){\line(0,1){10}}
      \put(83,272){$1$}
      \put(93,272){$3$}
      \put(72,273){$-$}
      \put(40,300){\line(1,0){20}}
      \put(60,300){\line(0,1){10}}
      \put(60,310){\line(-1,0){20}}
      \put(40,310){\line(0,-1){10}}
      \put(50,300){\line(0,1){10}}
      \put(43,302){$1$}
      \put(53,302){$2$}
      \put(32,303){$-$}
      \put(10,58){\vector(0,-1){16}}
      \put(12,49){\scriptsize{$(2,0)$}}
      \put(90,58){\vector(0,-1){16}}
      \put(92,49){\scriptsize{$(3,1)$}}
      \put(50,118){\vector(0,-1){16}}
      \put(52,109){\scriptsize{$(3,0)$}}
      \put(50,148){\vector(0,-1){16}}
      \put(52,139){\scriptsize{$(2,1)$}}
      \put(50,178){\vector(0,-1){16}}
      \put(52,169){\scriptsize{$(2,-2)$}}
      \put(50,208){\vector(0,-1){16}}
      \put(52,199){\scriptsize{$(3,-1)$}}
      \put(-30,118){\vector(0,-1){16}}
      \put(-28,109){\scriptsize{$(3,1)$}}
      \put(-30,148){\vector(0,-1){16}}
      \put(-28,139){\scriptsize{$(3,0)$}}
      \put(-30,178){\vector(0,-1){16}}
      \put(-28,169){\scriptsize{$(3,-1)$}}
      \put(-30,208){\vector(0,-1){16}}
      \put(-28,199){\scriptsize{$(3,-2)$}}
      \put(130,118){\vector(0,-1){16}}
      \put(132,109){\scriptsize{$(2,1)$}}
      \put(130,148){\vector(0,-1){16}}
      \put(132,139){\scriptsize{$(1,2)$}}
      \put(130,178){\vector(0,-1){16}}
      \put(132,169){\scriptsize{$(1,-3)$}}
      \put(130,208){\vector(0,-1){16}}
      \put(132,199){\scriptsize{$(2,-2)$}}
      \put(10,268){\vector(0,-1){16}}
      \put(12,259){\scriptsize{$(2,-1)$}}
      \put(90,268){\vector(0,-1){16}}
      \put(92,259){\scriptsize{$(3,-2)$}}
      \put(22,28){\vector(1,-1){16}}
      \put(31,21){\scriptsize{$(1,-3)$}}
      \put(-18,88){\vector(1,-1){16}}
      \put(-9,81){\scriptsize{$(2,-2)$}}
      \put(62,88){\vector(1,-1){16}}
      \put(71,81){\scriptsize{$(1,-3)$}}
      \put(22,238){\vector(1,-1){16}}
      \put(31,231){\scriptsize{$(3,-2)$}}
      \put(102,238){\vector(1,-1){16}}
      \put(111,231){\scriptsize{$(3,-1)$}}
      \put(62,298){\vector(1,-1){16}}
      \put(71,291){\scriptsize{$(2,-1)$}}
      \put(78,28){\vector(-1,-1){16}}
      \put(56,21){\scriptsize{$(2,0)$}}
      \put(118,88){\vector(-1,-1){16}}
      \put(96,81){\scriptsize{$(3,0)$}}
      \put(38,88){\vector(-1,-1){16}}
      \put(16,81){\scriptsize{$(3,1)$}}
      \put(-2,238){\vector(-1,-1){16}}
      \put(-23,232){\scriptsize{$(2,1)$}}
      \put(78,238){\vector(-1,-1){16}}
      \put(57,232){\scriptsize{$(1,2)$}}
      \put(38,298){\vector(-1,-1){16}}
      \put(17,292){\scriptsize{$(1,2)$}}
      \put(22,59){\vector(3,-1){55}}
      \put(46,54){\scriptsize{$(1,-3)$}}
      \put(-18,119){\vector(3,-1){55}}
      \put(6,114){\scriptsize{$(2,-2)$}}
      \put(62,119){\vector(3,-1){55}}
      \put(86,114){\scriptsize{$(1,-3)$}}
      \put(-18,149){\vector(3,-1){55}}
      \put(6,144){\scriptsize{$(2,-2)$}}
      \put(62,149){\vector(3,-1){55}}
      \put(86,144){\scriptsize{$(1,-3)$}}
      \put(38,179){\vector(-3,-1){55}}
      \put(6,176){\scriptsize{$(2,1)$}}
      \put(118,179){\vector(-3,-1){55}}
      \put(86,176){\scriptsize{$(1,2)$}}
      \put(38,209){\vector(-3,-1){55}}
      \put(6,206){\scriptsize{$(2,1)$}}
      \put(118,209){\vector(-3,-1){55}}
      \put(86,206){\scriptsize{$(1,2)$}}
      \put(78,269){\vector(-3,-1){55}}
      \put(46,266){\scriptsize{$(1,2)$}}
  \end{picture}
  \end{center}
  \caption{The Bethe-strap structure of 
  ${\cal T}_{2}(u)$ (\ref{t21-ex}) for $B(2|1)$:  
 The topmost  tableau corresponds to the {\em top term}.}
  \label{best3}
\end{figure}
 
We can prove (see Appendix A.1-A.3) 
the following Theorem, which
 is essential in the analytic Bethe ansatz.
\begin{theorem} \label{th-tate}
For   
$a\in {\bf Z}_{\ge 0}$, ${\cal T}^{a}(u)$  
((\ref{Tge1}) for $\lambda =\phi$, $\mu =(1^{a})$) 
is free of poles under the condition that
the BAEs {\rm (\ref{BAE})-(\ref{BAE4})} are valid
\footnote{
We consider the case that 
the solutions $\{ u_{j}^{(a)} \}$ of the 
BAEs (\ref{BAE})-(\ref{BAE4}) 
 have \symbol{96}generic' distribution: 
We assume that   
$u_{i}^{(a)}-u_{j}^{(a)} \ne (\alpha_{a}|\alpha_{a})$
 for any $i,j \in \{1,2,\dots , N_{a}\} $   
 and $a\in \{1,2,\dots, s+r \}$ ($i \ne j$)
 in BAEs (\ref{BAE})-(\ref{BAE4}). 
Moreover we assume that the color $b$ pole 
(see Appendix A.1-A.3) of ${\cal T}^{a}(u)$ 
 and the color $c$ pole 
 do not coincide each other if $b \ne c$. 
  We will need separate 
 consideration for the case where this assumption does not hold. 
 We also note that similar assumption was assumed in 
 \cite{T1,T2,T3,T4}.
}.
\end{theorem}
In proving Theorem \ref{th-tate}, we use the following lemmas.
\begin{lemma} \label{le-tate}
For  $r \in {\bf Z}_{\ge 2}$ and $b\in \{s+1,s+2, \dots,s+r-1\}$,   
\begin{eqnarray}
  \begin{array}{|c|l}\cline{1-1}
    b & _{v}  \\ \cline{1-1} 
    b+1 & _{v-2}  \\ \cline{1-1}
  \end{array}
  ,\qquad 
  \begin{array}{|c|l}\cline{1-1}
    \stackrel{\ }{\overline{b+1}} & _{v}  \\ \cline{1-1} 
    \stackrel{\ }{\overline{b}} & _{v-2}  \\ \cline{1-1}
  \end{array}
\end{eqnarray}
do not contain $Q_{b}$.
\end{lemma}
For $B(0|s)$ case, we use the following lemma: 
\begin{lemma} \label{le-yoko}
For  $b\in \{1,2, \dots,s-1\}$,   
\begin{eqnarray}
  \begin{array}{|c|c|}
   \multicolumn{2}{c}{\quad} \\ \hline 
     b & b+1 \\ \hline 
    \multicolumn{1}{c}{^{u}} & 
    \multicolumn{1}{c}{^{u+2}}
  \end{array}
  ,\qquad 
  \begin{array}{|c|c|}
   \multicolumn{2}{c}{\quad} \\ \hline 
     \overline{b+1} & \stackrel{\ }{\overline{b}} \\ \hline 
    \multicolumn{1}{c}{^{u}} & 
    \multicolumn{1}{c}{^{u+2}}
  \end{array} 
\end{eqnarray}
do not contain $Q_{b}$, and 
\begin{eqnarray}
  \begin{array}{|c|c|c|}
   \multicolumn{3}{c}{\quad} \\ \hline 
     s & 0 & \overline{s} \\ \hline 
    \multicolumn{1}{c}{^{u}} & 
    \multicolumn{1}{c}{^{u+2}} &
    \multicolumn{1}{c}{^{u+4}}
  \end{array}
\end{eqnarray}
does not contain $Q_{s}$.
\end{lemma}
Then owing to the relation (\ref{Jacobi-Trudi1}), 
$ {\cal T}_{\lambda \subset \mu}(u)$ for 
$B(r|s)$ is also free of poles under 
the condition that the BAEs (\ref{BAE})-(\ref{BAE4})
 are valid.	 
 Similarly, 
 owing to the relation (\ref{Jacobi-Trudi}), 
$ {\cal T}_{m}(u)$ for 
$D(r|s)$ is also free of poles under 
the condition that the BAE (\ref{BAE})
 is valid.
\eqreset
\section{Functional relations among DVFs}
Now we introduce the functional relations among DVFs.
 For $B(r|s)$ case, the following relation follows from 
 the determinant formulae 
 (\ref{Jacobi-Trudi1}) or (\ref{Jacobi-Trudi2}). 
\begin{eqnarray}
{\cal T}_{m}^{a}(u-1){\cal T}_{m}^{a}(u+1)
={\cal T}_{m-1}^{a}(u){\cal T}_{m+1}^{a}(u)+
{\cal T}_{m}^{a-1}(u){\cal T}_{m}^{a+1}(u) ,
\label{hirota}
\end{eqnarray}
where $m,a \in {\bf Z}_{\ge 1}$; 
${\cal T}_{m}^{0}(u)={\cal T}_{0}^{a}(u)=1$. 
This functional relation (\ref{hirota}) is a Hirota bilinear 
difference equation \cite{H} and 
can be proved  by using the Jacobi identity. 
 There is a constraint to (\ref{hirota}) 
follows from the relation 
(cf. \cite{DM,MRi} for $sl(r|s)$ case): \\ 
 {\em ${\cal T}_{\lambda \subset \mu}(u) =0$ if 
 $\lambda \subset \mu$ contains $a \times m$ rectangular subdiagram 
 ($a$: the number of row, $m$: the number of column) with 
$m \in {\bf Z}_{\ge 2s+2}$ and $a \in {\bf Z}_{\ge 2r+1}$.} \\ 
In particular, we have 
\begin{eqnarray}
{\cal T}_{m}^{a}(u)=0 \qquad {\rm if } \quad 
 m \in {\bf Z}_{\ge 2s+2} \quad {\rm and}
  \quad a \in {\bf Z}_{\ge 2r+1}.
\label{cons}
\end{eqnarray}
We also note that the determinant formula (\ref{Jacobi-Trudi}) 
 for $D(r|s)$  reduces to the following functional relation:
\begin{equation}
{\cal T}^{1}(u-1){\cal T}^{1}(u+1)
={\cal T}_{2}(u)+{\cal T}^{2}(u) ,
\end{equation}
if we set $m=2$.

In this section, we consider only $B(0|s)$ 
($s \in {\bf Z}_{\ge 1}$) case from now on. 
Now we redefine the function ${\cal T}_{\lambda \subset \mu}(u)$
 as follows:
\begin{eqnarray}
{\cal T}_{\lambda \subset \mu}(u):=
{\cal T}_{\lambda \subset \mu}(u)/
\{
\prod_{j=1}^{\mu_{1}^{\prime}}
{\cal F}_{\mu_{j}-\lambda_{j}}
(u-\mu_{1}+\mu_{1}^{\prime}+\mu_{j}+\lambda_{j}-2j+1)
\}, 
\end{eqnarray}
where 
\begin{eqnarray}
 && {\cal F}_{m}(u)=
\prod_{j=1}^{m-1}\phi(u-m+2j+1)\phi(u-2s-m+2j-2) \nonumber \\ 
 && \hspace{160pt} {\rm for} \quad m \in {\bf Z}_{\ge 2}, 
\end{eqnarray}
and 
\begin{eqnarray} 
{\cal F}_{1}(u)=1, \qquad  
{\cal F}_{0}(u)=\{\phi(u+1)\phi(u-2s-2)\}^{-1}. 
\end{eqnarray}
In particular, we have 
\begin{eqnarray}
{\cal T}_{m}(u)&:=&{\cal T}_{m}(u)/{\cal F}_{m}(u), \\ 
{\cal T}_{0}^{a}(u)&=&\prod_{j=1}^{a} {\cal T}_{0}(u+a-2j+1)
 \nonumber \\ 
 &=& \prod_{j=1}^{a} \phi(u+a-2j+2)\phi(u+a-2j-2s-1). 
\end{eqnarray}
There is remarkable duality for ${\cal T}_{m}(u)$. 
\begin{theorem}\label{dual-th}
For any $m \in \{0,1,\dots,2s+1 \}$, we have 
\begin{eqnarray}
 {\cal T}_{m}(u)={\cal T}_{2s-m+1}(u).
  \label{dual1}
\end{eqnarray}
\end{theorem}
{\em Outline of the proof}: 
At first, we consider the case
 that the vacuum parts are formally trivial. 
In proving the relation (\ref{dual1}), we use the following relations, 
which can be verified by direct computation. 
\begin{eqnarray}
  \begin{array}{|c|}
   \multicolumn{1}{c}{\quad} \\ \hline 
     \overline{a} \\ \hline 
    \multicolumn{1}{c}{^{u}}
  \end{array}
  & \times &
  \begin{array}{|c|c|c|c|}
   \multicolumn{4}{c}{\quad} \\ \hline 
     1 & 2 & \cdots \cdots & a \\ \hline 
    \multicolumn{1}{c}{^{u-2s-1}} & 
    \multicolumn{1}{c}{^{u-2s+1}} &
    \multicolumn{1}{c}{} &
    \multicolumn{1}{c}{^{u+2a-2s-3}}
  \end{array}
  \nonumber \\ 
  &=&
  \begin{array}{|c|c|c|c|}
   \multicolumn{4}{c}{\quad} \\ \hline 
    1 & 2 & \cdots \cdots & a-1 \\ \hline 
    \multicolumn{1}{c}{^{u-2s+1}} & 
    \multicolumn{1}{c}{^{u-2s+3}} &
    \multicolumn{1}{c}{} &
    \multicolumn{1}{c}{^{u+2a-2s-3}}
  \end{array}
  \label{modi1},
\end{eqnarray} 
\begin{eqnarray}
  \begin{array}{|c|}
   \multicolumn{1}{c}{\quad} \\ \hline 
     a \\ \hline 
    \multicolumn{1}{c}{^{u}}
  \end{array}
  & \times &
  \begin{array}{|c|c|c|c|}
   \multicolumn{4}{c}{\quad} \\ \hline 
     \stackrel{\ }{\overline{a}} & 
        \cdots \cdots & \overline{2} & 
         \stackrel{\ }{\overline{1}} \\ \hline 
    \multicolumn{1}{c}{^{u-2a+2s+3}} & 
    \multicolumn{1}{c}{} &
    \multicolumn{1}{c}{^{u+2s-1}} &
    \multicolumn{1}{c}{^{u+2s+1}}
  \end{array}
  \nonumber \\ 
  &=&
  \begin{array}{|c|c|c|c|}
   \multicolumn{4}{c}{\quad} \\ \hline 
     \stackrel{\ }{\overline{a-1}} & 
            \cdots \cdots & \overline{2} & 
              \stackrel{\ }{\overline{1}} \\ \hline 
    \multicolumn{1}{c}{^{u-2a+2s+3}} & 
    \multicolumn{1}{c}{} &
    \multicolumn{1}{c}{^{u+2s-3}} &
    \multicolumn{1}{c}{^{u+2s-1}}
  \end{array}
  \label{modi}
\end{eqnarray} 
and  
\begin{eqnarray}
\begin{array}{|c|c|c|c|c|c|c|c|c|}
   \multicolumn{9}{c}{\quad} \\ \hline 
     1 & 2 & \cdots  \cdots & s & 0 & 
     \stackrel{\ }{\overline{s}} & \cdots \cdots & 
     \overline{2} & \stackrel{\ }{\overline{1}} \\ \hline 
    \multicolumn{1}{c}{^{u-2s}} & 
    \multicolumn{1}{c}{^{u-2s+2}} &
    \multicolumn{1}{c}{} &
    \multicolumn{1}{c}{^{u-2}} &
    \multicolumn{1}{c}{^{u}} &
    \multicolumn{1}{c}{^{u+2}} &
    \multicolumn{1}{c}{} & 
    \multicolumn{1}{c}{^{u-2s-2}} &
    \multicolumn{1}{c}{^{u+2s}}
\end{array}
  =1 \label{const},
\end{eqnarray}
where $a \in \{1,2,\dots, s+1 \}$ 
\footnote{Here we assume 
$\framebox{$s+1$}=\framebox{$\overline{s+1}$}=\framebox{$0$}$.}; 
  the spectral parameter increases  (cf. (\ref{Tge1})) 
  as we go from the left to the right on each tableau. 
We will show that any term in ${\cal T}_{m}(u)$ coincides with 
a term in ${\cal T}_{2s-m+1}(u)$. 
We will consider the signs originated from 
the grading parameter (\ref{grading}) separately. 
Any term in ${\cal T}_{m}(u)$ can be expressed by a tableau 
$b \in B((m^{1}))$ such that 
$b(1,k)=i_{k}$ for $1\le k \le \alpha $ 
($1 \preceq i_{1} \prec \cdots \prec i_{\alpha} \preceq s$); 
$b(1,k)=\overline{j_{m-k+1}}$ for $\alpha +1\le k \le m $
 ($0 \preceq \overline{j_{m-\alpha}}
  \prec \cdots \prec \overline{j_{1}} \preceq \overline{1}$); 
  $\alpha \in {\bf Z}$. 
The term corresponding to this tableau is given as follows
\footnote{(\ref{mod0}) is $1$ if $m=0$.}: 
\begin{eqnarray}
\hspace{-20pt}  && \begin{array}{|c|c|c|c|c|c|}
   \multicolumn{6}{c}{\quad} \\ \hline 
     i_{1} & \cdots \cdots & i_{\alpha} & 
     \stackrel{\ }{\overline{j_{m-\alpha}}} & 
       \cdots \cdots & \overline{j_{1}} 
     \\ \hline 
    \multicolumn{1}{c}{^{u-m+1}} & 
    \multicolumn{1}{c}{} &
    \multicolumn{1}{c}{^{u-m+2\alpha -1}} &
    \multicolumn{1}{c}{^{u-m+2\alpha +1}} & 
    \multicolumn{1}{c}{} &
    \multicolumn{1}{c}{^{u+m-1}}
  \end{array}
  \label{mod0} \\ 
 \hspace{-20pt} &=&
  \begin{array}{|c|c|c|c|c|c|}
   \multicolumn{6}{c}{\quad} \\ \hline 
     i_{1} & \cdots \cdots & i_{\alpha} & 
     \stackrel{\ }{\overline{j_{m-\alpha}}} & 
       \cdots \cdots & \overline{j_{1}} \\ \hline 
    \multicolumn{1}{c}{^{u-m+1}} & 
    \multicolumn{1}{c}{} &
    \multicolumn{1}{c}{^{u-m+2\alpha -1}} &
    \multicolumn{1}{c}{^{u-m+2\alpha +1}} & 
    \multicolumn{1}{c}{} &
    \multicolumn{1}{c}{^{u+m-1}}
  \end{array}
  \nonumber \\ 
  \hspace{-20pt} && \times 
  \begin{array}{|c|c|c|c|c|c|c|c|c|}
   \multicolumn{9}{c}{\quad} \\ \hline 
     1 & 2 & \cdots \cdots  & s & 0 & 
     \overline{s} & \cdots \cdots & 
     \stackrel{\ }{\overline{2}} & \overline{1} \\ \hline 
    \multicolumn{1}{c}{^{u-m+2\alpha -2s}} & 
    \multicolumn{7}{c}{} &
    \multicolumn{1}{c}{^{u-m+2\alpha +2s}}
  \end{array}
  \label{mod2} \\ 
  \hspace{-20pt} &=& 
  \begin{array}{|c|c|c|}
   \multicolumn{3}{c}{\quad} \\ \hline 
     i_{1} & \cdots \cdots & i_{\alpha-1}  \\ \hline 
    \multicolumn{1}{c}{^{u-m+1}} & 
    \multicolumn{1}{c}{} &
    \multicolumn{1}{c}{^{u-m+2\alpha -3}} 
  \end{array}
  \times 
  \begin{array}{|c|c|c|}
   \multicolumn{3}{c}{\quad} \\ \hline  
  \stackrel{\ }{\overline{j_{m-\alpha-1}}} & 
     \cdots \cdots & \overline{j_{1}} \\ \hline 
    \multicolumn{1}{c}{^{u-m+2\alpha +3}} & 
    \multicolumn{1}{c}{} &
    \multicolumn{1}{c}{^{u+m-1}}
  \end{array}
  \nonumber \\ 
  \hspace{-20pt} && \times 
  \begin{array}{|c|}
   \multicolumn{1}{c}{\quad} \\ \hline  
  \stackrel{\ }{\overline{j_{m-\alpha}}} \\ \hline 
    \multicolumn{1}{c}{^{u-m+2\alpha +1}}
  \end{array}
  \times 
  \begin{array}{|c|c|c|c|}
   \multicolumn{4}{c}{\quad} \\ \hline 
     1 & 2 & \cdots \cdots & j_{m-\alpha}  \\ \hline 
    \multicolumn{1}{c}{^{u-m+2\alpha-2s}} & 
    \multicolumn{2}{c}{} &
    \multicolumn{1}{c}{^{u-m+2\alpha -2s+2j_{m-\alpha}-2}} 
  \end{array}
  \nonumber \\ 
  \hspace{-20pt} && \times 
  \begin{array}{|c|c|c|c|c|c|c|}
   \multicolumn{7}{c}{\quad} \\ \hline 
     j_{m-\alpha}+1 & \cdots \cdots & s & 0 & \overline{s} &
     \cdots \cdots & 
       \stackrel{\ }{\overline{i_{\alpha}+1}}  \\ \hline 
    \multicolumn{1}{c}{^{u-m+2\alpha-2s+2j_{m-\alpha}}} & 
    \multicolumn{5}{c}{} &
    \multicolumn{1}{c}{^{u-m+2\alpha +2s-2i_{\alpha}}} 
  \end{array}
  \nonumber \\ 
  \hspace{-20pt} && \times 
  \begin{array}{|c|}
   \multicolumn{1}{c}{\quad} \\ \hline  
  i_{\alpha} \\ \hline 
    \multicolumn{1}{c}{^{u-m+2\alpha -1}}
  \end{array}
  \times 
  \begin{array}{|c|c|c|c|}
   \multicolumn{4}{c}{\quad} \\ \hline 
     \stackrel{\ }{\overline{i_{\alpha}}} & 
       \cdots \cdots & \overline{2} & \overline{1} \\ \hline 
    \multicolumn{1}{c}{^{u-m+2\alpha-2i_{\alpha}+2s+2}} & 
    \multicolumn{2}{c}{} &
    \multicolumn{1}{c}{^{u-m+2\alpha +2s}} 
  \end{array}
  \label{mod3} \\ 
  \hspace{-20pt} &=& 
  \begin{array}{|c|c|c|}
   \multicolumn{3}{c}{\quad} \\ \hline 
     i_{1} & \cdots \cdots & i_{\alpha-1}  \\ \hline 
    \multicolumn{1}{c}{^{u-m+1}} & 
    \multicolumn{1}{c}{} &
    \multicolumn{1}{c}{^{u-m+2\alpha -3}} 
  \end{array}
  \times 
  \begin{array}{|c|c|c|}
   \multicolumn{3}{c}{\quad} \\ \hline  
  \stackrel{\ }{\overline{j_{m-\alpha-1}}} & 
    \cdots \cdots & \overline{j_{1}} \\ \hline 
    \multicolumn{1}{c}{^{u-m+2\alpha +3}} & 
    \multicolumn{1}{c}{} &
    \multicolumn{1}{c}{^{u+m-1}}
  \end{array}
  \nonumber \\ 
  \hspace{-20pt} && 
  \times 
  \begin{array}{|c|c|c|c|}
   \multicolumn{4}{c}{\quad} \\ \hline 
     1 & 2 & \cdots \cdots & j_{m-\alpha}-1  \\ \hline 
    \multicolumn{1}{c}{^{u-m+2\alpha-2s+2}} & 
    \multicolumn{2}{c}{} &
    \multicolumn{1}{c}{^{u-m+2\alpha -2s+2j_{m-\alpha}-2}} 
  \end{array}
  \nonumber \\ 
  \hspace{-20pt} && \times 
  \begin{array}{|c|c|c|c|c|c|c|}
   \multicolumn{7}{c}{\quad} \\ \hline 
     j_{m-\alpha}+1 & \cdots \cdots & s & 0 & \overline{s} &
     \cdots \cdots & 
       \stackrel{\ }{\overline{i_{\alpha}+1}}  \\ \hline 
    \multicolumn{1}{c}{^{u-m+2\alpha-2s+2j_{m-\alpha}}} & 
    \multicolumn{5}{c}{} &
    \multicolumn{1}{c}{^{u-m+2\alpha +2s-2i_{\alpha}}} 
  \end{array}
  \nonumber \\ 
  \hspace{-20pt} && \times  
  \begin{array}{|c|c|c|c|}
   \multicolumn{4}{c}{\quad} \\ \hline 
     \stackrel{\ }{\overline{i_{\alpha}-1}} & 
        \cdots \cdots & \overline{2} &
       \overline{1} \\ \hline 
    \multicolumn{1}{c}{^{u-m+2\alpha-2i_{\alpha}+2s+2}} & 
    \multicolumn{2}{c}{} &
    \multicolumn{1}{c}{^{u-m+2\alpha +2s-2}} 
  \end{array}
   \label{mod4} \\
 \hspace{-20pt} &=& 
  \cdots = \begin{array}{|c|c|c|c|c|c|}
   \multicolumn{6}{c}{\quad} \\ \hline 
     J_{1} & \cdots \cdots & J_{s+1-m+\alpha} & 
     \stackrel{\ }{\overline{I_{s-\alpha}}} & 
       \cdots \cdots & \overline{I_{1}} \\ \hline 
    \multicolumn{1}{c}{^{u+m-2s}} & 
    \multicolumn{1}{c}{} &
    \multicolumn{1}{c}{^{u-m+2\alpha }} &
    \multicolumn{1}{c}{^{u-m+2\alpha +2}} & 
    \multicolumn{1}{c}{} &
    \multicolumn{1}{c}{^{u-m+2s}}
  \end{array},
\label{mod5}
\end{eqnarray}
where $\{J_{k}\}= \{1,2,\dots, s,0 \} 
\setminus \{j_{1},j_{2},\dots,j_{m-\alpha}\}$
 ($1 \preceq J_{1} \prec \cdots \prec J_{s+1+\alpha -m} \preceq 0$); 
$\{\overline{I_{k}}\}= \{\overline{s},\overline{s-1},\dots, \overline{1} \}
 \setminus \{\overline{i_{\alpha}},\overline{i_{\alpha-1}},
 \dots,\overline{i_{1}} \}$
 ($\overline{s} \preceq \overline{I_{s-\alpha}}
  \prec \cdots \prec \overline{I_{1}} \preceq \overline{1}$). 
 (\ref{mod2}) follows from (\ref{const}); 
 (\ref{mod4}) follows from (\ref{modi1}) and (\ref{modi}). 
 After repetition of procedures similar to 
 (\ref{mod3})-(\ref{mod4}), we obtain (\ref{mod5}).  
 Apparently, (\ref{mod5}) is a term in ${\cal T}_{2s-m+1}(u)$. 
 Conversely, one can also show that any term in 
 ${\cal T}_{2s-m+1}(u)$ coincides 
 with a term in ${\cal T}_{m}(u)$. 

Noting the relation 
\begin{eqnarray}
\hspace{-30pt}&&\sum_{k=1}^{s+1-m+\alpha}p(J_{k})+
  \sum_{k=1}^{s-\alpha}p(\overline{I_{k}}) 
  \nonumber \\ 
\hspace{-30pt}&& \hspace{60pt}
=
\left(
 \sum_{k=1}^{s}p(k)+p(0)-\sum_{k=1}^{m-\alpha}p(j_{k})
\right)+
\left(
 \sum_{k=1}^{s}p(\overline{k})-
 \sum_{k=1}^{\alpha}p(\overline{i_{k}})
\right) \nonumber \\ 
\hspace{-30pt}&& \hspace{60pt}=
2s-\sum_{k=1}^{m-\alpha}p(j_{k})
-\sum_{k=1}^{\alpha}p(\overline{i_{k}}) 
 \nonumber \\ 
\hspace{-30pt}&& \hspace{60pt} 
 \equiv 
\sum_{k=1}^{m-\alpha}p(j_{k})
+\sum_{k=1}^{\alpha}p(\overline{i_{k}}) 
\quad {\rm mod} \quad 2, 
\end{eqnarray}
we find that 
 the overall sign for (\ref{mod0}) coincides with that for (\ref{mod5}). 

Finally, we comment on the vacuum parts. 
From now on, we assume that the vacuum parts are not trivial. 
Equivalence between the dress parts of ${\cal T}_{m}(u)$ 
and ${\cal T}_{2s-m+1}(u)$ has already been shown, 
so we have only to check that 
the vacuum part of 
\begin{eqnarray}
  \begin{array}{|c|c|c|c|c|c|}
   \multicolumn{6}{c}{\quad} \\ \hline 
     i_{1} & \cdots \cdots & i_{\alpha} & 
     \stackrel{\ }{\overline{j_{m-\alpha}}} & 
       \cdots \cdots & \overline{j_{1}} 
     \\ \hline 
    \multicolumn{1}{c}{^{u-m+1}} & 
    \multicolumn{1}{c}{} &
    \multicolumn{1}{c}{^{u-m+2\alpha -1}} &
    \multicolumn{1}{c}{^{u-m+2\alpha +1}} & 
    \multicolumn{1}{c}{} &
    \multicolumn{1}{c}{^{u+m-1}}
  \end{array}
  \ / {\cal F}_{m}(u) 
\end{eqnarray}
is equivalent to that of 
\begin{eqnarray}
\hspace{-10pt}
\begin{array}{|c|c|c|c|c|c|}
   \multicolumn{6}{c}{\quad} \\ \hline 
     J_{1} & \cdots \cdots & J_{s+1-m+\alpha} & 
     \stackrel{\ }{\overline{I_{s-\alpha}}} & 
       \cdots \cdots & \overline{I_{1}} 
     \\ \hline 
    \multicolumn{1}{c}{^{u+m-2s}} & 
    \multicolumn{1}{c}{} &
    \multicolumn{1}{c}{^{u-m+2\alpha }} &
    \multicolumn{1}{c}{^{u-m+2\alpha +2}} & 
    \multicolumn{1}{c}{} &
    \multicolumn{1}{c}{^{u-m+2s}}
  \end{array}
 \ / {\cal F}_{2s-m+1}(u). 
\end{eqnarray}
All we have to do is to check this  
by direct computation for the following 
four cases: 
(i) $i_{1}=1$ and $\overline{j_{1}}=\overline{1}$ 
($1 \prec J_{1}$ and $\overline{I_{1}} \prec \overline{1}$); 
(ii) $i_{1}=1$ and $\overline{j_{1}} \prec \overline{1}$ 
($J_{1}=1$ and $\overline{I_{1}} \prec \overline{1}$);
(iii) $1 \prec i_{1}$ and $\overline{j_{1}}=\overline{1}$ 
($1 \prec J_{1}$ and $\overline{I_{1}}=\overline{1}$);
(iv) $1 \prec i_{1}$ and $\overline{j_{1}} \prec \overline{1}$ 
($J_{1}=1$ and $\overline{I_{1}}=\overline{1}$). 
\rule{5pt}{10pt} \\ 
 Owing to the relation (\ref{Jacobi-Trudi2}), 
we can generalize the relation (\ref{dual1}) to  
\begin{eqnarray}
 {\cal T}_{m}^{a}(u)={\cal T}_{2s-m+1}^{a}(u) 
 \label{dual2} ,
\end{eqnarray}
where $a \in {\bf Z}_{\ge 1}$. 
Taking note on the relations (\ref{dual2}) and (\ref{cons}), 
 we shall rewrite the functional relation (\ref{hirota}) in a
  \symbol{96}canonical' form 
as the original $T$-system  for the simple Lie algebra \cite{KNS1}. 
Set 
$T_{m}^{(a)}(u)={\cal T}_{a}^{m}(u)$, 
$T_{2m}^{(s)}(u)={\cal T}_{s}^{m}(u)$ and  
$T_{0}^{(a)}(u)=T_{0}^{(s)}(u)=T_{m}^{(0)}(u)=1$
 for $a \in \{1,2,\dots, s-1\}$ and $m \in {\bf Z}_{\ge 1}$, 
where the subscript $(n,a)$ of $T_{n}^{(a)}(u)$  
corresponds to the Kac-Dynkin label $[b_{1}, b_{2},\dots ,b_{s}]$ 
 for $b_{i}=n\delta_{i a}$ (cf. (\ref{Kac-Dynkin-b0s})).
  Then we have 
\begin{eqnarray}
\hspace{-40pt} 
&& T_{m}^{(a)}(u-1)T_{m}^{(a)}(u+1)
=T_{m-1}^{(a)}(u)T_{m+1}^{(a)}(u)+
g_{m}^{(a)}(u)T_{m}^{(a-1)}(u)T_{m}^{(a+1)}(u) \nonumber \\ 
\hspace{-40pt}
&& \hspace{190pt} {\rm for } \qquad a \in \{1,2,\dots,s-2\}, 
\label{T-sys1}\\
\hspace{-40pt}
 && T_{m}^{(s-1)}(u-1)T_{m}^{(s-1)}(u+1)
=T_{m-1}^{(s-1)}(u)T_{m+1}^{(s-1)}(u) \nonumber \\ 
\hspace{-40pt}
&& \hspace{160pt} 
+g_{m}^{(s-1)}(u)T_{m}^{(s-2)}(u)T_{2m}^{(s)}(u) ,\\
\hspace{-40pt}
&& T_{2m}^{(s)}(u-1)T_{2m}^{(s)}(u+1)
=T_{2m-2}^{(s)}(u)T_{2m+2}^{(s)}(u)+
g_{2m}^{(s)}(u)T_{m}^{(s-1)}(u)T_{2m}^{(s)}(u), \label{T-sys3}
\end{eqnarray}
where $g_{m}^{(b)}(u)=
\{\prod_{j=1}^{m}{\cal T}_{0}(u+2j-m-1)\}^{\delta_{b,1}}$   
if $s\in {\bf Z}_{\ge 2}$; 
$g_{2m}^{(1)}(u)=\prod_{j=1}^{m}{\cal T}_{0}(u+2j-m-1)$ 
if $s=1$. 
Note that the function $g_{m}^{(b)}(u)$ obey the following relation 
\begin{eqnarray}
&& g_{m}^{(b)}(u+1)g_{m}^{(b)}(u-1)=
g_{m+1}^{(b)}(u)g_{m-1}^{(b)}(u) \quad {\rm if } \quad 
       s \in {\bf Z}_{\ge 2}, \nonumber \\
&& g_{2m}^{(1)}(u+1)g_{2m}^{(1)}(u-1)=
g_{2m+2}^{(1)}(u)g_{2m-2}^{(1)}(u) \quad {\rm if } \quad s=1.
\end{eqnarray} 
These functional relations (\ref{T-sys1})-(\ref{T-sys3}) 
will be $B(0|s)$ version of 
the $T$-system. Note that the subscript $n$ of 
$T_{n}^{(s)}(u)$ can take only even number (cf. (\ref{finite})). 
By construction, $T_{m}^{(a)}(u)$ can be expressed 
as a determinant over a 
matrix whose matrix elements are only the fundamental functions 
$T_{1}^{(1)}$, \dots, $T_{1}^{(s-1)}$, $T_{2}^{(s)}$
  and $g_{1}^{(1)}$ for $s\in {\bf Z}_{\ge 2}$; 
  $T_{2}^{(1)}$ and $g_{2}^{(1)}$ for $s=1$. 
  This can be summarized as follows:
\begin{theorem}
For $m\in {\bf Z}_{\ge 1}$, 
\begin{eqnarray}
&& T_{m}^{(a)}(u)= 
   {\rm det}_{1\le i,j \le m}({\cal T}_{a+i-j}(u+m-i-j+1)) 
\nonumber \\ && \hspace{150pt}
{\rm for} {\quad} a \in \{1,2, \dots ,s-1\}, \\
&& T_{2m}^{(s)}(u)=
   {\rm det}_{1\le i,j \le m}({\cal T}_{s+i-j}(u+m-i-j+1)) 
\end{eqnarray}
solves (\ref{T-sys1})-(\ref{T-sys3}). 
Here ${\cal T}_{a}(u)$ obeys the relation (\ref{dual1}) and 
the boundary condition 
\begin{eqnarray}
{\cal T}_{a}(u)=\left\{
\begin{array}{lll}
 0 &{\rm if} & a <0, \\ 
 g_{1}^{(1)}(u)& {\rm if} &   
   a=0 \quad {\rm and} \quad s \in {\bf Z}_{\ge 2}\\
 g_{2}^{(1)}(u)& {\rm if} & 
   a=0 \quad {\rm and} \quad s=1 \\ 
 T_{1}^{(a)}(u) & {\rm if} & a \in \{1,2, \dots ,s-1\} \\ 
 T_{2}^{(s)}(u) & {\rm if} & a=s ,
\end{array}
\right. 
\end{eqnarray}
where 
$g_{m}^{(a)}(u)=\{ \prod_{j=1}^{m}g_{1}^{(1)}
                         (u+2j-m-1)\}^{\delta_{a,1}}$ 
                         if  $s \in {\bf Z}_{\ge 2}$; 
$g_{2m}^{(1)}(u)=\prod_{j=1}^{m}g_{2}^{(1)}(u+2j-m-1)$ 
                         if  $ s=1 $. 
\end{theorem}
{\em Remark}: 
These functional realtions (\ref{T-sys1})-(\ref{T-sys3}) 
resemble to the ones for $A_{2s}^{(2)}$ \cite{KS2}. 
This resemblance will originate from resemblance between 
$B(0|s)^{(1)}$ and $A_{2s}^{(2)}$. 

There is a remarkable relation between 
the number ${\cal N}_{m}^{(a)}$ of the terms in 
$T_{m}^{(a)}(u)$ and the dimensionality (\ref{dim}) 
of the Lie superalgebra $B(0|s)$. 
 We conjecture that they are related each other as follows: 
\begin{eqnarray}
 {\cal N}_{m}^{(a)}&=& \sum 
  {\rm dim}V[k_{1},k_{2},\dots, k_{a},0,\dots,0]
   \quad {\rm if } \quad a \in \{1,2,\dots, s-1 \}, \nonumber \\ 
{\cal N}_{2m}^{(s)}&=& \sum {\rm dim}V[k_{1},k_{2},\dots, k_{s-1},2k_{s}]
 ,
\end{eqnarray}
where the summation is taken over non-negative 
 integers $\{ k_{j} \}$ 
such that $k_{1}+k_{2}+\cdots +k_{a} \le m $ and 
$k_{j} \equiv m \delta_{j a}$ mod $2$. 
\begin{table}
\begin{center}
 \begin{tabular}{ccccc} \hline 
 $m$                 & 1 & 2 & 3 & 4 \\ \hline 
 ${\cal N}_{m}^{(1)}$  & 5 & 15 & 35 & 70 \\  
 ${\cal N}_{2m}^{(2)}$ & 10 & 50 & 175 & 490 \\ \hline 
 \end{tabular} 
\end{center}
\caption{The number ${\cal N}_{m}^{(a)}$ of the terms 
 in $T_{m}^{(a)}(u)$ for $B(0|2)$.}
\label{num-osp14} 
\end{table}
\begin{table}
\begin{center}
 \begin{tabular}{cccc} \hline 
 $[b_{1},b_{2}]$ & ${\rm dim}V[b_{1},b_{2}]$ & 
 $[b_{1},b_{2}]$ & ${\rm dim}V[b_{1},b_{2}]$ \\ \hline
 0 0 & 1  & 0 2 & 10  \\ 
 1 0 & 5  & 0 4 & 35  \\ 
 2 0 & 14 & 0 6 & 84  \\ 
 3 0 & 30 & 2 2 & 81  \\ \hline 
  \end{tabular} 
\end{center}
\caption{The dimensionality of the module  
  $V[b_{1},b_{2}]$ for $B(0|2)$.}
\label{dim-osp14} 
\end{table}
For example, for $B(0|2)$ case, we have 
(cf. Table \ref{num-osp14} and Table \ref{dim-osp14}): 
\begin{eqnarray}
 {\cal N}_{1}^{(1)}&=& {\rm dim}V[1,0], \nonumber \\ 
 {\cal N}_{2}^{(1)}&=& {\rm dim}V[2,0]+{\rm dim}V[0,0], \nonumber \\ 
 {\cal N}_{3}^{(1)}&=& {\rm dim}V[3,0]+{\rm dim}V[1,0], \nonumber \\ 
 {\cal N}_{2}^{(2)}&=& {\rm dim}V[0,2], \nonumber \\ 
 {\cal N}_{4}^{(2)}&=& {\rm dim}V[0,4]+{\rm dim}V[2,0]
                              +{\rm dim}V[0,0], \nonumber \\ 
 {\cal N}_{6}^{(2)}&=& {\rm dim}V[0,6]+{\rm dim}V[2,2]
                              +{\rm dim}V[0,2].  
\end{eqnarray}
These relations seem to suggest a 
superization of the Kirillov-Reshetikhin formula 
\cite{KR}, which gives the multiplicity of occurrence of 
the irreducible representations of the Lie superalgebra 
in the Yangian module. 
\eqreset
\section{Summary and discussion}
In this paper, we have carried out an analytic Bethe ansatz 
 based on the Bethe ansatz equations (\ref{BAE})-(\ref{BAE4})
  with the distinguished
 simple root systems of the type II Lie superalgebras 
 $B(r|s)$ and $D(r|s)$. We have proposed 
 eigenvalue formulae of transfer matrices in 
 DVFs related to a class of tensor-like representations, 
 and shown their pole-freeness  under 
 the BAEs (\ref{BAE})-(\ref{BAE4}). 
  The key is the top term hypothesis 
   and the pole-freeness under the BAE.  
 A class of functional relations has been proposed for the DVFs. 
 In particular for $B(0|s)$ case, 
  remarkable duality among DVFs was found. 
 By using this, a complete set of functional relations 
 is written down for the DVFs labeled by rectangular Young 
 (super) diagrams.   
To the author' knowledge, this paper is the first 
 trial to construct {\em systematic } theory 
 of an analytic Bethe ansatz 
 related to fusion $B(r|s)$ and $D(r|s)$ vertex models. 
 
 In the present paper, we have executed an analytic Bethe ansatz 
 only for tensor-like representations.  
 As for spnorial representations, details are under investigation. 
 For example, in relation to the 
 64 dimensional typical representation of 
 $B(2|1)$, 
 we have confirmed the fact that the Bethe-strap generated by 
 the following top term
\footnote{Here we omit the vacuum part.}
\begin{eqnarray}
\frac{Q_{1}(u+\frac{5}{2})Q_{3}(u-\frac{1}{2})}
     {Q_{1}(u-\frac{5}{2})Q_{3}(u+\frac{1}{2})},
\end{eqnarray}
which carries $B(2|1)$ weight with the Kac-Dynkin label 
$(\frac{5}{2},0,1)$ consists of 64 terms. 

For $D(r|s)$ case, we have proposed DVFs labeled by 
   Young (super) diagrams with one row or one column. 
   It is tempting to extend these DVFs to 
   general Young (super) diagrams. 
   However, this will be a difficult task 
   since we lack of tableaux sum expressions of 
   DVFs labeled by general
    Young diagrams even for the non-superalgebra 
    $D_{r}$ case \cite{KS1}.   
    One way to bypass cumbersome tableaux sum expressions 
    is to construct a complete set of transfer matrix 
    functional relations (a hierarchy of $T$-system).  
    By solving it, we will be able to calculate DVFs.

It is an interesting problem to  derive TBA equations 
from our $T$-system (\ref{T-sys1})-(\ref{T-sys3}). 
This is accomplished by a similar procedure for 
 $sl(r|s)$ case \cite{JKS} (see also \cite{FK99}). 
  We are going to report this in 
 the near future.
\\
{\bf Acknowledgment} \\  
The author would like to thank Prof. A. Kuniba for encouragement. 
This work is supported by a Grant-in-Aid for 
JSPS Research   Fellows from the Ministry of Education, Science and 
Culture of Japan. 
\eqreset
\renewcommand{\theequation}{A.1.\arabic{equation}}
\section*{Appendix A.1  Outline of the proof of Theorem 
\ref{th-tate}: $B(r|s)$ ($r,s \in {\bf Z}_{\ge 1}$) case}
For simplicity, we assume that the vacuum parts 
are formally trivial from now on. 
We prove that ${\cal T}^{a}(u)$ 
is free of color $b$ poles, that is, 
$Res_{u=u_{k}^{(b)}+\cdots}{\cal T}^{a}(u)=0$ for any 
 $ b \in \{1,2,\dots, s+r \} $
 under the condition that the BAE (\ref{BAE}) is valid. 
 The function $\framebox{$c$}_{u}$  
 (\ref{z+}) with $c \in J $ has 
 color $b$ poles only for $c=b$, $b+1$, 
$\overline{b+1}$ or $\overline{b}$ if 
$b\in \{1,2,\dots,s+r-1 \} $; 
for $c=s+r$, $0$ or $\overline{s+r}$ if $b=s+r$, 
so we shall trace only 
\framebox{$b$}, \framebox{$b+1$}, 
\framebox{$\overline{b+1}$} or \framebox{$\overline{b}$}
 for $b\in\{1,2,\dots, s+r-1 \}$; 
\framebox{$s+r$}, \framebox{$0$} 
or \framebox{$\overline{s+r}$} for $b=s+r$.
 Let $S_{k}$ be the partial sum of ${\cal T}^{a}(u)$, 
 which contains 
$k$ boxes among \framebox{$b$}, \framebox{$b+1$}, 
\framebox{$\overline{b+1}$} or \framebox{$\overline{b}$}
 for $b \in \{1,2,\dots, s+r-1\}$; 
\framebox{$s+r$}, \framebox{$0$} 
or \framebox{$\overline{s+r}$} for $b=s+r$.
 Evidently, $S_{0}$ does not have 
color $b$ poles. 

 Now we examine $S_{1}$ which is a summation  
 of the tableaux (with sign) of the form 
\begin{equation} 
\begin{array}{|c|}\hline
    \xi \\ \hline 
    \eta   \\ \hline 
   \zeta \\ \hline
\end{array},
\end{equation}
where \framebox{$\xi$} and \framebox{$\zeta$} are columns with 
total length $a-1$ and they do not involve $Q_{b}$. 
\framebox{$\eta$} is \framebox{$b$}, \framebox{$b+1$}, 
\framebox{$\overline{b+1}$} or \framebox{$\overline{b}$}
 for $b \in \{1,2,\dots, s+r-1\}$; 
\framebox{$s+r$}, \framebox{$0$} 
or \framebox{$\overline{s+r}$} for $b=s+r$. 
 Thanks to the relations (\ref{res1})-(\ref{res8}), 
 $S_{1}$ is free of color $b$ poles under the BAE (\ref{BAE}). 
 Hereafter we consider $S_{k}$  for $k\ge 2$. \\ 
$\bullet$ 
The case
\footnote{This is void for $B(r|1)$ case.}
 $b \in \{1,2,\dots,s-1 \}$: 
$S_{k} (k\ge 2)$ is a  
summation of the tableaux (with sign) of the form 
\begin{eqnarray}
&& 
\sum_{n_{1}=0}^{k_{1}}
\sum_{n_{2}=0}^{k_{2}} 
\begin{array}{|c|}\hline 
    \xi \\ \hline
    E_{1n_{1}}    \\ \hline 
    \eta \\ \hline 
    E_{2n_{2}}  \\ \hline 
    \zeta \\ \hline 
\end{array}
 \nonumber \\ 
&& =
\left(
\sum_{n_{1}=0}^{k_{1}} 
\begin{array}{|c|}\hline 
    E_{1n_{1}}    \\ \hline 
\end{array}
\right)
\left(
\sum_{n_{2}=0}^{k_{2}}
\begin{array}{|c|}\hline 
    E_{2n_{2}}    \\ \hline 
\end{array}
\right)
\times 
\begin{array}{|c|}\hline 
    \xi \\ \hline 
\end{array}
\times 
\begin{array}{|c|}\hline 
    \eta \\ \hline 
\end{array}
\times 
\begin{array}{|c|}\hline 
    \zeta \\ \hline 
\end{array}
\label{tableauxk2}, 
\end{eqnarray} 
where \framebox{$\xi$}, \framebox{$\eta$} and 
\framebox{$\zeta$} are columns with total 
length $a-k$, which do not contain \framebox{$b$}, 
\framebox{$b+1$}, \framebox{$\overline{b+1}$} and 
\framebox{$\overline{b}$}; 
\framebox{$E_{1n_{1}}$} is a column 
\footnote{We assume that 
$\framebox{$E_{10}$}=
\begin{array}{|c|l}\cline{1-1} 
    b+1 & _{v} \\ \cline{1-1} 
   \vdots & \\ \cline{1-1} 
    b+1 & _{v-2k_{1}+2}\\ \cline{1-1} 
 \end{array}
$
 and  
 $\framebox{$E_{1 k_{1}}$}=
\begin{array}{|c|l}\cline{1-1} 
    b & _{v} \\ \cline{1-1} 
   \vdots & \\ \cline{1-1} 
    b & _{v-2k_{1}+2}\\ \cline{1-1} 
 \end{array}
$. \\ \ }
of the form: 
\begin{eqnarray}
 \begin{array}{|c|l}\cline{1-1} 
    b   & _v \\ \cline{1-1} 
    \vdots & \\ \cline{1-1} 
    b & _{v-2n_{1}+2}\\ \cline{1-1} 
    b+1 & _{v-2n_{1}} \\ \cline{1-1} 
   \vdots & \\ \cline{1-1} 
    b+1 & _{v-2k_{1}+2}\\ \cline{1-1} 
 \end{array} 
&=&\frac{Q_{b-1}(v-b+1-2n_{1})Q_{b}(v-b+2)}
      {Q_{b-1}(v-b+1)Q_{b}(v-b-2n_{1}+2)}  
\label{tableauxk3} \\ 
&& \times  
 \frac{Q_{b}(v-b-2k_{1})Q_{b+1}(v-b+1-2n_{1})}
      {Q_{b}(v-b-2n_{1})Q_{b+1}(v-b+1-2k_{1})},
\nonumber
\end{eqnarray}
where $v=u+h_{1}$: $h_{1}$ is some shift parameter 
and \framebox{$E_{2 n_{2}}$} is a column
\footnote{We assume that 
$\framebox{$E_{20}$}=
\begin{array}{|c|l}\cline{1-1} 
    \stackrel{\ }{\overline{b}} & _{w} \\ \cline{1-1} 
   \vdots & \\ \cline{1-1} 
    \stackrel{\ }{\overline{b}} & _{w-2k_{2}+2}\\ \cline{1-1} 
 \end{array}
$
 and  
 $\framebox{$E_{2 k_{2}}$}=
\begin{array}{|c|l}\cline{1-1} 
    \stackrel{\ }{\overline{b+1}} & _{w} \\ \cline{1-1} 
   \vdots & \\ \cline{1-1} 
    \stackrel{\ }{\overline{b+1}} 
    & _{w-2k_{2}+2}\\ \cline{1-1} 
 \end{array}
$.}
 of the form:
\begin{eqnarray}
 \begin{array}{|c|l}\cline{1-1} 
    \stackrel{\ }{\overline{b+1}} & _w \\ \cline{1-1} 
    \vdots & \\ \cline{1-1} 
    \stackrel{\ }{\overline{b+1}} & _{w-2n_{2}+2}\\ \cline{1-1} 
    \stackrel{\ }{\overline{b}} & _{w-2n_{2}} \\ \cline{1-1} 
   \vdots & \\ \cline{1-1} 
    \stackrel{\ }{\overline{b}} & _{w-2k_{2}+2}\\ \cline{1-1} 
 \end{array} 
&=&\frac{Q_{b-1}(w-2s+2r+b-2n_{2})}
      {Q_{b-1}(w-2s+2r+b-2k_{2})}  
\label{tableauxk5} \\ 
&& \times 
\frac{Q_{b}(w-2s+2r+b-2k_{2}-1)}{Q_{b}(w-2s+2r+b-2n_{2}-1)}
\nonumber \\
&& \times  
 \frac{Q_{b}(w-2s+2r+b+1)Q_{b+1}(w-2s+2r+b-2n_{2})}
 {Q_{b}(w-2s+2r+b-2n_{2}+1)Q_{b+1}(w-2s+2r+b)},
\nonumber
\end{eqnarray} 
where 
$w=u+h_{2}$: $h_{2}$ is some shift parameter; 
$k=k_{1}+k_{2}$
\footnote{We assume that $\framebox{$E_{1 n_{1}}$}=1$
 (resp. $\framebox{$E_{2 n_{2}}$}=1$) 
for $k_{1}=0$ (resp. $k_{2}=0$). In this case, 
\framebox{$E_{i n_{i}}$} does 
not have poles.}. 

For $b \in \{1,2,\dots, s-1\}$, 
\framebox{$E_{1 n_{1}}$} has color $b$ poles at
 $u=-h_{1}+b+2n_{1}+u_{p}^{(b)}$ and 
 $u=-h_{1}+b+2n_{1}-2+u_{p}^{(b)}$ 
  for $1 \le n_{1} \le k_{1}-1$; at  $u=-h_{1}+b+u_{p}^{(b)}$ 
for $n_{1}=0$ ; at $u=-h_{1}+b+2k_{1}-2+u_{p}^{(b)}$
 for $n_{1}=k_{1}$
 \footnote{
We  assume that these poles at 
$u=-h_{1}+b+2n_{1}+u_{i}^{(b)}$,
  and 
 $u=-h_{1}+b+2n_{1}-2+u_{q}^{(b)}$ 
 do not coincide each other 
 for any $i,q \in \{1,2,\dots , N_{b}\} $: 
 namely $u_{i}^{(b)}-u_{q}^{(b)} \ne 2$.}.  
The color $b$ residues at 
$u=-h_{1}+b+2n_{1}+u_{p}^{(b)}$  
 in \framebox{$E_{1 n_{1}}$} and \framebox{$E_{1 \> n_{1}+1}$}
 cancel each other under the BAE (\ref{BAE}). 
 Thus, under the BAE
  (\ref{BAE}), $\sum_{n_{1}=0}^{k_{1}}\framebox{$E_{1 n_{1}}$}$ 
 is free of color $b$ poles
 (see Figure \ref{part-bs}).
\begin{figure}
    \setlength{\unitlength}{1.5pt}
    \begin{center}
    \begin{picture}(270,40) 
     \put(0,0){$
       \framebox{$E_{1 0}$} 
       \stackrel{0}{\longleftarrow }
       \framebox{$E_{1 1}$} 
       \stackrel{2}{\longleftarrow } 
       \cdots 
       \stackrel{2n-4}{\longleftarrow }
       \framebox{$E_{1 n-1}$} 
       \stackrel{2n-2}{\longleftarrow }
       \framebox{$E_{1 n}$} 
       \stackrel{2n}{\longleftarrow } 
       \framebox{$E_{1 n+1}$}
       \stackrel{2n+2}{\longleftarrow } 
       \cdots 
       \stackrel{2k_{1}-2}{\longleftarrow }
       \framebox{$E_{1 k_{1}}$}
     $}  
    \end{picture}
  \end{center}
  \caption{Partial Bethe-strap structure of  
 $ E_{1 n}$ 
    for color $b$ poles ($1\le b \le s-1$):  
 The number $n$ on the arrow denotes the common color $b$ pole 
 $-h_{1}+b+n+u_{k}^{(b)}$ 
 of the pair of the tableaux connected by the arrow.   
 This common pole vanishes under the BAE (\ref{BAE}).}
  \label{part-bs}
\end{figure}
 
 \framebox{$E_{2 n_{2}}$} has color $b$ poles at
 $u=-h_{2}+2s-2r-b+2n_{2}-1+u_{p}^{(b)}$ and 
 $u=-h_{2}+2s-2r-b+2n_{2}+1+u_{p}^{(b)}$ 
  for $1 \le n_{2} \le k_{2}-1$; at  
  $u=-h_{2}+2s-2r-b+1+u_{p}^{(b)}$ 
for $n_{2}=0$ ; at 
$u=-h_{2}+2s-2r-b+2k_{2}-1+u_{p}^{(b)}$
 for $n_{2}=k_{2}$. 
 The color $b$ residues at 
$u=-h_{2}+2s-2r-b+2n_{2}+1+u_{p}^{(b)}$
 in \framebox{$E_{2 n_{2}}$} and \framebox{$E_{2, n_{2}+1}$}
 cancel each other under the BAE (\ref{BAE}). 
 Thus, under the BAE
  (\ref{BAE}), $\sum_{n_{2}=0}^{k_{2}}\framebox{$E_{2, n_{2}}$}$ 
 is free of color $b$ poles. 
  So is $S_{k}$. \\ 
$\bullet$ 
The case $ b=s $ : $S_{k} (k\ge 2)$ is a 
summation of the tableaux (with sign) of the form 
\begin{eqnarray}
&& \begin{array}{|c|}\hline 
    D_{11}    \\ \hline 
    \eta \\ \hline 
    D_{21}  \\ \hline
\end{array}
-\begin{array}{|c|}\hline 
    D_{11}    \\ \hline 
    \eta  \\ \hline 
    D_{22}  \\ \hline 
\end{array}
-\begin{array}{|c|}\hline
    D_{12}   \\ \hline 
    \eta  \\ \hline 
    D_{21}  \\ \hline
\end{array}
+\begin{array}{|c|}\hline 
    D_{12}  \\ \hline 
    \eta  \\ \hline 
    D_{22}  \\ \hline 
\end{array} \nonumber \\ 
&&=(
\begin{array}{|c|}\hline 
    D_{11}  \\ \hline 
\end{array}
-\begin{array}{|c|}\hline 
    D_{12}  \\ \hline 
\end{array}
)(
\begin{array}{|c|}\hline 
    D_{21}  \\ \hline 
\end{array}
-\begin{array}{|c|}\hline 
    D_{22}  \\ \hline 
\end{array}
)
\begin{array}{|c|}\hline 
    \eta \\ \hline 
\end{array}
\end{eqnarray} 
where \framebox{$\eta$} is a column with 
length $a-k$, which does not contain \framebox{$s$}, 
\framebox{$s+1$}, \framebox{$\overline{s+1}$} and 
\framebox{$\overline{s}$}; 
\framebox{$D_{11}$} is a column 
\footnote{We assume that 
$\framebox{$D_{11}$}=\framebox{$s+1$}_{v}$ if $k_{1}=1$.}
of the form: 
\begin{equation}
\begin{array}{|c|l}\cline{1-1}
    s & _v  \\ \cline{1-1} 
    \vdots & \\  \cline{1-1}
    s & _{v-2k_{1}+4} \\ \cline{1-1} 
    s+1 & _{v-2k_{1}+2} \\ \cline{1-1} 
\end{array}
= \frac{Q_{s-1}(v-s-2k_{1}+3)Q_{s}(v-s+2)Q_{s+1}(v-s-2k_{1}+1)}
       {Q_{s-1}(v-s+1)Q_{s}(v-s-2k_{1}+2)Q_{s+1}(v-s-2k_{1}+3)}; 
\label{tableauxk1-1}
\end{equation}
\framebox{$D_{12}$} is a column 
of the form:  
\begin{equation}
\begin{array}{|c|l}\cline{1-1}
    s & _v \\ \cline{1-1}  
   \vdots & \\ \cline{1-1} 
    s & _{v-2k_{1}+4}\\ \cline{1-1}
    s & _{v-2k_{1}+2}\\ \cline{1-1} 
\end{array}
=\frac{Q_{s-1}(v-s-2k_{1}+1)Q_{s}(v-s+2)}
       {Q_{s-1}(v-s+1)Q_{s}(v-s-2k_{1}+2)}
\label{tableauxk1-2}, 
\end{equation}
where $v=u+h_{1}$: $h_{1}$ is some shift parameter; 
\framebox{$D_{21}$} is a column
\footnote{We assume that 
$\framebox{$D_{21}$}=\framebox{$\overline{s+1}$}_{w}$
 if $k_{2}=1$.}
 of the form: 
\begin{eqnarray}
\begin{array}{|c|l}\cline{1-1}
    \stackrel{\ }{\overline{s+1}} & _w  \\ \cline{1-1} 
    \stackrel{\ }{\overline{s}} & _{w-2} \\ \cline{1-1}
    \vdots & \\  \cline{1-1} 
    \stackrel{\ }{\overline{s}} & _{w-2k_{2}+2} \\ \cline{1-1} 
\end{array}
&=&\frac{Q_{s-1}(w-s+2r-2)}{Q_{s-1}(w-s+2r-2k_{2})} 
\label{tableauxk1-3} \\ 
&& \times 
\frac{Q_{s}(w-s+2r-2k_{2}-1)Q_{s+1}(w-s+2r)}
     {Q_{s}(w-s+2r-1)Q_{s+1}(w-s+2r-2)}; 
\nonumber 
\end{eqnarray}
\framebox{$D_{22}$} is a column of the form:
\begin{equation}
\begin{array}{|c|l}\cline{1-1}
    \stackrel{\ }{\overline{s}} & _w \\ \cline{1-1} 
    \stackrel{\ }{\overline{s}} & _{w-2} \\ \cline{1-1}
    \vdots & \\ \cline{1-1} 
    \stackrel{\ }{\overline{s}} & _{w-2k_{2}+2}\\ \cline{1-1} 
\end{array}
=\frac{Q_{s-1}(w-s+2r)Q_{s}(w-s+2r-2k_{2}-1)}
      {Q_{s-1}(w-s+2r-2k_{2})Q_{s}(w-s+2r-1)} 
       \label{tableauxk1-4}, 
\end{equation}
where $w=u+h_{2}$: $h_{2}$ is some shift parameter; 
$k=k_{1}+k_{2}$
 \footnote{Here we discussed the case for 
 $k_{1}\ge 1 $ and $k_{2}\ge 1 $; 
 the case for $k_{1}=0 $ or $k_{2}=0 $ can be treated similarly.}. 
Obviously, the color $b=s$ residues at $v=s+2k_{1}-2+u_{j}^{(s)}$ 
 in (\ref{tableauxk1-1}) and 
 (\ref{tableauxk1-2}) cancel each other
 under the BAE (\ref{BAE}). 
And the color $b=s$ residues
 at $w=s-2r+1+u_{j}^{(s)}$ 
 in (\ref{tableauxk1-3}) and
 (\ref{tableauxk1-4}) cancel each other
 under the BAE (\ref{BAE}).  
 Thus $S_{k}$ does not 
 have color $s$ poles under the BAE (\ref{BAE}). \\ 
$\bullet$ 
The case
\footnote{This is void for $B(1|s)$ case.}
 $b \in \{s+1,s+2,\dots, s+r-1\}$ : 
Owing to the admissibility conditions, 
we have only to consider $S_{k}$ for 
$k=2,3,4$. 

$S_{2}$ is a 
summation of the tableaux (with sign) of the form 
\begin{eqnarray}
&& 
\begin{array}{|c|}\hline 
    \xi    \\ \hline
    b    \\ \hline 
    b+1 \\ \hline 
    \zeta^{\prime}  \\ \hline
\end{array}
+
\begin{array}{|c|}\hline 
    \xi^{\prime}    \\ \hline
    \stackrel{\ }{\overline{b+1}} \\ \hline
    \stackrel{\ }{\overline{b}}   \\ \hline
    \zeta  \\ \hline
\end{array}
\end{eqnarray}
and 
\begin{eqnarray}
&& 
\begin{array}{|c|}\hline 
    \xi    \\ \hline
    b    \\ \hline 
    \eta \\ \hline 
    \stackrel{\ }{\overline{b}} \\ \hline
    \zeta    \\ \hline
\end{array}
+
\begin{array}{|c|}\hline 
    \xi    \\ \hline
    b    \\ \hline 
    \eta \\ \hline 
    \stackrel{\ }{\overline{b+1}} \\ \hline
    \zeta    \\ \hline
\end{array}
+
\begin{array}{|c|}\hline 
    \xi    \\ \hline
    b+1    \\ \hline 
    \eta \\ \hline 
    \stackrel{\ }{\overline{b}} \\ \hline
    \zeta    \\ \hline
\end{array}
+
\begin{array}{|c|}\hline 
    \xi    \\ \hline
    b+1    \\ \hline 
    \eta \\ \hline 
    \stackrel{\ }{\overline{b+1}} \\ \hline
    \zeta    \\ \hline
\end{array}
 \nonumber \\ 
&&=
\begin{array}{|c|}\hline 
    \xi \\ \hline 
\end{array}
(
\begin{array}{|c|}\hline 
    b  \\ \hline 
\end{array}
+
\begin{array}{|c|}\hline 
    b+1  \\ \hline 
\end{array}
)
\begin{array}{|c|}\hline 
    \eta \\ \hline 
\end{array}
(
\begin{array}{|c|}\hline 
    \stackrel{\ }{\overline{b}}  \\ \hline 
\end{array}
+
\begin{array}{|c|}\hline 
    \stackrel{\ }{\overline{b+1}} \\ \hline 
\end{array}
)
\begin{array}{|c|}\hline 
    \zeta \\ \hline 
\end{array},
\end{eqnarray} 
where 
\{ \framebox{$\xi$}, \framebox{$\eta$}, \framebox{$\zeta$} \}, 
\{ \framebox{$\xi$}, \framebox{$\zeta^{\prime}$} \} and 
\{ \framebox{$\xi^{\prime}$}, \framebox{$\zeta$} \}
 are  columns with total 
length $a-2$, which do not contain \framebox{$b$}, 
\framebox{$b+1$}, \framebox{$\overline{b+1}$} and 
\framebox{$\overline{b}$}.  
Thus, owing to Lemma \ref{le-tate},
the relations (\ref{res3}) and (\ref{res6}), $S_{2}$ does not 
 have color $b$ poles under the BAE (\ref{BAE}). 
 
$S_{3}$ is a 
summation of the tableaux (with sign) of the form 
\begin{eqnarray}
&& 
\begin{array}{|c|}\hline 
    \xi    \\ \hline
    b    \\ \hline 
    \eta \\ \hline 
    \stackrel{\ }{\overline{b+1}} \\ \hline
    \stackrel{\ }{\overline{b}} \\ \hline
    \zeta    \\ \hline
\end{array}
+
\begin{array}{|c|}\hline 
    \xi    \\ \hline
    b+1    \\ \hline 
    \eta \\ \hline 
    \stackrel{\ }{\overline{b+1}} \\ \hline
    \stackrel{\ }{\overline{b}} \\ \hline
    \zeta    \\ \hline
\end{array}
+
\begin{array}{|c|}\hline 
    \xi    \\ \hline
    b    \\ \hline
    b+1    \\ \hline 
    \eta^{\prime} \\ \hline 
    \stackrel{\ }{\overline{b}} \\ \hline
    \zeta    \\ \hline
\end{array}
+
\begin{array}{|c|}\hline 
    \xi    \\ \hline
    b    \\ \hline
    b+1    \\ \hline 
    \eta^{\prime} \\ \hline 
    \stackrel{\ }{\overline{b+1}} \\ \hline
    \zeta    \\ \hline
\end{array}
,
\end{eqnarray} 
where \{ \framebox{$\xi$}, \framebox{$\eta$}, \framebox{$\zeta$} \} 
 and \{ \framebox{$\xi$}, \framebox{$\eta^{\prime }$},
  \framebox{$\zeta$} \} 
are  columns with 
total length $a-3$, which do not contain \framebox{$b$}, 
\framebox{$b+1$}, \framebox{$\overline{b+1}$} and 
\framebox{$\overline{b}$}.
Thus, owing to the relations (\ref{res3}), (\ref{res6}) and 
Lemma \ref{le-tate}, $S_{3}$ does not 
 have color $b$ poles under the BAE (\ref{BAE}). 

$S_{4}$ is a 
summation of the tableaux (with sign) of the form 
\begin{eqnarray}
&& 
\begin{array}{|c|}\hline 
    \xi    \\ \hline
    b    \\ \hline 
    b+1    \\ \hline
    \eta \\ \hline 
    \stackrel{\ }{\overline{b+1}} \\ \hline
    \stackrel{\ }{\overline{b}} \\ \hline
    \zeta    \\ \hline
\end{array}
,
\end{eqnarray} 
where \framebox{$\xi$}, \framebox{$\eta$} and 
\framebox{$\zeta$} are  columns with 
total length $a-4$, which do not contain \framebox{$b$}, 
\framebox{$b+1$}, \framebox{$\overline{b+1}$} and 
\framebox{$\overline{b}$}.
Thus, owing to the 
Lemma \ref{le-tate}, $S_{4}$ does not 
 have color $b$ poles under the BAE (\ref{BAE}).
 \\ 
$\bullet$ 
The case $b=s+r$ :  
$S_{k}$ ($k\ge 2$) is a 
summation of the tableaux (with sign) of the form 
\begin{eqnarray}
\begin{array}{|c|l}\cline{1-1} 
    \xi &   \\ \cline{1-1}
    0  & _{v}  \\ \cline{1-1} 
    0  & _{v-2} \\ \cline{1-1}
    \vdots &  \\ \cline{1-1}
    0  & _{v-2k+4} \\ \cline{1-1}
    0 & _{v-2k+2} \\ \cline{1-1}
    \zeta   & \\ \cline{1-1}
\end{array}
+
\begin{array}{|c|l}\cline{1-1} 
    \xi &   \\ \cline{1-1}
    s+r  & _{v}  \\ \cline{1-1} 
    0  & _{v-2} \\ \cline{1-1}
    \vdots &  \\ \cline{1-1}
    0  & _{v-2k+4} \\ \cline{1-1}
    0 & _{v-2k+2} \\ \cline{1-1}
    \zeta   & \\ \cline{1-1}
\end{array}
&+&
\begin{array}{|c|l}\cline{1-1} 
    \xi &   \\ \cline{1-1}
    0  & _{v}  \\ \cline{1-1} 
    0  & _{v-2} \\ \cline{1-1}
    \vdots &  \\ \cline{1-1}
    0  & _{v-2k+4} \\ \cline{1-1}
    \stackrel{\ }{\overline{s+r}} & _{v-2k+2} \\ \cline{1-1}
    \zeta   & \\ \cline{1-1}
\end{array}
+
\begin{array}{|c|l}\cline{1-1} 
    \xi &   \\ \cline{1-1}
    s+r  & _{v}  \\ \cline{1-1} 
    0  & _{v-2} \\ \cline{1-1}
    \vdots &  \\ \cline{1-1}
    0  & _{v-2k+4} \\ \cline{1-1}
    \stackrel{\ }{\overline{s+r}} & _{v-2k+2} \\ \cline{1-1}
    \zeta  &  \\ \cline{1-1}
\end{array}
 \nonumber \\ 
&& =
A(v)B(v) \times \framebox{$\xi$} \times \framebox{$\eta$},
\end{eqnarray} 
where $v=u+h_{3}$: $h_{3}$ is some shift parameter; 
\framebox{$\xi$} and 
\framebox{$\zeta$} are columns with total 
length $a-k$, which do not contain \framebox{$s+r$}, 
\framebox{$0$} and 
\framebox{$\overline{s+r}$};
\begin{eqnarray}
 A(v)&=&\frac{Q_{s+r}(v-s+r+1)}{Q_{s+r}(v-s+r)} 
  \nonumber \\ 
 && +\frac{Q_{s+r-1}(v-s+r+1)Q_{s+r}(v-s+r-1)}
       {Q_{s+r-1}(v-s+r-1)Q_{s+r}(v-s+r)}, \\ 
B(v)&=& \frac{Q_{s+r}(v-s+r-2k)}{Q_{s+r}(v-s+r-2k+1)} 
\nonumber \\ 
 && +\frac{Q_{s+r-1}(v-s+r-2k)Q_{s+r}(v-s+r-2k+2)}
       {Q_{s+r-1}(v-s+r-2k+2)Q_{s+r}(v-s+r-2k+1)}.\nonumber 
\end{eqnarray}
One can check $A(v)$ and $B(v)$ are free of color $s+r$
 poles under the BAE (\ref{BAE}). 
 Thus,  $S_{k}$ does not 
 have color $s+r$ poles under the BAE (\ref{BAE}). 
 \rule{5pt}{10pt} \\ 
 {\em Remark}: There is another proof for
  Theorem \ref{th-tate} by the determinant formula 
  (\ref{Jacobi-Trudi2}): 
  for $b \in \{ 1,2,\dots, s-1 \}$, we prove 
  ${\cal T}_{m}(u)$ is free of color $b$ poles, 
  and then the pole-freeness of ${\cal T}^{a}(u)$ follows 
  from (\ref{Jacobi-Trudi2}); while 
  for $b \in \{ s,s+1, \dots, s+r \}$, we prove 
  ${\cal T}^{a}(u)$ is free of color $b$ poles in the 
  same way as the above-mentioned proof.  
  An advantage of this another proof 
  is that we do not encounter an 
  awkward expression like (\ref{tableauxk2}). 
  We note that similar idea is also applicable for 
  $sl(r|s)$ case \cite{T1,T2,T3}. 
\eqreset
\renewcommand{\theequation}{A.2.\arabic{equation}}
\section*{Appendix A.2  Outline of the proof of Theorem 
\ref{th-tate}: $B(0|s)$ ($s \in {\bf Z}_{\ge 1}$) case}
 We will show that ${\cal T}_{m}(u)$ 
is free of color $b$ poles, namely, 
$Res_{u=u_{k}^{(b)}+\cdots}{\cal T}_{m}(u)=0$ for any 
 $ b \in \{1,2,\dots, s \} $
 under the condition that the BAE 
 (\ref{BAE1})-(\ref{BAE4}) is valid. 
 The function $\framebox{$c$}_{u}$  
 (\ref{z+}) with $c \in J $ has 
 color $b$ poles only for $c=b$, $b+1$, 
$\overline{b+1}$ or $\overline{b}$ if 
$b\in \{1,2,\dots,s-1 \} $; 
for $c=s$, $0$ or $\overline{s}$ if $b=s$, 
so we shall trace only 
\framebox{$b$}, \framebox{$b+1$}, 
\framebox{$\overline{b+1}$} or \framebox{$\overline{b}$}
 for $b\in\{1,2,\dots, s-1 \}$; 
\framebox{$s$}, \framebox{$0$} 
or \framebox{$\overline{s}$} for $b=s$.
 Let $S_{k}$ be the partial sum of ${\cal T}_{m}(u)$, 
 which contains 
$k$ boxes among \framebox{$b$}, \framebox{$b+1$}, 
\framebox{$\overline{b+1}$} or \framebox{$\overline{b}$}
 for $b\in\{1,2,\dots, s-1 \}$; 
\framebox{$s$}, \framebox{$0$} 
or \framebox{$\overline{s}$} for $b=s$. 
 Apparently, $S_{0}$ does not have 
color $b$ poles. 

 Now we examine $S_{1}$, which is a summation  
 of the tableaux (with sign) of the form 
\begin{equation} 
\begin{array}{|c|c|c|}\hline
    \xi &  \eta & \zeta 
    \\ \hline 
\end{array}
\end{equation}
where \framebox{$\xi$} and \framebox{$\zeta$} are rows with 
total length $m-1$ and they do not involve $Q_{b}$. 
\framebox{$\eta$} is \framebox{$b$}, \framebox{$b+1$}, 
\framebox{$\overline{b+1}$} or \framebox{$\overline{b}$}
 for $b\in\{1,2,\dots, s-1 \}$; 
\framebox{$s$}, \framebox{$0$} 
or \framebox{$\overline{s}$} for $b=s$. 
 Owing to the relations (\ref{res1-b0s})-(\ref{res4-b0s}), 
 $S_{1}$ is free of color $b$ poles under the BAEs 
 (\ref{BAE1})-(\ref{BAE4}). 
 From now on, we consider $S_{k}$  for $k\ge 2$. \\ 
$\bullet$ 
The case
\footnote{This is void for $B(0|1)$ case.}
 $b\in\{1,2,\dots, s-1 \}$: Owing to the admissibility
 conditions, 
we have only to consider $S_{2}$, $S_{3}$ and $S_{4}$. 

$S_{2}$ is a  
summation of the tableaux (with sign) of the form 
\begin{eqnarray}
 \begin{array}{|c|c|c|c|}\hline 
   \xi & b & b+1 & \eta^{\prime} \\
    \hline 
 \end{array}
 +
 \begin{array}{|c|c|c|c|}\hline 
   \xi^{\prime} & \stackrel{\ }{\overline{b+1}} & 
               \overline{b} & \zeta
     \\ \hline 
 \end{array}
\end{eqnarray}
and 
\begin{eqnarray}
 && 
 \begin{array}{|c|c|c|c|c|}\hline 
   \xi & b & \eta & \stackrel{\ }{\overline{b}} & \zeta
    \\ \hline 
 \end{array}
+
\begin{array}{|c|c|c|c|c|}\hline 
   \xi & b & \eta & \stackrel{\ }{\overline{b+1}} & \zeta
    \\ \hline 
 \end{array} 
 \nonumber \\ 
&& +
 \begin{array}{|c|c|c|c|c|}\hline 
   \xi & b+1 & \eta & \stackrel{\ }{\overline{b}} & \zeta
    \\ \hline 
 \end{array}
+
 \begin{array}{|c|c|c|c|c|}\hline 
   \xi & b+1 & \eta & \stackrel{\ }{\overline{b+1}} & \zeta
    \\ \hline 
 \end{array} 
 \nonumber \\ 
&& =
 \begin{array}{|c|}\hline 
   \xi \\ \hline 
 \end{array} 
 (
 \begin{array}{|c|}\hline 
   b \\ \hline 
 \end{array}
 +
 \begin{array}{|c|}\hline 
   b+1 \\ \hline 
 \end{array}
 )
 \begin{array}{|c|}\hline 
   \eta \\ \hline 
 \end{array}
 (
 \begin{array}{|c|}\hline 
   \stackrel{\ }{\overline{b}} \\ \hline 
 \end{array}
 +
 \begin{array}{|c|}\hline 
   \stackrel{\ }{\overline{b+1}} \\ \hline 
 \end{array}
 )
 \begin{array}{|c|}\hline 
   \zeta \\ \hline 
 \end{array}
\end{eqnarray}
where (\framebox{$\xi$}, \framebox{$\eta^{\prime}$}),
(\framebox{$\xi^{\prime}$}, \framebox{$\zeta$}) and 
(\framebox{$\xi$}, \framebox{$\eta$}, \framebox{$\zeta$}) 
 are rows with total 
length $m-2$, which do not contain \framebox{$b$}, 
\framebox{$b+1$}, \framebox{$\overline{b+1}$} and 
\framebox{$\overline{b}$}. 
Thus, owing to Lemma \ref{le-yoko},
the relations (\ref{res1-b0s}) and (\ref{res4-b0s}),
 $S_{2}$ does not 
 have color $b$ poles under the BAE (\ref{BAE1}) and (\ref{BAE2}). 

 $S_{3}$ is a  
summation of the tableaux (with sign) of the form 
\begin{eqnarray}
 && 
 \begin{array}{|c|c|c|c|c|c|}\hline 
   \xi & b & b+1 & \eta & \stackrel{\ }{\overline{b}} & \zeta
    \\ \hline 
 \end{array}
+
\begin{array}{|c|c|c|c|c|c|}\hline 
   \xi & b & b+1 & \eta & \stackrel{\ }{\overline{b+1}} & \zeta
    \\ \hline 
 \end{array} 
 \nonumber \\ 
&& =
 \begin{array}{|c|c|c|c|}\hline 
   \xi & b & b+1 & \eta \\ \hline 
 \end{array} 
 (
 \begin{array}{|c|}\hline 
   \stackrel{\ }{\overline{b}} \\ \hline 
 \end{array}
 +
 \begin{array}{|c|}\hline 
   \stackrel{\ }{\overline{b+1}} \\ \hline 
 \end{array}
 )
 \begin{array}{|c|}\hline 
   \zeta \\ \hline 
 \end{array}
\end{eqnarray}
and 
\begin{eqnarray}
 && 
 \begin{array}{|c|c|c|c|c|c|}\hline 
   \xi & b & \eta^{\prime} & \stackrel{\ }{\overline{b+1}} 
     & \overline{b} & \zeta
    \\ \hline 
 \end{array}
+
\begin{array}{|c|c|c|c|c|c|}\hline 
   \xi & b+1 & \eta^{\prime} & \stackrel{\ }{\overline{b+1}}
      & \overline{b} & \zeta
    \\ \hline 
 \end{array} 
 \nonumber \\ 
&& =
 \begin{array}{|c|}\hline 
   \xi  \\ \hline 
 \end{array} 
 (
 \begin{array}{|c|}\hline 
   b \\ \hline 
 \end{array}
 +
 \begin{array}{|c|}\hline 
   b+1 \\ \hline 
 \end{array}
 )
 \begin{array}{|c|c|c|c|}\hline 
   \eta^{\prime} & \stackrel{\ }{\overline{b+1}} 
     & \overline{b} & \zeta \\ \hline 
 \end{array}
\end{eqnarray}
where 
(\framebox{$\xi $}, \framebox{$\eta$}, \framebox{$\zeta$}) and 
(\framebox{$\xi$}, \framebox{$\eta^{\prime}$}, \framebox{$\zeta$}) 
 are rows with total 
length $m-3$, which do not contain \framebox{$b$}, 
\framebox{$b+1$}, \framebox{$\overline{b+1}$} and 
\framebox{$\overline{b}$}. 
Thus, owing to Lemma \ref{le-yoko},
the relations (\ref{res1-b0s}) and (\ref{res4-b0s}), $S_{3}$ does not 
 have color $b$ poles under the BAE (\ref{BAE1}) and (\ref{BAE2}).

$S_{4}$ is a  
summation of the tableaux (with sign) of the form 
\begin{eqnarray}
 && 
 \begin{array}{|c|c|c|c|c|c|c|}\hline 
   \xi & b & b+1 & \eta & \stackrel{\ }{\overline{b+1}} 
     & \overline{b} & \zeta
    \\ \hline 
 \end{array}
\end{eqnarray}
where 
 \framebox{$\xi$}, \framebox{$\eta$} and \framebox{$\zeta$} 
 are rows with total 
length $m-4$, which do not contain \framebox{$b$}, 
\framebox{$b+1$}, \framebox{$\overline{b+1}$} and 
\framebox{$\overline{b}$}. 
Thus, owing to Lemma \ref{le-yoko}, $S_{4}$ does not 
 have color $b$ poles. 
 
$\bullet$ The case $ b=s $ : 
Owing to the admissibility
 conditions, 
we have only to consider $S_{2}$ and $S_{3}$. 

$S_{2}$ is a 
summation of the tableaux (with sign) of the form 
\begin{eqnarray}
 \hspace{-20pt} && 
 \begin{array}{|c|c|c|c|}\hline 
   \xi & s & 0 & \eta 
    \\ \hline 
 \end{array}
 =\frac{Q_{s-1}(v-s-1)Q_{s}(v-s+3)}{Q_{s-1}(v-s+1)Q_{s}(v-s+1)}
 \times 
 \begin{array}{|c|}\hline 
    \xi 
    \\ \hline 
 \end{array}
 \times 
 \begin{array}{|c|}\hline 
    \eta 
    \\ \hline 
 \end{array},
 \\ 
 \hspace{-20pt} && 
 \begin{array}{|c|c|c|c|}\hline 
   \xi & s & \stackrel{\ }{\overline{s}} & \eta 
    \\ \hline 
 \end{array}
 =\frac{Q_{s-1}(v-s-1)Q_{s}(v-s+2)}{Q_{s-1}(v-s+1)Q_{s}(v-s)}
  \nonumber \\ 
 \hspace{-20pt}&& \hspace{70pt} \times 
\frac{Q_{s-1}(v-s+2)Q_{s}(v-s-1)}{Q_{s-1}(v-s)Q_{s}(v-s+1)}
\times 
 \begin{array}{|c|}\hline 
    \xi 
    \\ \hline 
 \end{array}
 \times 
 \begin{array}{|c|}\hline 
    \eta 
    \\ \hline 
 \end{array}, \\
 \hspace{-20pt}  && 
 \begin{array}{|c|c|c|c|}\hline 
   \xi & 0 & \stackrel{\ }{\overline{s}} & \eta 
    \\ \hline 
 \end{array}
 =\frac{Q_{s-1}(v-s+2)Q_{s}(v-s-2)}{Q_{s-1}(v-s)Q_{s}(v-s)}
 \times 
 \begin{array}{|c|}\hline 
    \xi 
    \\ \hline 
 \end{array}
 \times 
 \begin{array}{|c|}\hline 
    \eta 
    \\ \hline 
 \end{array}, 
\end{eqnarray}
where 
 \framebox{$\xi$} and \framebox{$\eta$}  
 are rows with total length $m-2$, which do not contain  
\framebox{$s$}, \framebox{$0$} and 
\framebox{$\overline{s}$}; $v=u+h$: $h$ is a shift parameter. 
 The color $s$ residues at $u=u_{k}^{(s)}+s-1-h$ of 
 the functions 
$
\begin{array}{|c|c|c|c|}\hline 
   \xi & s & 0 & \eta 
    \\ \hline 
 \end{array}
$ and 
$
\begin{array}{|c|c|c|c|}\hline 
   \xi & s & \stackrel{\ }{\overline{s}} & \eta 
    \\ \hline 
 \end{array}
$ 
cancel each other under the BAE (\ref{BAE3}) or (\ref{BAE4}).
The color $s$ residues at $u=u_{k}^{(s)}+s-h$ of 
 the functions 
$
\begin{array}{|c|c|c|c|}\hline 
   \xi & s & \stackrel{\ }{\overline{s}} & \eta 
    \\ \hline 
 \end{array}
$ 
and 
$
 \begin{array}{|c|c|c|c|}\hline 
   \xi & 0 & \stackrel{\ }{\overline{s}} & \eta 
    \\ \hline 
 \end{array}
$
cancel each other under the BAE (\ref{BAE3}) or (\ref{BAE4}). 
Thus, $S_{2}$ does not 
 have color $s$ poles under the BAE (\ref{BAE3}) or (\ref{BAE4}).
 
$S_{3}$ is a  
summation of the tableaux (with sign) of the form 
\begin{eqnarray}
 \begin{array}{|c|c|c|c|c|}\hline 
   \xi & s & 0 & \stackrel{\ }{\overline{s}} & \eta 
    \\ \hline 
 \end{array}
\end{eqnarray}
where 
 \framebox{$\xi$} and \framebox{$\eta$}  
 are rows with total 
length $m-3$, which do not contain  
\framebox{$s$}, \framebox{$0$} and 
\framebox{$\overline{s}$}. 
Thus, owing to Lemma \ref{le-yoko}, $S_{3}$ does not 
 have color $s$ poles. 
 
Then $ {\cal T}_{m}(u)$ is free of poles under 
the condition that the BAEs 
(\ref{BAE1})-(\ref{BAE4}) are valid; 
owing to the relation (\ref{Jacobi-Trudi2}), 
this also hold true for 
$ {\cal T}_{\lambda \subset \mu}(u)$. In particular, 
 the pole-freeness of ${\cal T}^{a}(u)$ follows 
 immediately.
\eqreset
\renewcommand{\theequation}{A.3.\arabic{equation}}
\section*{Appendix A.3  Outline of the proof of Theorem 
\ref{th-tate}: $D(r|s)$ 
($r \in {\bf Z}_{\ge 2}$, $s \in {\bf Z}_{\ge 1}$) case}
We prove that ${\cal T}^{a}(u)$ 
is free of color $b$ poles, that is, 
$Res_{u=u_{k}^{(b)}+\cdots}{\cal T}^{a}(u)=0$ for any 
 $ b \in \{1,2,\dots, s+r \} $
 under the condition that the BAE (\ref{BAE}) is valid. 
 The function $\framebox{$c$}_{u}$  
 (\ref{z++}) with $c \in J $ has 
 color $b$ poles only for $c=b$, $b+1$, 
$\overline{b+1}$ or $\overline{b}$ if 
$b\in \{1,2,\dots,s+r-1 \} $; 
for $c=s+r-1$, $s+r$, $\overline{s+r}$ or $\overline{s+r-1}$ 
if $b=s+r$, 
so we shall trace only 
\framebox{$b$}, \framebox{$b+1$}, 
\framebox{$\overline{b+1}$} or \framebox{$\overline{b}$}
 for $b\in\{1,2,\dots, s+r-1 \}$; 
 \framebox{$s+r-1$}, \framebox{$s+r$}, \framebox{$\overline{s+r}$} 
 or \framebox{$\overline{s+r-1}$} 
 for $b=s+r$.
 Let $S_{k}$ be the partial sum of ${\cal T}^{a}(u)$, 
 which contains 
$k$ boxes among \framebox{$b$}, \framebox{$b+1$}, 
\framebox{$\overline{b+1}$} or \framebox{$\overline{b}$}
 for $b\in \{1,2,\dots,s+r-1 \} $; 
\framebox{$s+r-1$}, \framebox{$s+r$}, \framebox{$\overline{s+r}$}
 or \framebox{$\overline{s+r-1}$} 
for $b=s+r$.
 Apparently, $S_{0}$ does not have 
color $b$ poles. 

 Next we consider $S_{1}$, which is a summation  
 of the tableaux (with sign) of the form 
\begin{equation} 
\begin{array}{|c|}\hline
    \xi \\ \hline 
    \eta   \\ \hline 
   \zeta \\ \hline
\end{array},
\end{equation}
where \framebox{$\xi$} and \framebox{$\zeta$} are columns with 
total length $a-1$ and they do not contain $Q_{b}$. 
\framebox{$\eta$} is \framebox{$b$}, \framebox{$b+1$}, 
\framebox{$\overline{b+1}$} or \framebox{$\overline{b}$}
 for $b\in \{1,2,\dots,s+r-1 \} $; 
\framebox{$s+r-1$}, \framebox{$s+r$}, \framebox{$\overline{s+r}$} 
 or \framebox{$\overline{s+r-1}$} for $b=s+r$. 
 Owing to the relations (\ref{res1-d})-(\ref{res8-d}), 
 $S_{1}$ is free of color $b$ poles under the BAE (\ref{BAE}). 
 From now on we consider $S_{k}$  for $k\ge 2$. \\ 
$\bullet$ 
The case $b\in \{1,2,\dots,s+r-2 \} $: 
The proof is similar to $B(r|s)$ ($r\in {\bf Z}_{\ge 1}$) case, 
so we omit it. \\ 
$\bullet$ 
The case $b=s+r-1$ or $b=s+r$:  
$S_{2n}$ ($k=2n$, $n \in {\bf Z}_{\ge 2}$ 
\footnote{The case $n=1$ can be treated similarly.}) is a 
summation of the tableaux (with signs) of the form 
\begin{eqnarray}
&&
\begin{array}{|c|}\hline 
    \xi  \\ \hline
    s+r-1  \\ \hline 
    \stackrel{\ }{\overline{s+r}} \\ \hline 
    s+r  \\ \hline
    \vdots \\  \hline
    \stackrel{\ }{\overline{s+r}} \\ \hline
    s+r  \\ \hline
    \stackrel{\ }{\overline{s+r-1}}  \\ \hline
    \zeta  \\ \hline
\end{array}
+
\begin{array}{|c|}\hline 
    \xi \\ \hline
    s+r-1  \\ \hline 
    \stackrel{\ }{\overline{s+r}} \\ \hline 
    s+r  \\ \hline
    \vdots \\ \hline
    \stackrel{\ }{\overline{s+r}} \\ \hline
    s+r  \\ \hline
    \stackrel{\ }{\overline{s+r}}  \\ \hline
    \zeta  \\ \hline
\end{array}
+
\begin{array}{|c|}\hline 
    \xi    \\ \hline
    s+r   \\ \hline
    \stackrel{\ }{\overline{s+r}} \\ \hline 
    s+r   \\ \hline
    \vdots \\ \hline
    \stackrel{\ }{\overline{s+r}} \\ \hline
    s+r \\ \hline
    \stackrel{\ }{\overline{s+r-1}} \\ \hline
    \zeta  \\ \hline
\end{array}
+
\begin{array}{|c|l}\cline{1-1} 
    \xi & \\ \cline{1-1}
    s+r & _{v} \\ \cline{1-1} 
    \stackrel{\ }{\overline{s+r}}  & _{v-2}  \\ \cline{1-1} 
    s+r  & _{v-4} \\ \cline{1-1}
    \vdots &  \\ \cline{1-1}
    \stackrel{\ }{\overline{s+r}}  & _{v-4n+6} \\ \cline{1-1}
    s+r & _{v-4n+4} \\ \cline{1-1}
    \stackrel{\ }{\overline{s+r}} & _{v-4n+2} \\ \cline{1-1}
    \zeta  &  \\ \cline{1-1}
\end{array}
\nonumber \\ 
&&=
A(v)B(v) \times \framebox{$\xi$} \times \framebox{$\zeta$}
\label{even-nul}
\end{eqnarray} 
and 
\begin{eqnarray}
&&
\begin{array}{|c|}\hline 
    \xi  \\ \hline
    \stackrel{\ }{\overline{s+r}}  \\ \hline 
    s+r  \\ \hline
    \stackrel{\ }{\overline{s+r}} \\ \hline 
    s+r  \\ \hline
    \vdots \\  \hline
    \stackrel{\ }{\overline{s+r}} \\ \hline
    s+r  \\ \hline
    \stackrel{\ }{\overline{s+r}}  \\ \hline
    s+r  \\ \hline
    \zeta  \\ \hline
\end{array}
+
\begin{array}{|c|}\hline 
    \xi \\ \hline
    s+r-1  \\ \hline 
    s+r    \\ \hline
    \stackrel{\ }{\overline{s+r}} \\ \hline 
    s+r  \\ \hline
    \vdots \\ \hline
    \stackrel{\ }{\overline{s+r}} \\ \hline
    s+r  \\ \hline
    \stackrel{\ }{\overline{s+r}}  \\ \hline
    s+r  \\ \hline
    \zeta  \\ \hline
\end{array}
+
\begin{array}{|c|}\hline 
    \xi    \\ \hline
    \stackrel{\ }{\overline{s+r}} \\ \hline 
    s+r   \\ \hline
    \stackrel{\ }{\overline{s+r}} \\ \hline 
    s+r   \\ \hline
    \vdots \\ \hline
    \stackrel{\ }{\overline{s+r}} \\ \hline
    s+r \\ \hline
    \stackrel{\ }{\overline{s+r}} \\ \hline
    \stackrel{\ }{\overline{s+r-1}} \\ \hline
    \zeta  \\ \hline
\end{array}
+
\begin{array}{|c|l}\cline{1-1} 
    \xi & \\ \cline{1-1}
    s+r-1 & _{v} \\ \cline{1-1} 
    s+r  & _{v-2} \\ \cline{1-1}
    \stackrel{\ }{\overline{s+r}} & _{v-4}  \\ \cline{1-1} 
    s+r  & _{v-6} \\ \cline{1-1}
    \vdots &  \\ \cline{1-1}
    \stackrel{\ }{\overline{s+r}}  & _{v-4n+8} \\ \cline{1-1}
    s+r & _{v-4n+6} \\ \cline{1-1}
    \stackrel{\ }{\overline{s+r}}  & _{v-4n+4} \\ \cline{1-1}
    \stackrel{\ }{\overline{s+r-1}} & _{v-4n+2} \\ \cline{1-1}
    \zeta  &  \\ \cline{1-1}
\end{array}
\nonumber \\ 
&&=
C(v)D(v) \times \framebox{$\xi$} \times \framebox{$\zeta$},
\end{eqnarray} 
where $v=u+h_{1}$: $h_{1}$ is some shift parameter; 
\framebox{$\xi$} and 
\framebox{$\zeta$} are columns with total 
length $a-2n$, which do not contain \framebox{$s+r-1$}, 
\framebox{$s+r$}, \framebox{$\overline{s+r}$} and 
\framebox{$\overline{s+r-1}$};
\begin{eqnarray}
A(v)&=&\frac{Q_{s+r-1}(v-s+r+1)}{Q_{s+r-1}(v-s+r-1)} 
  \nonumber \\ 
 && +\frac{Q_{s+r-2}(v-s+r)Q_{s+r-1}(v-s+r-3)}
       {Q_{s+r-2}(v-s+r-2)Q_{s+r-1}(v-s+r-1)}, \nonumber \\ 
B(v)&=& \frac{Q_{s+r-1}(v-s+r-4n-1)}{Q_{s+r-1}(v-s+r-4n+1)} 
\nonumber \\ 
 && +\frac{Q_{s+r-2}(v-s+r-4n)Q_{s+r-1}(v-s+r-4n+3)}
          {Q_{s+r-2}(v-s+r-4n+2)Q_{s+r-1}(v-s+r-4n+1)},
          \nonumber \\ 
 C(v)&=&\frac{Q_{s+r}(v-s+r+1)}{Q_{s+r}(v-s+r-1)} 
  \nonumber \\ 
 && +\frac{Q_{s+r-2}(v-s+r)Q_{s+r}(v-s+r-3)}
       {Q_{s+r-2}(v-s+r-2)Q_{s+r}(v-s+r-1)}, \\ 
D(v)&=& \frac{Q_{s+r}(v-s+r-4n-1)}{Q_{s+r}(v-s+r-4n+1)} 
\nonumber \\ 
 && +\frac{Q_{s+r-2}(v-s+r-4n)Q_{s+r}(v-s+r-4n+3)}
          {Q_{s+r-2}(v-s+r-4n+2)Q_{s+r}(v-s+r-4n+1)}.
          \nonumber 
\end{eqnarray} 
Apparently, $A(v)$ and $B(v)$ (resp. $C(v)$ and $D(v)$)  
do not contain $Q_{s+r}$ (resp. $Q_{s+r-1}$). 
One can also check $A(v)$ and $B(v)$ 
(resp. $C(v)$ and $D(v)$) are free of color $s+r-1$ 
(resp. $s+r$) poles under the BAE (\ref{BAE}).  

$S_{2n+1}$ ($k=2n+1$, $n \in {\bf Z}_{\ge 2}$ 
\footnote{The case $n=1$ can be treated similarly.}) is a 
summation of the tableaux (with signs) of the form 
\begin{eqnarray}
&&
\begin{array}{|c|}\hline 
    \xi  \\ \hline
    s+r \\ \hline  
    \stackrel{\ }{\overline{s+r}} \\ \hline 
    s+r  \\ \hline
    \vdots \\  \hline
    \stackrel{\ }{\overline{s+r}} \\ \hline
    s+r  \\ \hline
    \stackrel{\ }{\overline{s+r}}  \\ \hline
    s+r  \\ \hline
    \zeta  \\ \hline
\end{array}
+
\begin{array}{|c|}\hline 
    \xi \\ \hline
    s+r  \\ \hline 
    \stackrel{\ }{\overline{s+r}} \\ \hline 
    s+r  \\ \hline
    \vdots \\ \hline
    \stackrel{\ }{\overline{s+r}} \\ \hline
    s+r  \\ \hline
    \stackrel{\ }{\overline{s+r}}  \\ \hline
    \stackrel{\ }{\overline{s+r-1}}  \\ \hline
    \zeta  \\ \hline
\end{array}
+
\begin{array}{|c|}\hline 
    \xi    \\ \hline
    s+r-1   \\ \hline
    \stackrel{\ }{\overline{s+r}} \\ \hline 
    s+r   \\ \hline
    \vdots \\ \hline
    \stackrel{\ }{\overline{s+r}} \\ \hline
    s+r \\ \hline
    \stackrel{\ }{\overline{s+r}} \\ \hline
    s+r \\ \hline 
    \zeta  \\ \hline
\end{array}
+
\begin{array}{|c|l}\cline{1-1} 
    \xi & \\ \cline{1-1}
    s+r-1 & _{v} \\ \cline{1-1} 
    \stackrel{\ }{\overline{s+r}} & _{v-2}  \\ \cline{1-1} 
    s+r  & _{v-4} \\ \cline{1-1}
    \vdots &  \\ \cline{1-1}
    \stackrel{\ }{\overline{s+r}} & _{v-4n+6} \\ \cline{1-1}
    s+r & _{v-4n+4} \\ \cline{1-1}
    \stackrel{\ }{\overline{s+r}} & _{v-4n+2} \\ \cline{1-1}
    \stackrel{\ }{\overline{s+r-1}} & _{v-4n} \\ \cline{1-1}
    \zeta  &  \\ \cline{1-1}
\end{array}
\nonumber \\ 
&&=
E(v)F(v) \times \framebox{$\xi$} \times \framebox{$\zeta$}
\end{eqnarray}
and 
\begin{eqnarray}
&&
\begin{array}{|c|}\hline 
    \xi  \\ \hline 
    \stackrel{\ }{\overline{s+r}} \\ \hline 
    s+r \\ \hline  
    \stackrel{\ }{\overline{s+r}} \\ \hline 
    s+r  \\ \hline
    \vdots \\  \hline
    \stackrel{\ }{\overline{s+r}} \\ \hline
    s+r  \\ \hline
    \stackrel{\ }{\overline{s+r}} \\ \hline
    \zeta  \\ \hline
\end{array}
+
\begin{array}{|c|}\hline 
    \xi \\ \hline
    s+r-1  \\ \hline 
    s+r  \\ \hline 
    \stackrel{\ }{\overline{s+r}} \\ \hline 
    s+r  \\ \hline
    \vdots \\ \hline
    \stackrel{\ }{\overline{s+r}} \\ \hline
    s+r  \\ \hline
    \stackrel{\ }{\overline{s+r}}  \\ \hline
    \zeta  \\ \hline
\end{array}
+
\begin{array}{|c|}\hline 
    \xi    \\ \hline
    \stackrel{\ }{\overline{s+r}} \\ \hline 
    s+r   \\ \hline
    \stackrel{\ }{\overline{s+r}} \\ \hline 
    s+r   \\ \hline
    \vdots \\ \hline
    \stackrel{\ }{\overline{s+r}} \\ \hline
    s+r \\ \hline
    \stackrel{\ }{\overline{s+r-1}} \\ \hline 
    \zeta  \\ \hline
\end{array}
+
\begin{array}{|c|l}\cline{1-1} 
    \xi & \\ \cline{1-1}
    s+r-1 & _{v} \\ \cline{1-1} 
    s+r & _{v-2} \\ \cline{1-1} 
    \stackrel{\ }{\overline{s+r}} & _{v-4}  \\ \cline{1-1} 
    s+r  & _{v-6} \\ \cline{1-1}
    \vdots &  \\ \cline{1-1}
    \stackrel{\ }{\overline{s+r}} & _{v-4n+4} \\ \cline{1-1}
    s+r & _{v-4n+2} \\ \cline{1-1}
    \stackrel{\ }{\overline{s+r-1}} & _{v-4n} \\ \cline{1-1}
    \zeta  &  \\ \cline{1-1}
\end{array}
\nonumber \\ 
&&=
G(v)H(v) \times \framebox{$\xi$} \times \framebox{$\zeta$}
\end{eqnarray}
where $v=u+h_{2}$: $h_{2}$ is some shift parameter; 
\framebox{$\xi$} and 
\framebox{$\zeta$} are columns with total 
length $a-2n-1$, which do not contain \framebox{$s+r-1$}, 
\framebox{$s+r$}, \framebox{$\overline{s+r}$} and 
\framebox{$\overline{s+r-1}$};
\begin{eqnarray}
E(v)&=&\frac{Q_{s+r-1}(v-s+r+1)}{Q_{s+r-1}(v-s+r-1)} 
  \nonumber \\ 
 && +\frac{Q_{s+r-2}(v-s+r)Q_{s+r-1}(v-s+r-3)}
       {Q_{s+r-2}(v-s+r-2)Q_{s+r-1}(v-s+r-1)}, \nonumber \\ 
F(v)&=& \frac{Q_{s+r}(v-s+r-4n-3)}{Q_{s+r}(v-s+r-4n-1)} 
\nonumber \\ 
 && +\frac{Q_{s+r-2}(v-s+r-4n-2)Q_{s+r}(v-s+r-4n+1)}
          {Q_{s+r-2}(v-s+r-4n)Q_{s+r}(v-s+r-4n-1)},
          \nonumber \\ 
 G(v)&=&\frac{Q_{s+r}(v-s+r+1)}{Q_{s+r}(v-s+r-1)} 
  \nonumber \\ 
 && +\frac{Q_{s+r-2}(v-s+r)Q_{s+r}(v-s+r-3)}
       {Q_{s+r-2}(v-s+r-2)Q_{s+r}(v-s+r-1)}, \\ 
H(v)&=& \frac{Q_{s+r-1}(v-s+r-4n-3)}{Q_{s+r-1}(v-s+r-4n-1)} 
\nonumber \\ 
 && +\frac{Q_{s+r-2}(v-s+r-4n-2)Q_{s+r-1}(v-s+r-4n+1)}
          {Q_{s+r-2}(v-s+r-4n)Q_{s+r-1}(v-s+r-4n-1)}.
          \nonumber 
\end{eqnarray} 
Apparently, $E(v)$ and $H(v)$ (resp. $F(v)$ and $G(v)$)  
do not contain $Q_{s+r}$ (resp. $Q_{s+r-1}$). 
One can also check $E(v)$ and $H(v)$ 
(resp. $F(v)$ and $G(v)$) are free of color $s+r-1$ 
(resp. $s+r$) poles under the BAE (\ref{BAE}). 

 Thus,  $S_{k}$ have neither 
 color $s+r-1$ poles nor 
 color $s+r$ poles under the BAE (\ref{BAE}). 
\eqreset
\renewcommand{\theequation}{B.\arabic{equation}}
\section*{Appendix B Generating series for 
${\cal T}^{a}(u)$ and ${\cal T}_{m}(u)$}
The functions ${\cal T}^{a}(u)$ 
and ${\cal T}_{m}(u)$ 
 ($a,m \in {\bf Z }$; $u \in {\bf C }$) 
 are determined by the following 
 non-commutative generating series.  \\ 
$B(r|s)$ case: 
\begin{eqnarray}
\hspace{-30pt} &&   (1+\framebox{$\overline{1}$}X)^{-1}\cdots 
 (1+\framebox{$\overline{s}$}X)^{-1} (1+\framebox{$\overline{s+1}$}X)
\cdots (1+\framebox{$\overline{s+r}$}X)(1-\framebox{$0$}X)^{-1}
  \nonumber \\
\hspace{-30pt} && \times  (1+\framebox{$s+r$}X) \cdots (1+\framebox{$s+1$}X)
 (1+\framebox{$s$}X)^{-1} \cdots (1+\framebox{$1$}X)^{-1}
   \nonumber \\        
\hspace{-30pt}  && \hspace{40pt} =\sum_{a=-\infty}^{\infty} 
         {\cal T}^{a}(u+a-1)X^{a},
   \end{eqnarray}
\begin{eqnarray}
\hspace{-30pt} &&   (1-\framebox{$1$}X)\cdots 
 (1-\framebox{$s$}X) (1-\framebox{$s+1$}X)^{-1}
\cdots (1-\framebox{$s+r$}X)^{-1}(1+\framebox{$0$}X)
  \nonumber \\
\hspace{-30pt} && \times  (1-\framebox{$\overline{s+r}$}X)^{-1} 
\cdots (1-\framebox{$\overline{s+1}$}X)^{-1}
 (1-\framebox{$\overline{s}$}X) \cdots (1-\framebox{$\overline{1}$}X)
   \nonumber \\        
\hspace{-30pt}  && \hspace{40pt} =\sum_{m=-\infty}^{\infty} 
         {\cal T}_{m}(u+m-1)X^{m}.
\end{eqnarray}   
$D(r|s)$ case: 
\begin{eqnarray}
\hspace{-30pt} &&   (1+\framebox{$\overline{1}$}X)^{-1}\cdots 
 (1+\framebox{$\overline{s}$}X)^{-1} (1+\framebox{$\overline{s+1}$}X)
\cdots (1+\framebox{$\overline{s+r}$}X)
\nonumber \\ 
\hspace{-30pt} && \times 
(1-\framebox{$s+r$}X\framebox{$\overline{s+r}$}X)^{-1}
  \nonumber \\
\hspace{-30pt} && \times  (1+\framebox{$s+r$}X) \cdots (1+\framebox{$s+1$}X)
 (1+\framebox{$s$}X)^{-1} \cdots (1+\framebox{$1$}X)^{-1}
   \nonumber \\        
\hspace{-30pt}  && \hspace{40pt} =\sum_{a=-\infty}^{\infty} 
         {\cal T}^{a}(u+a-1)X^{a},
\end{eqnarray}
\begin{eqnarray}
\hspace{-30pt} &&   (1-\framebox{$1$}X)\cdots 
 (1-\framebox{$s$}X) (1-\framebox{$s+1$}X)^{-1}
\cdots (1-\framebox{$s+r-1$}X)^{-1} \nonumber \\ 
\hspace{-30pt} && \times 
[(1-\framebox{$s+r$}X)^{-1}+
(1-\framebox{$\overline{s+r}$}X)^{-1}-1]
  \nonumber \\
\hspace{-30pt} && \times  (1-\framebox{$\overline{s+r-1}$}X)^{-1} 
\cdots (1-\framebox{$\overline{s+1}$}X)^{-1}
 (1-\framebox{$\overline{s}$}X) \cdots (1-\framebox{$\overline{1}$}X)
   \nonumber \\        
\hspace{-30pt}  && \hspace{40pt} =\sum_{m=-\infty}^{\infty} 
         {\cal T}_{m}(u+m-1)X^{m}.
\end{eqnarray}  
Here $X$ is a shift operator $X=e^{2\partial_{u}}$. 
In particular,  we have ${\cal T}^{0}(u)=1$; 
${\cal T}_{0}(u)=1$; ${\cal T}^{a}(u)=0$ for $a<0$; 
 ${\cal T}_{m}(u)=0$ for $m<0$. 
              

\newpage
\begin{thebibliography}{80}
        \bibitem{Ka}
        Kac V 1977 {\it Adv. Math. } {\bf 26} 8

        \bibitem{KulSk}
        Kulish P P and Sklyanin E K 1982 
        {\it J. Sov. Math} {\bf 19} 1596
        
        \bibitem{Kul}
        Kulish P P 1986 {\it J. Sov. Math.} {\bf 35} 2648 
        
        \bibitem{BS}
        Bazhanov V V and Shadrikov A G 1987  
        {\it Theor. Math. Phys.} {\bf 73} 1302
        
        \bibitem{KuR}
        Kulish P P and Reshetikhin N Yu 1989 
        {\it Lett. Math. Phys.} {\bf 18} 143 
        
        \bibitem{DFI}
        Deguchi T, Fujii A and Ito K 1990  
        {\it Phys. Lett.} B {\bf 238} 242
        
        \bibitem{Sa} Saleur H 1990 {\it Nucl. Phys.} B 
        {\bf 336} 363
        
        \bibitem{ZBG}
        Zhang R B, Bracken A J and Gould M D 1991 
        {\it Phys. Lett. B} {\bf 257} 133 
        
        \bibitem{MR94}
        Martins M J and Ramos P B 1994 
        {\it J. Phys. A: Math. Gen.} {\bf 27} L703 
        
        \bibitem{EK}
        Essler F H L and Korepin V E 1992 
       {\it Phys. Rev. } {\bf 46} B 9147 
       
        \bibitem{EKS}
        Essler F H L, Korepin V E and Schoutens K 1992  
        {\it Phys. Rev. Lett. } {\bf 68} 2960
           
        \bibitem{FK}
        Foerster A and Karowski M 1993 
        {\it Nucl. Phys.} B {\bf 396} 611            
      
        \bibitem{Ma}
        Maassarani Z 1995 
        {\it J. Phys. A: Math. Gen.} {\bf 28} 1305 
        
        \bibitem{RM}
        Ramos P B and Martins M J 1996 
        {\it Nucl. Phys.} B {\bf 474} 678   
         
        \bibitem{PF}
        Pfannm\"{u}ller M P and Frahm H 1996  
        {\it Nucl. Phys.} B {\bf 479} 575    
        
        \bibitem{MR}
        Martins M J and Ramos P B 1997 {\it Nucl. Phys.} B 
        {\bf 500} 579. 
            
        \bibitem{ZB}
        Zhou Y K and Batchelor M T 1997 {\it Nucl. Phys.} 
        B {\bf 490} 576 

        \bibitem{KRS}
        Kulish P P, Reshetikhin N Yu and Sklyanin E K
        1981 {\it Lett. Math. Phys.} {\bf 5} 393 

        \bibitem{T1}
        Tsuboi Z 1997
        {\it J. Phys. A: Math. Gen. } {\bf 30} 7975 
        
        \bibitem{T2}
        Tsuboi Z 1998 {\it Physica } A {\bf 252 } 565
         
        \bibitem{T3}
        Tsuboi Z 1998 
        {\it J. Phys. A: Math. Gen. } {\bf 31} 5485
        
        \bibitem{T4}
        Tsuboi Z  1999 
        {\it Physica } A {\bf 267} 173 

        \bibitem{R1}
        Reshetikhin N Yu 1983 {\it Sov. Phys. JETP} {\bf 57} 691  
        
        \bibitem{R2}
        Reshetikhin N Yu 1987 {\it Lett. Math. Phys.} {\bf 14} 235 
                
        \bibitem{BR}
        Bazhanov V V and Reshetikhin N Yu 1990  
        {\it J. Phys. A Math. Gen. } {\bf 23 }1477  
        
        \bibitem{KS1}
        Kuniba A and Suzuki J 1995 {\it Commun. Math. Phys.} 
        {\bf 173} 225        
         
        \bibitem{KOS}
        Kuniba A, Ohta Y and Suzuki J 1995 
        {\it J. Phys. A: Math. Gen.} {\bf 28} 6211  
        
        \bibitem{S2}
        Suzuki J 1994 {\it Phys. Lett. } A {\bf 195} 190  
        
        \bibitem{KS2}
        Kuniba A and Suzuki J 1995 
        {\it J. Phys. A: Math. Gen.} {\bf 28} 711
                              
%
%
%
        \bibitem{RW}
        Reshetikhin N Yu and Wiegmann P B 1987 
        {\it Phys. Lett. } B {\bf 189} 125 
        
        \bibitem{OW}
        Ogievetsky E I and Wiegmann P B 1986
        {\it  Phys. Lett.} B {\bf 168} 360  
                                          
        \bibitem{KNS1}
        Kuniba A, Nakanishi T and Suzuki J 1994 
         {\it Int. J. Mod. Phys.} A {\bf 9} 5215                                                                     
        \bibitem{KP}
        Kl\"{u}mper A and Pearce P 1992 
        {\it Physica } A {\bf 183 } 304 
                        
        \bibitem{KNS2}
        Kuniba A, Nakanishi T and Suzuki J 1994 
         {\it Int. J. Mod. Phys.} A {\bf 9} 5267 
         
        \bibitem{KNH}
        Kuniba A, Nakamura S and Hirota R 1996 
         {\it J. Phys. A: Math. Gen.} {\bf 29} 1759    
         
        \bibitem{TK}
        Tsuboi Z and Kuniba A 1996 
        {\it J. Phys. A Math. Gen.} {\bf 29} 7785    
        
        \bibitem{KLWZ} 
        Krichever I, Lipan O, Wiegmann P and Zabrodin A 1997 
         {\it Commun. Math. Phys. } {\bf 188} 267
        
        \bibitem{T5}
        Tsuboi Z 1997 
        {\it J. Phys. Soc. Jpn.} {\bf 66} 3391
                  
        \bibitem{Ka2}
        Kac V 1978 {\it Lecture Notes in Mathematics } 
        {\bf 676} 597 
        
        \bibitem{BB}
        Balantekin A B and  Bars I 1981 
         {\it J. Math. Phys.} {\bf 22} 1149 
        
        \bibitem{FJ}
        Farmer R J and Jarvis P D 1984
         {\it J. Phys. A: Math. Gen.} {\bf 17} 2365 
         
        \bibitem{MSS}
        Morel B, Sciarrino A and Sorba P 
        1985 {\it J. Phys. A: Math. Gen.} {\bf 18} 1597
         
        \bibitem{Je}
        Van der Jeugt J 1996 {\it J. Math. Phys.} {\bf 37} 4176. 
        
        \bibitem{D}
        Drinfel'd V G 1988 {\it Sov. Math. Dokl} {\bf 36} 212  
                
        \bibitem{Y1}
        Yamane H 1994  {\it Publ. RIMS, Kyoto Univ.} {\bf 30} 15 
        
        \bibitem{Y2}
        Yamane H 1996 {\it Preprent} q-alg/9603015 
        
        \bibitem{K2}
        Kuniba A 1998 {\it Preprint} 
        {\it Bethe ansatz and difference equations} 
         in Japanese 

        \bibitem{KN1}
        Kashiwara M and Nakashima T 1994 
         {\it J. Algebra} {\bf  165} 295
        
        \bibitem{KN2}
        Nakashima T 1993 
        {\it Commun. Math. Phys.} {\bf  154} 215
        
        \bibitem{FR}
        Frenkel E and Reshetikhin N 1996 
         {\it Commun. Math. Phys.} {\bf 178} 237 
            
        \bibitem{H}
        Hirota R 1981 {\it J. Phys. Soc. Jpn.} 
        {\bf 50} 3785     
        
        \bibitem{DM}
        Deguchi T and Martin P P 1992 {\it Int. J. Mod. Phys. A} 
        {\bf 7} {\it Suppl. 1A} 165

        \bibitem{MRi}
        Martin P and Rittenberg V 1992 
        {\it Int. J. Mod. Phys. A} {\bf 7} 
        {\it Suppl. 1B} 707
         
        \bibitem{KR} 
        Kirillov A N and Reshetikhin N Yu  
        1990 {\it J. Sov. Math.} {\bf 52} 3156.                 
                                 
        \bibitem{JKS}
        J\"{u}ttner G, Kl\"{u}mper A and Suzuki J 1998 
        {\it Nucl. Phys.} B {\bf 512} 581
        
        \bibitem{FK99}
        Fujii A and Kl\"{u}mper A 1999 
        {\it Nucl. Phys.} B {\bf 546} 751
         
 \end{thebibliography}
\end{document}